\documentclass[smallextended]{svjour3}

\usepackage[export]{adjustbox}
\usepackage[utf8]{inputenc}
\usepackage{mathpazo}
\usepackage{amsmath}
\usepackage[round]{natbib}
\usepackage[margin=1in]{geometry}
\usepackage{amssymb}
\usepackage{standalone}
\usepackage{graphicx}
\usepackage{overpic}
\usepackage{epstopdf}

\usepackage[normalem]{ulem}
\usepackage[rgb,dvipsnames]{xcolor}
\usepackage{tikz}
\usepackage{bm}
\usepackage[caption=false]{subfig}
\usepackage{booktabs}
\usepackage[math]{cellspace}
\AtBeginDocument{
    \heavyrulewidth=.08em
    \lightrulewidth=.05em
    \cmidrulewidth=.03em
    \belowrulesep=.65ex
    \belowbottomsep=0pt
    \aboverulesep=.4ex
    \abovetopsep=0pt
    \cmidrulesep=\doublerulesep
    \cmidrulekern=.5em
    \defaultaddspace=.5em
}

\definecolor{MyBlue}{rgb}  {0.1,0.1,0.9}
\definecolor{MyRed}{rgb}   {0.9,0.1,0.1}
\definecolor{MyGreen}{rgb} {0.05,0.4,0.05}
\definecolor{burntorange}{rgb}{0.8, 0.33, 0.0}
\definecolor{NeilMagenta}{rgb}{0.8, 0.1, 0.8}

\newcommand{\ak}[1]{{#1}}

\renewcommand{\vec}[1]{\bm{#1}}
\newcommand{\abs}[1]{\left | {#1} \right |}
\newcommand{\diff}[2]{\frac{\mathrm{d} #1}{\mathrm{d} #2}}
\newcommand{\pdiff}[2]{\frac{\partial #1}{\partial #2}}
\newcommand{\pddiff}[2]{\frac{\partial^2 #1}{\partial #2^2}}
\newcommand{\dt}[1]{\pdiff{#1}{t}}
\newcommand{\lapbel}{\nabla^2_{\Omega(t)}}
\newcommand{\grad}{\nabla_{\Omega(t)}}
\newcommand{\intd}{\mathop{}\!\mathrm{d}}

\newcommand \beq{\begin{eqnarray}}
\newcommand \eeq{\end{eqnarray}}
\newcommand \beqno{\begin{eqnarray*}}
\newcommand \eeqno{\end{eqnarray*}}
\newcommand \bit{\begin{itemize}}
\newcommand \eit{\end{itemize}}

\begin{document}
\title{Concentration-Dependent Domain Evolution in Reaction-Diffusion Systems}
\titlerunning{Concentration-Dependent Domain Evolution in Reaction-Diffusion Systems}
\author{Andrew L. Krause\footnote{Corresponding author \email{andrew.krause@durham.ac.uk}} \and Eamonn A. Gaffney \and Benjamin J. Walker}
\authorrunning{A. L. Krause \and E. A. Gaffney \and B. J. Walker}

\institute{
    A. L. Krause 
    \at Mathematical Sciences Department, Durham University, Upper Mountjoy Campus, Stockton Rd, Durham DH1 3LE, United Kingdom \\
    E. A. Gaffney
    \at Wolfson Centre for Mathematical Biology, Mathematical Institute, University of Oxford, Andrew Wiles Building, Radcliffe Observatory Quarter, Woodstock Road, Oxford, OX2 6GG, United Kingdom \\
    B. J. Walker
    \at Department of Mathematics, University College London, London, WC1H 0AY, United Kingdom
    }

\date{Received: date / Accepted: date}

\maketitle

\begin{abstract}
Pattern formation has been extensively studied in the context of evolving (time-dependent) domains in recent years, with domain growth implicated in ameliorating problems of pattern robustness and selection, in addition to more realistic modelling in developmental biology. Most work to date has considered prescribed domains evolving as given functions of time, but not  the scenario of concentration-dependent dynamics, which is also highly  relevant in a developmental setting. Here, we study such concentration-dependent domain evolution for reaction-diffusion systems to elucidate fundamental aspects of these more complex models. We pose a general form of one-dimensional domain evolution, and extend this to $N$-dimensional manifolds under mild constitutive assumptions in lieu of developing a full tissue-mechanical model. In the 1D case, we are able to extend linear stability analysis around homogeneous equilibria, though this is of limited utility in understanding complex pattern dynamics in fast growth regimes. We numerically demonstrate a variety of dynamical behaviours in 1D and 2D planar geometries, giving rise to several new phenomena, especially near regimes of critical bifurcation boundaries such as peak-splitting instabilities. For sufficiently fast growth and contraction, concentration-dependence can have an enormous impact on the nonlinear dynamics of the system both qualitatively and quantitatively. We highlight crucial differences between 1D evolution and higher dimensional models, explaining obstructions for linear analysis and underscoring the importance of careful constitutive choices in defining domain evolution in higher dimensions. We raise important questions in the modelling and analysis of biological systems, in addition to numerous mathematical questions that appear tractable in the one-dimensional setting, but are vastly more difficult for higher-dimensional models.

\end{abstract}
 

\maketitle

\section{Introduction}

In proposing his chemical theory of morphogenesis, Turing was clear about the simplifications made to idealize this theory of pattern formation to its core mechanism of a diffusion-driven instability \citep{turing1952chemical}. In addition to the enormous experimental and theoretical literature exploring this mechanism, an important avenue of research has been extending Turing's simple theory to ever-more-realistic scenarios incorporating extensions of reaction-diffusion models. For example, there is now a wide literature studying pattern formation in stochastic Turing systems \citep{woolley2011stochastic, erban2019stochastic, adamer2020coloured}, mechanical and mechano-chemically coupled models \citep{murray1984cell,murray1984generation, oster1985model,murray2004mathematical, vaughan2013modified}, gene-expression time delays in reaction-diffusion systems \citep{gaffney2006gene, seirin2010influence, sargood2022fixed}, cross-diffusion and other generalized transport mechanisms \citep{ritchie_hyperbolic_2020}, reaction-diffusion patterning on manifolds and networks \citep{plaza2004effect, mccullen2016pattern,ide2016turing, krause2019influence}, larger numbers of morphogens \citep{diego2018key, scholes2019comprehensive},  and a host of other generalizations that explore Turing's basic insight regarding diffusion-driven pattern formation in increasingly complicated settings; see \citep{krause_near_2021} for a broad review. Such extensions relax assumptions that Turing originally made, and add nuance to the core idea of a diffusion-driven instability leading to pattern formation.

One such assumption is the idea that reaction-diffusion processes `pre-pattern' morphogens, which then influence cells downstream to induce changes in cell fate and, hence, spatial organisation of tissue structure. The role of spatial heterogeneity in reaction-diffusion systems has been explored extensively in recent years \citep{maini1995hierarchical, page2003pattern, page2005complex, green2015positional, krause2018heterogeneity, krause_WKB}, plausibly capturing hierarchical pattern formation observed experimentally. There is an inherent separation of timescales needed to justify such an approach to heirarchical pattern formation, in turn motivating studies where this assumption is relaxed with instead  tissue restructuring and morphogen dynamics occurring concomitantly on  similar timescales  in development.

 Further related to timescales of reaction-diffusion signalling and tissue restructuring is the role of domain growth, which has also been heavily studied \citep{crampin1999reaction, crampin2002mode, crampin2002pattern, plaza2004effect, van_gorder_growth_2019}. It is likely that domain growth and restructuring are not downstream of morphogen patterning in many cases, but concomitant processes instead \citep{boehm2010role}. One major insight from these models incorporating domain growth is some amelioration of robustness problems inherent to Turing-type patterning, whereby marginally different initial conditions may evolve to quantitatively different numbers of pattern elements \citep{maini2012turing}. \cite{crampin1999reaction} showed that domain growth could instead lead to a predictable sequence of spike-doubling in 1D reaction-diffusion models, and \cite{ueda2012mathematical} and others have confirmed that this spike-doubling phenomena occurs in generic \ak{1D} systems  in certain circumstances, such as when the growth is sufficiently slow. While these studies have given important insights into how growth impacts pattern formation, they all consider cases where the domain evolution is explicitly prescribed, and hence cannot account for the feedback between signalling and domain evolution. 

There is also a robust chemical setting exploring reaction-diffusion patterning in analogues of growing domains using the photosensitive CDIMA reaction and related chemical systems; see \citep{konow2021insights} for a recent review.   \cite{liu2022control} explored a model of this reaction system in terms of a `wave of competence' to pattern formation, and made comparisons to its use as a model of a growing domain. One key point raised in the work by \cite{liu2022control} was that the boundary conditions in such a model likely do not correspond to a moving boundary with simple Neumann or Dirichlet conditions, but possibly something more intricate involving the current concentration on either side of the system. 

Following the original formulation of local domain growth given by \cite{crampin1999reaction}, several authors have considered models of concentration-dependent growth  \citep{dillon1999mathematical, neville2006interactions, baker2007mechanism, seirin2011dynamics}. These studies highlighted a number of important issues in modelling the feedback between domain growth and signalling dynamics, notably the importance of dilution of the domain impacting the structure of pattern elements, as well as the importance of careful constitutive choices made in understanding the complex interplay between domain growth and pattern formation. In most of these studies, the focus  was primarily on the slow-growth regime that was qualitatively comparable to prescribed growth scenarios, with only small apparent impacts of concentration-dependence on the overall domain evolution in comparison to prescribed-growth scenarios. \cite{neville2006interactions}, following work by \cite{ward1997mathematical}, did explore faster growth regimes, implicating dilution effects in impacting concentration profiles and, hence, leading to more complex interactions. We also mention that \citet{seirin2011dynamics} explored the impact of gene-expression time delays in concentration-dependent growth dynamics, which had a nontrivial impact on the timescales and ability for systems to admit spatial patterns. 

There is also a large literature developing numerical methods for these kinds of PDE models. \ak{In more than one spatial dimension in particular, there are several different choices of numerical approach for growing domains including moving finite-element meshes \citep{barreira2011surface, dziuk2013finite}, phase-field approaches \citep{tauriello2013coupling, tam2022pattern}, and arbitrary Lagrangian-Eulerian frameworks \citep{mackenzie2021conservative}.} \ak{Outside of the applications in pattern formation amd morphogenesis, there is a growing literature on concentration-dependent domain growth and restructuring in oncology, cell polarity, chemotaxis, and other areas \citep{chen2014tumor, macdonald2016computational}.}

Another common approach for concentration-dependent growth is to consider models of Stefan-like moving boundary problems, which depend on the local concentrations at the boundary \citep{du2010spreading, sharma2016global, hadeler2016stefan, bao2018free, el2019revisiting, sharma2021global, murphy2021travelling, jepson2022travelling}. Such models have been studied intensively in terms of existence theory, travelling waves, and their applications in ecology, epidemiology, and the spreading of cells in developmental and oncological settings. More recently, such models have been explored in terms of reaction-diffusion patterning of a moving boundary \citep{tam2022pattern}, reminiscent of work on wave-initiated patterning \citep{myerscough1992analysis, krause2020non} but with the importance that the domain is not fixed. In some cases these moving-boundary problems can be seen as a special case of highly localized domain evolution, as shown in the apical growth case of \citet{crampin2002pattern}, though it is not always the case that one can find an equivalence between these two classes of models, especially in two or more spatial dimensions.

At the tissue-scale, all of the models above use PDE-based models derived from conservation of mass and focus on the morphogen dynamics, while tending to discount the cells themselves (noting the exceptions that \citet{ward1997mathematical, neville2006interactions} do discuss the importance of mitosis and cell death in these interactions). In comparison, there is now a large computational literature using cell-based and tissue-scale models from other formalisms, such as vertex-based and cellular-Potts models \citep{osborne2017comparing, sharpe2017computer, metzcar2019review, fletcher2022seven}. Section 7 of \citep{groves2020mitogens} provides an overview of concentration-dependent growth models of the morphogen Shh, as well as some discussion of how likely it is that patterning and growth are concomitant processes. These models, and particularly approaches fitting them to data about cell movement and tissue morphogenesis such as \citet{spiess2022approximated}, are valuable for matching simpler theoretical insights regarding domain restructuring and morphogen signalling. 

Here, we consider a simple framework of concentration-dependent growth within the more classical PDE-based approaches in order to elucidate basic theoretical properties of feedbacks between domain evolution and morphogen patterning. We follow the ideas of \citep{crampin1999reaction, baker2007mechanism, seirin2011dynamics} where we model a local volume element of the tissue as either expanding or contracting in time, depending on morphogen concentration.  We present a version of such a model for a general $N$-dimensional manifold in Section \ref{Model_Sect}, as in the prescribed growth models of \citep{plaza2004effect, van_gorder_growth_2019}, showing why local volume evolution is insufficient to fully characterize the evolution of either the domain or the concentrations for $N>1$, necessitating further constitutive assumptions. In the 1D case, we use the framework of \citep{van_gorder_growth_2019, van2020turing} to perform a linear stability analysis of homogeneous base states in Section \ref{LinStab_Sect}, deriving an extra term arising due to concentration-dependence. We numerically explore the 1D model in a variety of less-explored regimes in Section \ref{1DNumerics_Sect}, namely those involving domain contraction and rapid local evolution, where dilution leads to complex interplays of pattern structure and domain evolution, mediated both by the size of the domain but also by the impact of highly localized dilution. We then pose a simple constitutive model for $N$-dimensional manifolds that is both simple and plausible. We explore this model numerically in 2D planar geometries, demonstrating simple but important lessons in the constitutive choices made, as well as striking behaviours of patterns even in simple concentration-dependent settings. We close with a discussion of our results in Section \ref{Discussion_Sect}, highlighting in particular the rich and unexplored dynamics of these systems. One important insight described here is the tractability of the 1D system to mathematical exploration and how this contrasts the difficulties in both modelling and analysis for any higher-dimensional generalizations.

\section{A Simple Model of Concentration-Dependent Domain Evolution}\label{Model_Sect}

We formulate a general approach to modelling morphogen-dependent growth for $N$-dimensional manifold domains with boundaries, and describe concrete versions of this in one-dimensional and planar domains. We follow the general notation from \citep{krause2019influence,van_gorder_growth_2019}, though refer to \citep{crampin1999reaction,crampin2002pattern,lee2011dynamics} for an equivalent discussion using different terminology/notation.

We consider $m$ morphogen concentrations $\vec{u} = (u_1,u_2,\dots,u_m)$ on a compact evolving domain $\Omega(t) \subset \mathbb{R}^N$, focusing on the cases $m=1$ and $m=2$. We assume that this domain is bounded by a sufficiently smooth simple closed hypersurface $\partial \Omega(t)$ for all time $t$. By considering conservation of mass in a non-dimensionalised setting and moving to a Lagrangian frame, we find that these morphogens satisfy
\begin{equation}\label{RD_Eqnu}
    \dt{\vec{u}} = \vec{D}\lapbel{} \vec{u} - \vec{u}\left(\grad{} \cdot \vec{a}\right) + \vec{f}(\vec{u}),
\end{equation}
  where $\vec{D} = \textrm{diag}(D_1, D_2, \dots D_m)\in \mathbb{R}^{m\times m}$ is a diagonal matrix of diffusion coefficients, $\vec{f}(\vec{u})\in \mathbb{R}^m$ the vector of reaction kinetics, $\vec{a}\in \mathbb{R}^N$ the material flow defining the domain evolution, and $\lapbel{}$ and $\grad{} \cdot $ are, respectively, the Laplace-Beltrami\footnote{The governing equations can be posed for general $N$-dimensional manifolds, though we will only pursue numerical simulations for flat one- and two-dimensional domains.} and divergence operators on the domain $\Omega(t)$. The divergence of $\vec{a}$ represents a dilution of concentration during domain growth. We will assume no-flux conditions at any domain boundary, and specify initial concentrations $\vec{u}(0,\vec{X})$ where $\vec{X} = (X_1, X_2,\dots,X_n)^T \in \Omega(0)$ are the initial Lagrangian coordinates. 
  We will write the spatial derivatives in terms of a metric tensor $\vec{G} = (G^{ij})$, written in the Lagrangian coordinates $\vec{X}$, and denote the inverse by $\vec{H} = \vec{G}^{-1}$. In these coordinates, we have that the Laplace-Beltrami operator can be written as
  \begin{equation}\label{Laplace-Beltrami}
      \lapbel{} u = \frac{1}{\sqrt{|\det \vec{G}|}}\sum_{i,j=1}^n \pdiff{}{X_i}\left (\sqrt{|\det \vec{G}|} H^{ij} \pdiff{u}{X_j}\right),
  \end{equation}
  and the dilution coefficient as
  \begin{equation}\label{dilution}
      \grad{} \cdot \vec{a} = \dt{}\log\left (\sqrt{|\det \vec{G}|}\right ).
  \end{equation}
  We note that $\mu = \sqrt{|\det \vec{G}|}$ is the coefficient of the volume form, and in Lagrangian coordinates $\vec{X}$ can be thought of as the local expansion or contraction of the domain. We will assume that the domain evolves according to a local isotropic expansion or contraction at a rate $S(t,\vec{u})$, that is,
  \begin{equation}\label{growth_law}
      \grad{} \cdot \vec{a} = \dt{}\log(\mu)= S(t,\vec{u}).
  \end{equation}
  As we are mapping from the initial domain $\Omega(0)$ to $\Omega(t)$, we take $\mu(0,\vec{X}) =1$. Note that, for ease of presentation, we are taking for granted that there is a global coordinate system used to express $\vec{G}$ locally, though in principle there is no obstruction to defining the transport terms in more complicated situations that require multiple coordinate charts etc. We will also assume that the mappings remain sufficiently smooth, so that, in particular, $\sqrt{|\det \vec{G}|}>0$ for all time $t$. We assume that the concentrations satisfy Neumann boundary conditions along all boundaries throughout this work. We will primarily be interested in cases where $S$ depends only on the concentrations, rather than explicitly on time $t$, but we include this dependence to recover well-studied models of prescribed growth with $S=S(t)$.

  \subsection{One-Dimensional Model}\label{1DModel}
  
  As noted by \cite{lee2011dynamics}, in one spatial dimension Equations \eqref{RD_Eqnu}-\eqref{growth_law} can be made into a closed system by fixing a stationary point in the Eulerian frame (and the choice of such a point does not influence the dynamics). We can see this by considering the Eulerian coordinate $x(t)\in\Omega(t)$, which is related to the Lagrangian $X\in\Omega(0)=[0,L]$ (for some initial domain length $L>0$) by the scalar metric\footnote{The last equality here is only unique up to a sign -- the choice of the positive sign preserves orientation between the Lagrangian and Eulerian frames.}
  \begin{equation}\label{1DMetric}
      \mu(t,X) = \sqrt{|G(t,X)|} =  \pdiff{x(t)}{X}.
  \end{equation}
  One can then integrate \eqref{1DMetric} to determine how material points move. The reaction-diffusion system \eqref{RD_Eqnu} then takes the form
  \begin{equation}\label{RD_Eqnu1D}
    \dt{\vec{u}} = \frac{\vec{D}}{\mu} \pdiff{}{X}\left (\frac{1}{\mu} \pdiff{\vec{u}}{X} \right ) - \vec{u}S(t,\vec{u}) + \vec{f}(\vec{u}),
\end{equation}
  with the growth dynamics given by
  \begin{equation}\label{growth_law_1D}
      \dt{}\log(\mu) = S(t,\vec{u}).
  \end{equation}
  We can integrate \eqref{growth_law_1D} and use \eqref{1DMetric} to find that material points are given by,
  \begin{equation}\label{Eulerian1D}
      x(t,X) = \int_0^X\mu(t,y)\intd{y} = \int_0^X \exp\left (\int_0^t S(s,\vec{u}(s,y)) \intd{s}\right )\intd{y},
  \end{equation}
  where we are fixing the Eulerian domain to have the same zero, i.e.~$x(t,0)=0$, so that $\Omega(t) = [0, x(t,L)]$. Equations \eqref{RD_Eqnu1D}-\eqref{growth_law_1D} can then be solved on the fixed Lagrangian coordinates, $X \in [0,L]$, (independently of fixing an Eulerian point) and visualized on the Eulerian domain given by \eqref{Eulerian1D}. While in theory all of the dynamics is encoded in the Lagrangian equations alone, in practice the Eulerian domain is also computed to help simulate these equations, as they can become numerically ill-posed over moderate timescales on the Lagrangian domain due to rapid separation or clustering of material points -- see Section \ref{1DNumerics_Sect} for details.
  
  \subsection{$N>1$-Dimensional Model}\label{2D_Model_Sect}
  In two spatial dimensions, Equations \eqref{RD_Eqnu}-\eqref{growth_law} do not provide a closed system that uniquely determines the concentrations and domain evolution. While \eqref{growth_law} can be solved for $\mu$, this does not determine the full metric tensor $\vec{G}$, which is needed to interpret the Laplace-Beltrami operator in \eqref{RD_Eqnu}, as well as to determine how material points move. Equivalently, there is not a unique flow $\vec{a}$ which satisfies \eqref{dilution}. Here, we will make some of the simplest possible constitutive assumptions in order to derive a model on a compact, simply connected $N$-dimensional manifold. We then focus our analysis on a two-dimensional planar domain as the simplest example of this. Our assumptions will be of a kinematic nature, and will neglect a more detailed mechanical consideration of growing or deforming tissue.
  
  We assume that the local flow is irrotational  as a simple  constitutive constraint, together with the assumption that the flow has no tangential component at the boundaries of the domain. These two constraints are  fully consistent with the assumption that the flow is generated {\it only} by point sources of density scaling with $S$  and that cells constituting the domain cannot move along the edges of the domain (Sections 2.4, 2.5, \citet{Bachelor1967}).

  With these constitutive constraints, and given  sufficient smoothness, we have by the Helmholtz Decomposition Theorem that the flow $\vec{a}$ is a conservative vector field, so that there is some scalar potential $\phi$ such that $\vec{a} = \grad{}\phi$. Combining this assumption with \eqref{growth_law}, we find that $\phi$ satisfies a Poisson equation on $\Omega(t)$ given by
  \begin{equation}\label{Poisson}
      \lapbel{}\phi = S(t,\vec{u})
  \end{equation}
  and, by using \eqref{Laplace-Beltrami}, we can write this as an equation on the Lagrangian domain $\Omega(0)$ in terms of the metric tensor $\vec{G}$ in the coordinates $\vec{X}$. To complete the specification of $\vec{a}$, we need a suitable boundary condition for \eqref{Poisson}. Given the further assumption that the flow at domain boundaries is purely normal, i.e.~$\vec{a}\cdot \vec{t}=0$ on $\partial \Omega(t)$ for any unit tangent vector $\vec{t}$, we have $\vec{t}\cdot \grad{}\phi=0$, so that $\phi$ is constant along $\partial \Omega(t)$. As the potential $\phi$ is only defined up to a spatially uniform function of time, without loss of generality we set 
  \begin{equation} \phi = 0, ~~~ 
  \bm x \in \partial\Omega(t) \label{BC0}
  \end{equation} 
 where   $\partial \Omega(t)$ denotes the boundary. Furthermore, $\partial \Omega(t)$ will be determined by the material points from $\partial \Omega(0)$, so we can interpret this as a homogeneous Dirichlet condition on $\phi$ in the Lagrangian frame as well.
  
  Finally, in order to relate Eulerian and Lagrangian points, we can write out the flow as
  \begin{equation}\label{Eulerian2DFlow}
      \dt{\vec{x}} = \vec{a} = \grad{}\phi.
\end{equation}
In the $N=2$ planar setting, this can be expanded as
\begin{align}\label{Eulerian2Dx1}
    \dt{x_1} &= \phi_{X_1}\left(H^{11}\pdiff{x_1}{X_1} + H^{12}\pdiff{x_1}{X_2}\right) + \phi_{X_2}\left(H^{21}\pdiff{x_1}{X_1} + H^{22}\pdiff{x_1}{X_2}\right),\\
    \dt{x_2} &=   \phi_{X_1}\left(H^{11}\pdiff{x_2}{X_1} + H^{12}\pdiff{x_2}{X_2}\right) + \phi_{X_2}\left(H^{21}\pdiff{x_2}{X_1} + H^{22}\pdiff{x_2}{X_2}\right),\label{Eulerian2Dx2}
\end{align}
where the subscripts denote partial derivatives and $\vec{x} = (x_1,x_2)^T$ are the Eulerian coordinates. Equations \eqref{Eulerian2Dx1}-\eqref{Eulerian2Dx2} consist of a hyperbolic system of first-order partial differential equations, so we do not prescribe boundary conditions and instead compute the flow of material points directly.  We can write the inverse metric tensor, $\vec{H}$, in terms of derivatives of the Eulerian coordinates as
\begin{equation}
    H^{11} = \frac{1}{\mu^2}\left | \pdiff{\vec{x}}{X_2}\right |^2, \quad H^{12} = H^{21} =  -\frac{1}{\mu^2}\pdiff{\vec{x}}{X_1} \cdot \pdiff{\vec{x}}{X_2},\quad H^{22} = \frac{1}{\mu^2}\left | \pdiff{\vec{x}}{X_1}\right |^2,
\end{equation}
  and the coefficient of the volume form as
  \begin{equation}\label{Volume2D}
    \mu = \sqrt{ \abs{\pdiff{\vec{x}}{X_1}}^2\abs{\pdiff{\vec{x}}{X_2}}^2 - \left(\pdiff{\vec{x}}{X_1}\cdot\pdiff{\vec{x}}{X_2}\right)^2}.
  \end{equation}
  
  Equations \eqref{Poisson}-\eqref{Volume2D} form a closed system for the Eulerian coordinates as functions of the Lagrangian coordinates, and hence components of the metric tensor, once values of $\vec{u}$ are provided to determine $S(t,\vec{u})$. That is, Equations \eqref{RD_Eqnu} and \eqref{Poisson}-\eqref{Volume2D} form a closed system for the evolution of concentrations on an evolving domain, written entirely in the Lagrangian reference frame. While we will only consider  finite planar 2D domains for numerical simplicity, our formulation works for a general manifold with boundary. 

  \section{Linear Instability Analysis}\label{LinStab_Sect}
  
  Here, we provide a linear instability analysis of homogeneous base states (generalizing homogeneous equilibria) for \eqref{RD_Eqnu}-\eqref{growth_law} in the case of $m=2$ and $N=1$, describing the obstacles to generalizing this analysis at the end. Due to the non-autonomous nature of an evolving domain, there are different choices for defining a notion of a base state in order to study linear instabilities \citep{van_gorder_growth_2019}. Here, we use a natural generalization of a homogeneous base state $(\vec{u}^*(t), \mu^*(t))$, which evolves as
\begin{equation}\label{mu*}
    \diff{\mu^*}{t} = S(t,\vec{u}^*)\mu^*,
\end{equation}
  \begin{equation}\label{u*}
    \diff{\vec{u}^*}{t} =  
  - S(t,\vec{u}^*)\vec{u}^* + \vec{f}(\vec{u}^*),
  \end{equation}
where we use $\vec{f} = (f,g)^T$ as a vector of reaction kinetics. As in \cite{van_gorder_growth_2019}, we take $\vec{u}^*(0)$ to satisfy $\vec{f}(\vec{u}^*(0))=\vec{0}$ to agree with the linear stability analysis on static manifolds, and take $\vec{\mu}^*(0)=1$. We perturb the system \eqref{RD_Eqnu}-\eqref{growth_law} by writing\footnote{In fact, one can derive the base state system \eqref{u*} by assuming that we are interested in perturbations of this form and collecting the $O(1)$ terms. } $\vec{u} = \vec{u}^*(t) + \varepsilon \vec{U}^*(t,\vec{x})$ and $\mu = \mu^*(t) + \varepsilon \nu(t,\vec{x})$ for $|\varepsilon| \ll 1$. Substituting these into \eqref{RD_Eqnu1D}-\eqref{growth_law_1D} and retaining only terms up to order $\varepsilon$, we find that the perturbations satisfy
\begin{equation}\label{mu_li}
    \dt{\nu} = S(t,\vec{u}^*)\nu + \left(\nabla_{\vec{u}}S  \cdot \vec{U}\right ),
\end{equation}
\begin{equation}\label{full_u_li}
    \dt{\vec{U}} =  \frac{\vec{D}}{(\mu^*)^2} \pddiff{\vec{U}}{X}
  - S(t, \vec{u}^*)\vec{U} - \vec{K}\vec{U}+ \vec{J}\vec{U},
  \end{equation}
where we define
\begin{equation*}
    (\nabla_{\vec{u}}S)_i = \pdiff{S}{u_i}(t,\vec{u}^*), \quad K_{ij} = \pdiff{S}{u_j}(t,\vec{u}^*)u_i^*, \quad J_{ij} = \pdiff{f_i}{u_j}(\vec{u}^*), \quad i,j=1,2.  
\end{equation*}

We note that \eqref{full_u_li} does not depend on $\nu$, as terms involving $\nu$ will be multiplied by gradients of $\vec{u}^*$, which will vanish as these are spatially homogeneous. Hence, linear perturbations in the manifold evolution do not play a role in the stability of homogeneous equilibria. Therefore, we can neglect \eqref{mu_li}, and instead study just the linear system \eqref{full_u_li} subject to the base states defined by \eqref{mu*}-\eqref{u*}. We can also see that \eqref{u*} does not depend on $\mu^*$, so that we can directly solve \eqref{mu*} in terms of the function $\vec{u}^*$. If $S$ does not depend explicitly on time $t$, we find that the base state of the manifold, and its growth or contraction, is spatially uniform, in particular with the base state of the concentrations evolving autonomously via equation \eqref{u*}.

This system can be solved via the method introduced in \citep{van_gorder_growth_2019}. More directly, in \citep{van2020turing} the authors study linear systems of exactly the form \eqref{full_u_li} if we define the linearized kinetics as $\vec{M} = {\vec{J}} - \vec{K} - S(t,\vec{u}^*)\vec{I}$, where $\vec{I}$ is the two-by-two identity matrix, and write $\mu^*(t) = \exp(\int_0^tS(s,\vec{u}^*(s))\intd{s})$. The key insight is that the time-dependence of the Laplace-Beltrami operator can be completely factored out, so that one can take a Lagrangian decomposition of $\vec{U}$:
\begin{equation}
    \vec{U}(t,{X}) = \sum_{k=0}^\infty \psi_k({X})\vec{V}_k(t), \quad  \pddiff{\vec{\psi}_k}{X} = {\rho_k}\psi_k(\vec{X}) ,
\end{equation}
where the constant spatial eigenvalues ${\rho_k} = (k \pi /L)^2$ correspond to the standard Laplacian on an interval $[0,L]$ with associated eigenfunctions $\psi_k(X) = \cos(k \pi x/L)$. Perturbations then grow according to the non-autonomous ODE system
\begin{equation}
    \dt{\vec{V}_k} = -\rho_k(\mu^*(t))^{-2}\vec{D}\vec{V}_k + \vec{M}(t)\vec{V}_k.
\end{equation}

Theorem 2.1 of \cite{van2020turing} then gives a differential inequality that implies linear instability of a given mode over an interval of time, which we state here in our notation.
\begin{theorem}{\citep{van2020turing}}
Let $\mathcal{I}_k\subset [0,\infty)$, and assume that $\vec{M}$ is not diagonal for any $t \in \mathcal{I}_k$. Then, the perturbations associated with the $k$th mode, $\vec{V}_k$, grow exponentially if
\begin{multline}
    \det(\vec{M}(t))-\frac{D_1M_{11}(t) + D_2M_{22}(t)}{(\mu^*(t))^2}\rho_k + \frac{D_1D_2}{(\mu^*(t))^4}\rho_k^2 \\ < \max \left \{M_{12}(t)\diff{}{t}\left (\frac{M_{11}(t)(\mu^*(t))^2-D_1\rho_k}{M_{12}(t)(\mu^*(t))^2} \right), M_{21}(t)\diff{}{t}\left (\frac{M_{22}(t)(\mu^*(t))^2-D_2\rho_k}{M_{21}(t)(\mu^*(t))^2} \right)   \right\}
\end{multline}
is satisfied for all $t \in \mathcal{I}_k$.
\end{theorem}

This criterion gives a local-in-time condition for perturbations to be linearly unstable. Of course, pattern formation requires such instabilities to grow to sufficient amplitudes to be observable against the possibly complex behaviour of the spatially uniform base state given by $\vec{u}^*(t)$. In particular, sufficiently rapid variations of $\vec{u}^*$ in time can prevent patterns from forming due to the interval of instability associated with a particular mode, $\mathcal{I}_k$, being too small. Typically when patterns do form, they may stay within the linearly predicted regime but deviate from the maximally unstable mode due to nonlinear effects such as peak splitting \citep{crampin1999reaction,ueda2012mathematical}. We do note that a new term appears in our instability condition due to concentration-dependence, namely $\vec{K}$. One can show that, for sufficiently slow growth regimes, this instability criterion (and a stability condition for $k=0$) reduce to a quasi-static correction of the usual Turing conditions, which was computed in a different manner by \cite{madzvamuse2010stability}.

Additionally, in the concentration-dependent setting, these linear stability results are even more limited as they only hold for short times, when the growth is effectively uniform. Local growth of the domain can lead to substantial differences from such an assumption due to nonlinear effects (e.g.~local dilution interacting with kinetics). Finally, this analysis cannot be repeated for $N>1$ dimensional domains, as was done for coordinate-dilational growth in \citep{van_gorder_growth_2019}, because our constitutive assumptions for the flow do not allow the eigenfunctions to be separable due to non-uniformity in the growth even for constant $S$, as will be shown in Figure \ref{NonUniform2DFig}.

\section{One-Dimensional Numerical Results}\label{1DNumerics_Sect}
We now explore the 1D model numerically, seeking to examine the impact of concentration-dependent growth compared to explicitly pre-defined forms of growth on the pattern-forming behaviour of classical reaction-diffusion models.  In particular, the dilation rate $S$ will be specified on a case by case basis below. Furthermore, writing $\bm{u} = (u,v)$ and $\bm{f} = (f,g)$, we  will consider the Schnakenberg kinetics   \citep{schnakenberg1979simple,murray2004mathematical}:
\begin{equation}\label{Schnack}
    f(u,v) = a - u + u^2v, \quad g(u,v) = b - u^2v,
\end{equation}
 the Gierer-Meinhardt kinetics \citep{gierer1972theory}:
\begin{equation}\label{GM}
    f(u,v) = a - \frac{u^2}{v} -bu, \quad g(u,v) = u^2-cv,
\end{equation}
and the FitzHugh-Nagumo kinetics \citep{fitzhugh1955mathematical,nagumo1962active}:
\begin{equation}\label{FHN}
    f(u,v) = c\left(u-\frac{u^3}{3}+v-i_0\right), \quad g(u,v) = \frac{(a-u-bv)}{c}.
\end{equation}
 In all cases we consider positive parameters, i.e.~$a,b,c,i_0>0$.

We will also demonstrate the impact of concentration-dependent growth on the dynamics of scalar reaction diffusion models using the (nondimensionalized) logistic \citep{fisher1937wave, murray2007mathematical} and bistable \citep{chafee1974bifurcation, keener2021biology} kinetics, respectively given by
\begin{equation}\label{Fisher}
    f(u) = u(1-u),
\end{equation}
\begin{equation}\label{Bistable}
    f(u) = u(1-u^2).
\end{equation}

We simulate \eqref{RD_Eqnu1D}-\eqref{growth_law_1D} using a simple method-of-lines finite-difference scheme in MATLAB. In particular, the diffusion term is discretized as,
\begin{equation}
    \frac{1}{\mu}\pdiff{}{X}\left(\frac{1}{\mu}\pdiff{u}{X} \right) \approx \frac{1}{2\mu_i}\left(\frac{1}{\mu_i}(u_{i+1}+u_{i-1}-2u_i)+\frac{1}{\mu_{i+1}}(u_{i+1}-u_i) +\frac{1}{\mu_{i-1}}(u_{i-1}-u_i)\right),
\end{equation}
where $u_i$ and $\mu_i$ represent these variables at the $i$th grid point. The resulting system of ODEs is integrated in time using the MATLAB function \texttt{ode15s}, which implements a variable-step, variable-order solver \citep{Shampine1997}. Relative and absolute tolerances are taken to be $10^{-11}$ and, unless otherwise mentioned, $N_s=10^4$ equispaced grid points are used. For the kinetics \eqref{Schnack}-\eqref{FHN}, initial data are taken as a perturbation of the homogeneous steady state of the form $u(0,x) = u^*(1+\eta(x)), v(0,x) = v^*(1+\xi(x))$, where $\eta$ and $\xi$ are independently and identically normally distributed random variables with zero mean and variance $10^{-2}$. We note that the kinetics \eqref{FHN} may have up to three homogeneous equilibria, but we only consider parameters where there is one real equilibrium.

Due to the rapid (often exponential) separation of material points, each simulation is broken up into several iterations where, at the end of one iteration, the final Eulerian domain given by \eqref{Eulerian1D} becomes the new Lagrangian domain. This new computational domain is then uniformly discretised, and $\mu$ is set to unity throughout the domain. Interpolating the morphogen concentrations onto this new computational domain, this overall procedure ensures relative uniformity of the mesh compared to a single fixed Lagrangian domain. Numerical solution of the recast Lagrangian equations proceeds until the next discrete epoch, at which point the computational domain is once again redefined and remeshed. Here, we choose the iteration size in an ad hoc nature for each considered example, though in principle one could utilise some metric of mesh quality to adaptively select these lengths. Refinements of this iteration procedure had no impact on any of the reported simulations. Maximum timestep controls and grid point refinements were also used to check convergence in select simulations. Additionally, a COMSOL LiveLink implementation was also implemented for the 1D model, incorporating mesh refinement between iterations, using a comparable method to that detailed in Section \ref{2D_Numerics} for the two-dimensional simulations. These different numerical methods gave quantitatively comparable solutions on sufficiently fine meshes. All code and associated documentation can be found at \citep{Krause_Concentration_Dependent_Growth_2022}.

\subsection{Travelling Waves in Diffusive Logistic Equations}\label{TW_Sect}

\begin{figure}
      \begin{tabular}{ccc}
    \subfloat[$S = 0.005u$]{\includegraphics[width=0.32\textwidth]{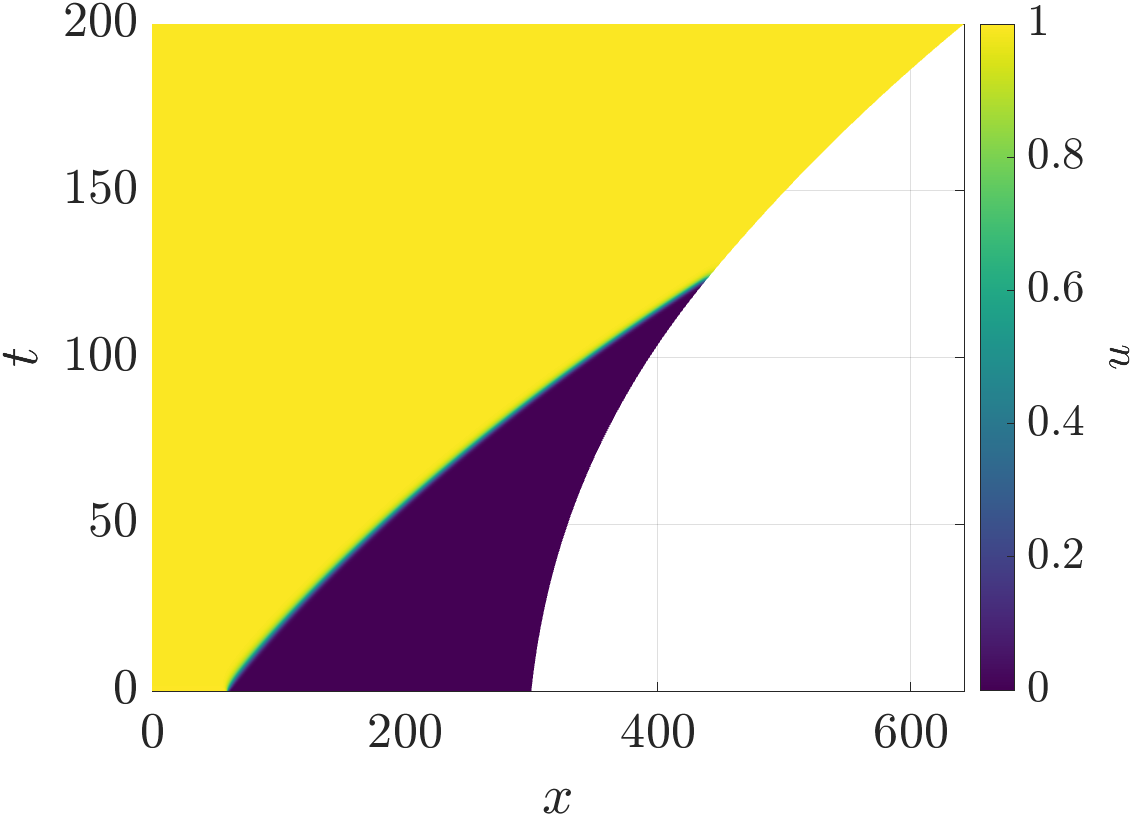}} &
    \subfloat[$S = 0.005(1-u)$]{\includegraphics[width=0.32\textwidth]{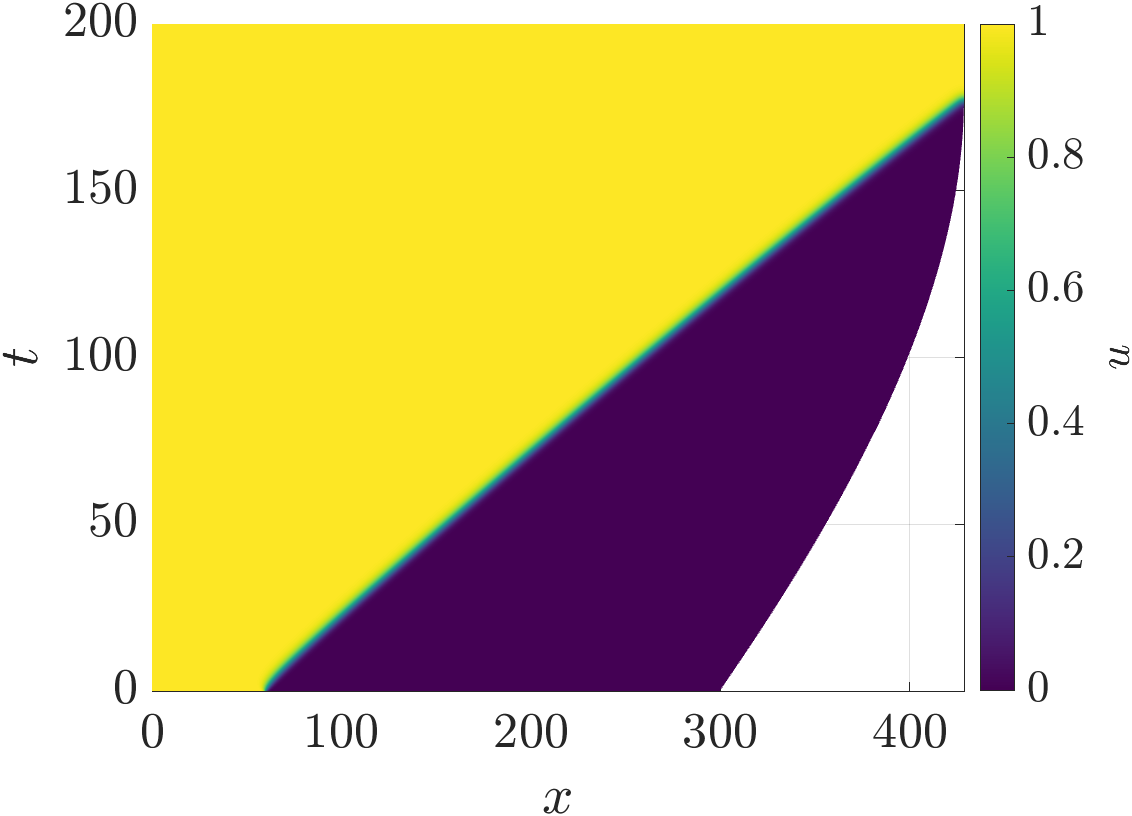}} &
    \subfloat[$S = 0.01(1-2u)$]{\includegraphics[width=0.32\textwidth]{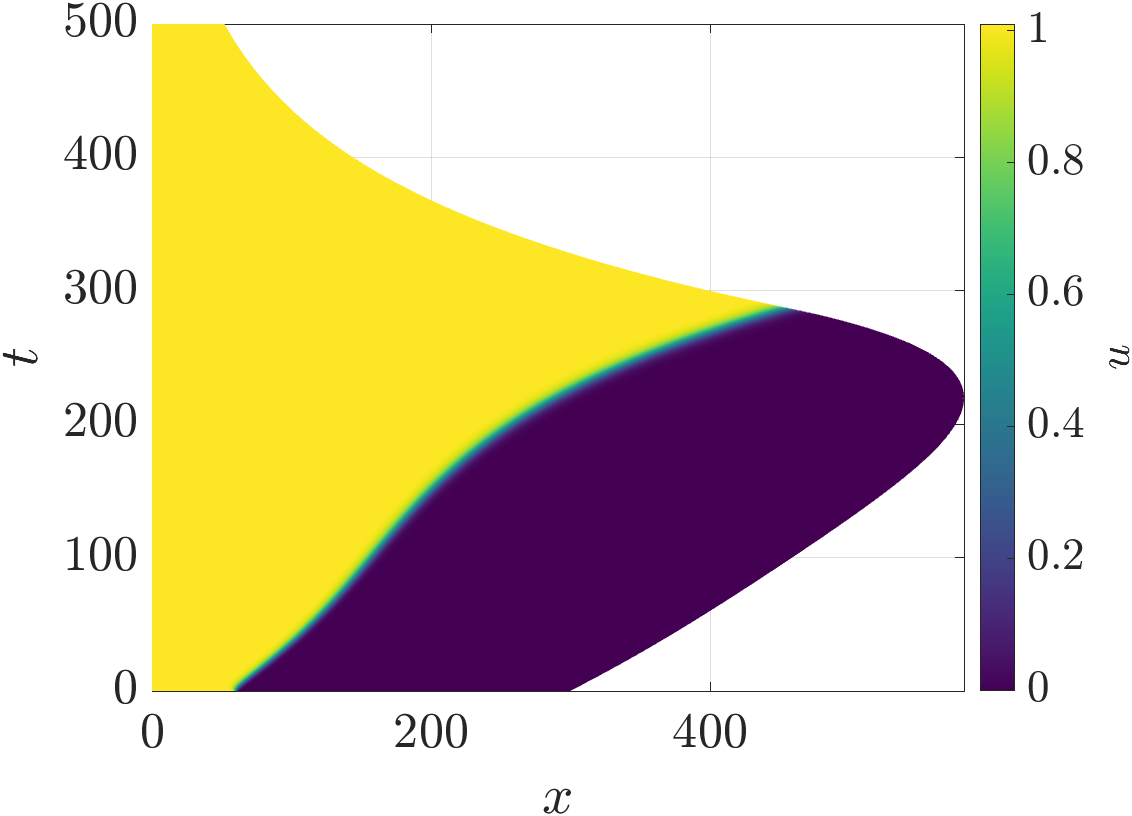}}\\
  \end{tabular}
    \caption{Values of $u$ from 1D simulations of the logistic kinetics \eqref{Fisher} under different growth scenarios. In all simulations, the initial domain length is $L=30$, with $D_1=1$. The initial condition is taken as $u(0,x) = (1+\tanh(L/5-x))/2$, so that the leftmost $20\%$ of the domain has the initial value $u \approx 1$ and the rest has the value $u \approx 0$.}
    \label{TWFig}
\end{figure}

To begin our exploration of concentration-dependent growth, we first consider the scalar case of $m=1$ and the logistic growth model given by the kinetics \eqref{Fisher}. This model has been a paradigm of emergent travelling waves due to the interactions of diffusion and nonlinearity; see Chapter 13 of \citep{murray2007mathematical} for an overview of this and related models. We save a detailed discussion of wave-type behaviour in these models for future work, giving instead just three examples of how concentration-dependent growth can change the structure of the typical constant-speed travelling wave observed in this model. 

We show these three examples in Fig.~\ref{TWFig}. For panel (a), we observe exponential growth, though at an increasing rate as the region for which $S(u) \approx 0.005$ grows as the wave travels across the domain. After $t \approx 120$ time units, the domain growth saturates to a constant exponential rate. In contrast, panel (b) shows an example where the region of positive growth decreases as the wave advances, eventually halting around $t \approx 180$. We remark that this solution is approaching (exponentially as $u$ tends to $1$) a homogeneous equilibrium of the model on a fix domain size, where both $u$ and $\mu$ reach a fixed value in time.  Finally, panel (c) gives the most exotic dynamics, where the domain is contracting whenever $u \approx 1$, and the domain is growing whenever $u \approx 0$. This leads to a transient period of growth as the wave moves, changing the proportion of the domain that is growing, until growth ceases and the domain begins contracting. Once the entire domain has approximately reached the $u=1$ steady state, it henceforth contracts at a fixed exponential rate.

There is a nontrivial impact of the growth on the apparent speed of the wave front, which is consistent with other models of uniform growth on travelling waves \citep{landman2003mathematical}. One has to be careful in defining a wave speed in this case, 
in part due to how one defines distances in evolving domains. This highlights one important feature of choosing $x = X = 0$ as a fixed point in the Eulerian-Lagrangian mapping, as it makes the rightmost point move the most, though in reality the domain growth and contraction is occurring locally throughout the domain. We leave further exploration of these ideas to future work and, instead, focus on pattern formation in the rest of the paper. In all subsequent 1D figures, we will instead fix the midpoint of the domain in our visualizations.

\subsection{Pattern Formation in Schnakenberg Systems}

\begin{figure}
      \begin{tabular}{cc}
    \subfloat[$S = 0.003919$]{\includegraphics[width=0.33\textwidth]{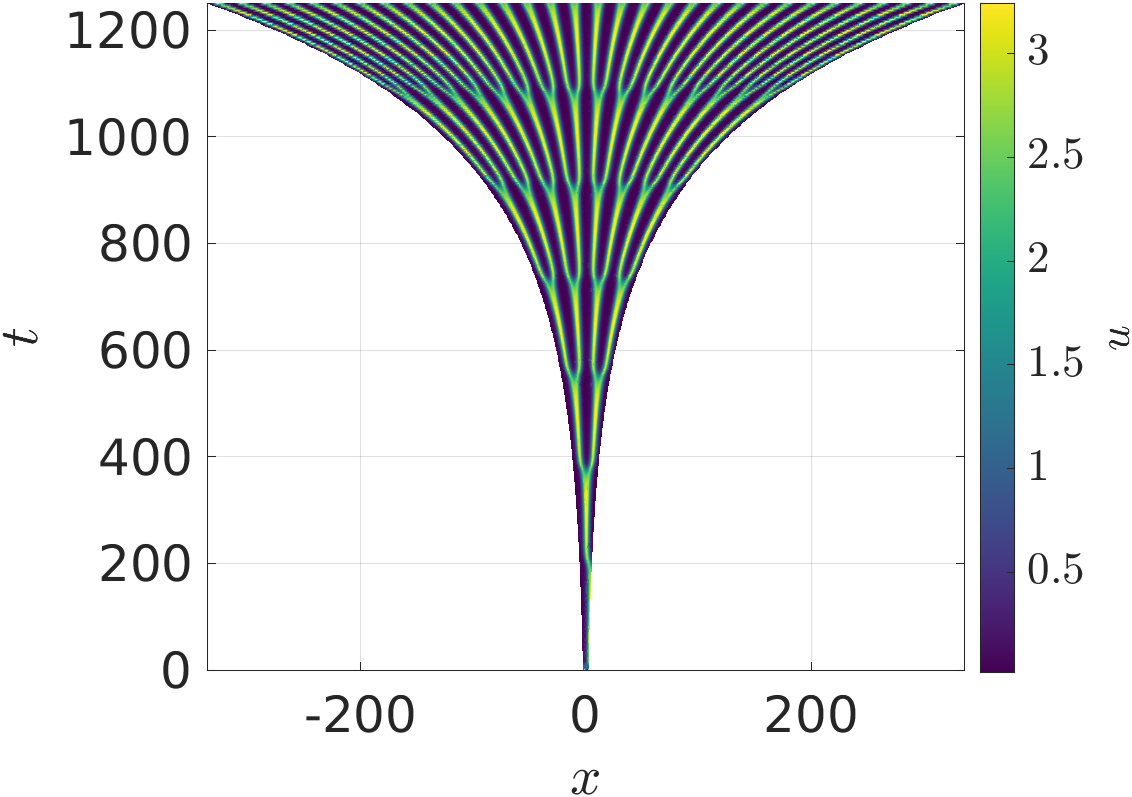}} &
    \subfloat[$S = 0.049082$]{\includegraphics[width=0.33\textwidth]{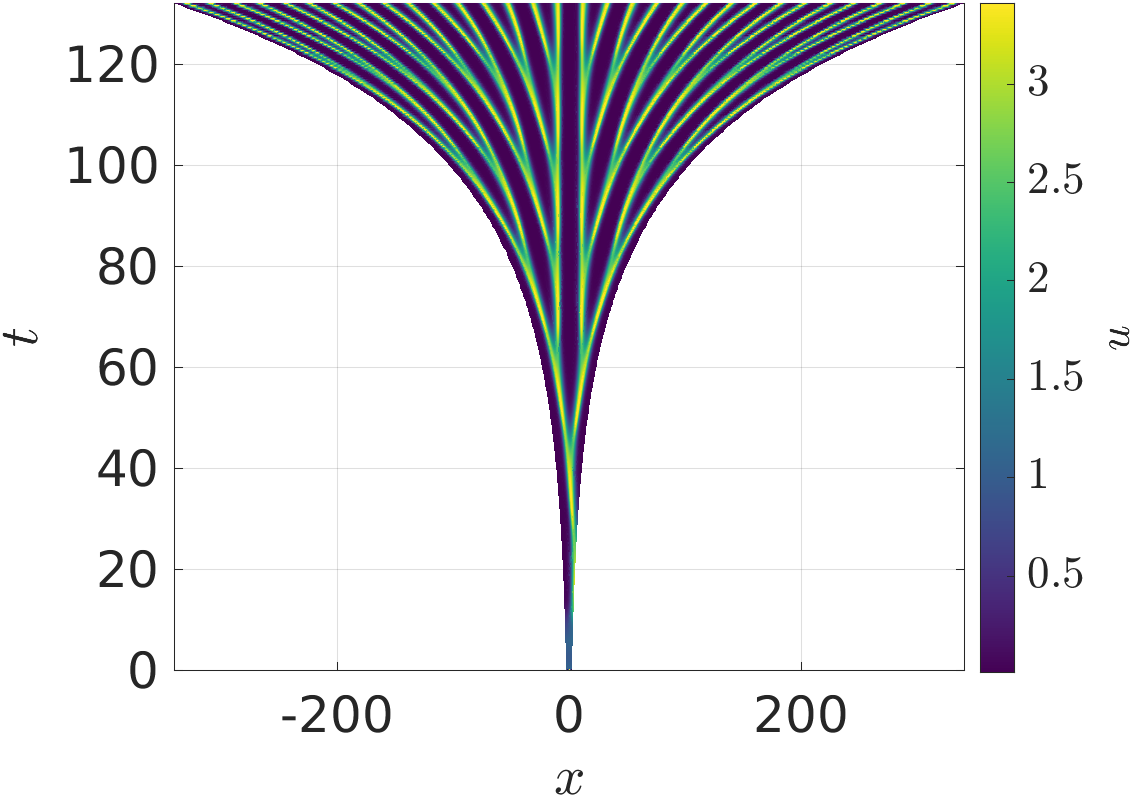}}
    \subfloat[$S = 0.042761$]{\includegraphics[width=0.33\textwidth]{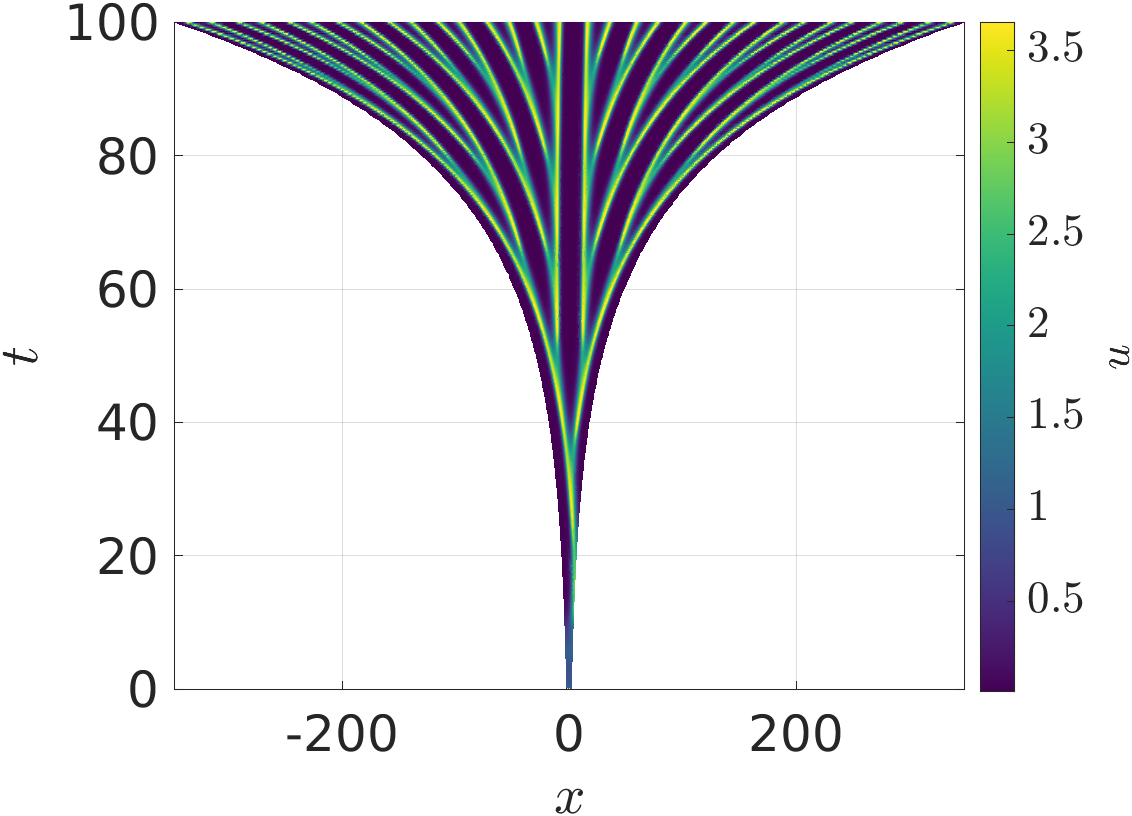}}\\
    \subfloat[$S = 0.005(1+\tanh(100(u-1.2)))$]{\includegraphics[width=0.33\textwidth]{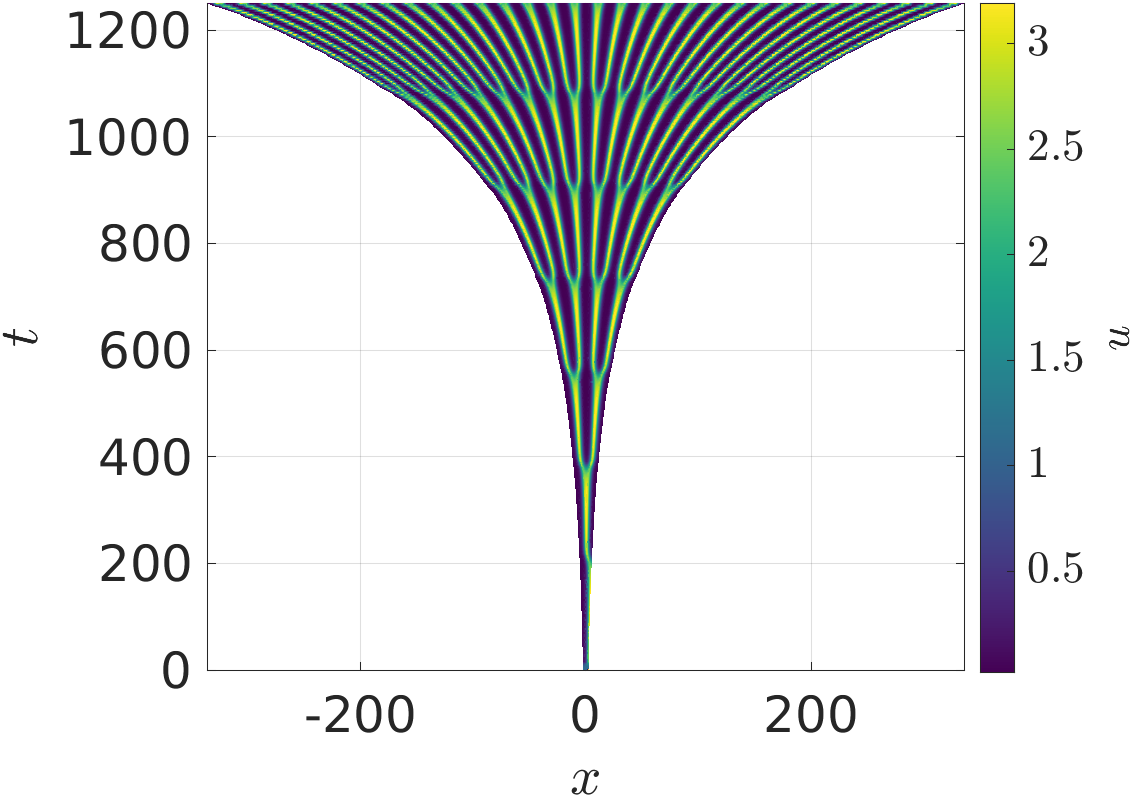}} &
    \subfloat[$S = 0.05(1+\tanh(100(u-1.2)))$]{\includegraphics[width=0.33\textwidth]{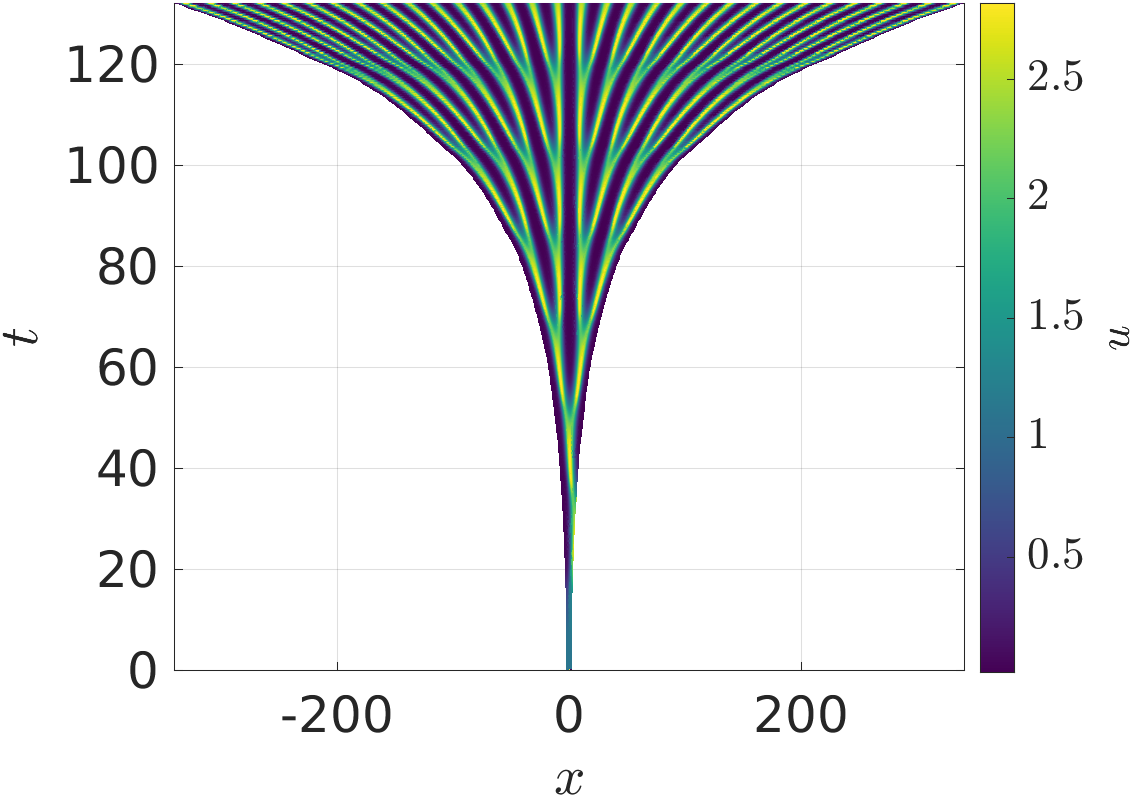}}
    \subfloat[$S = 0.25(1+\tanh(100(u-1.2)))$]{\includegraphics[width=0.33\textwidth]{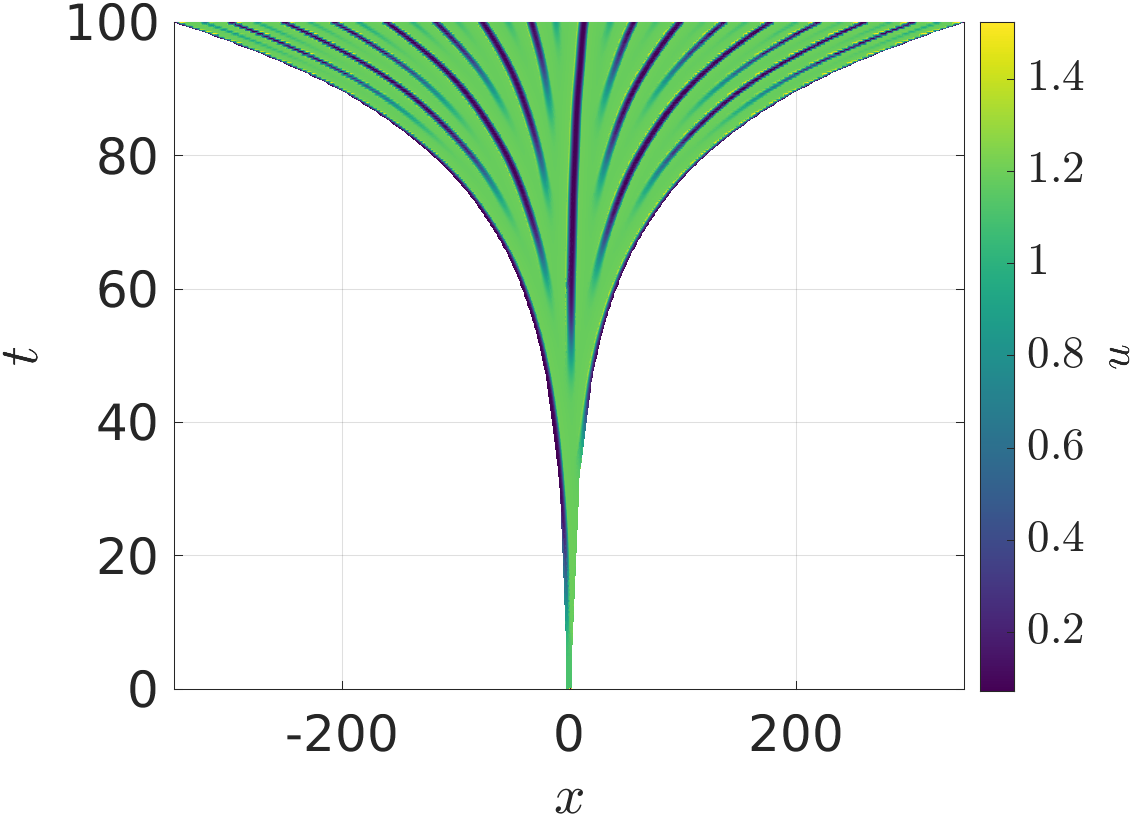}} \\
    \subfloat[$S = 0.02u$]{\includegraphics[width=0.33\textwidth]{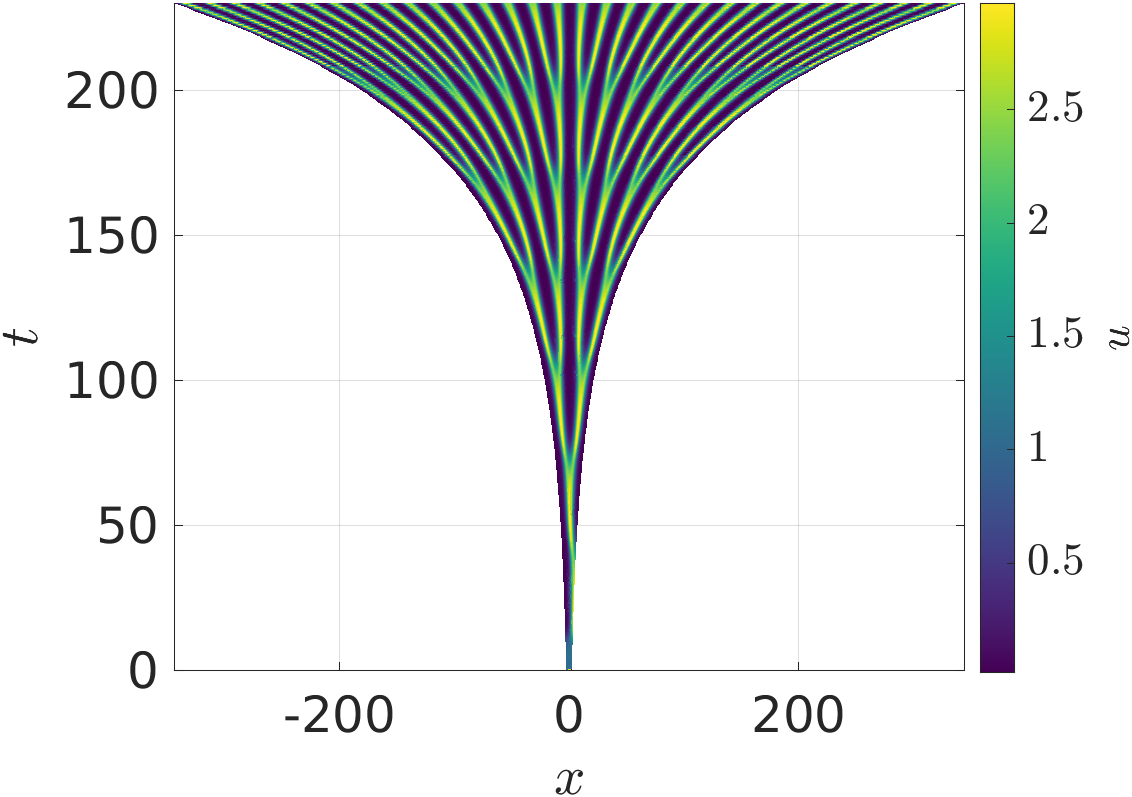}} &
    \subfloat[$S = 0.1u$]{\includegraphics[width=0.33\textwidth]{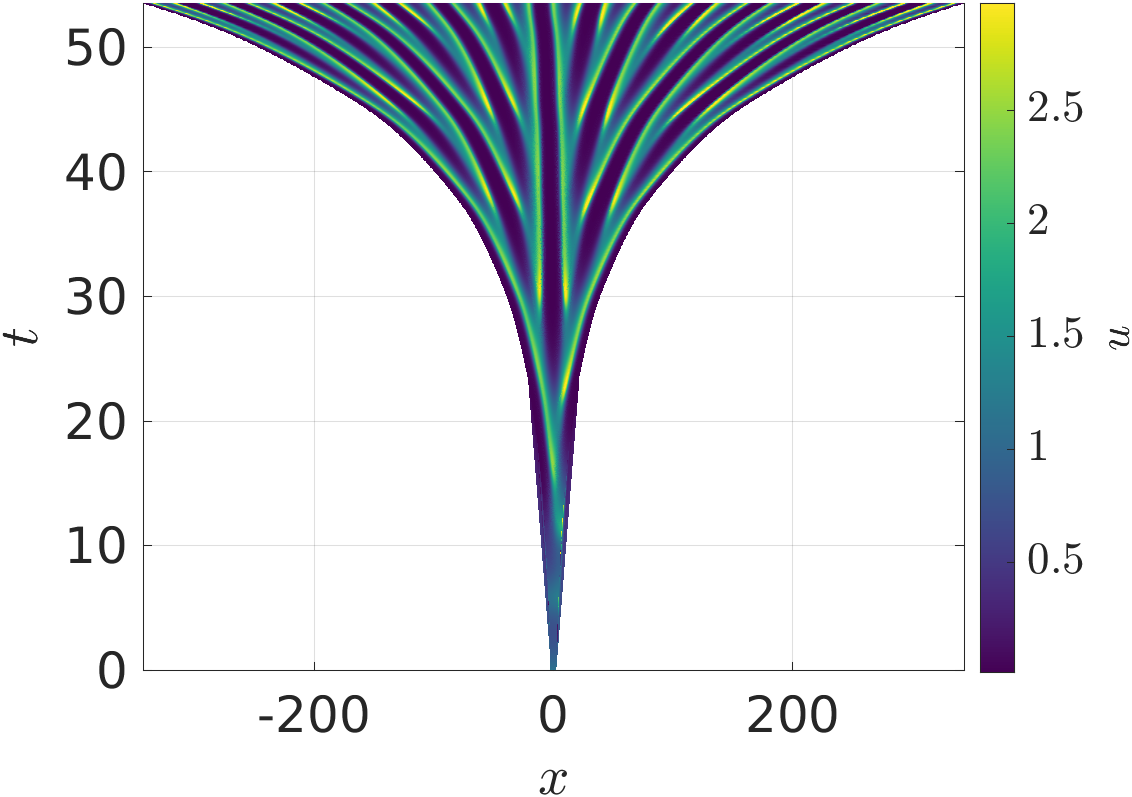}}
    \subfloat[$S = 0.15u$]{\includegraphics[width=0.33\textwidth]{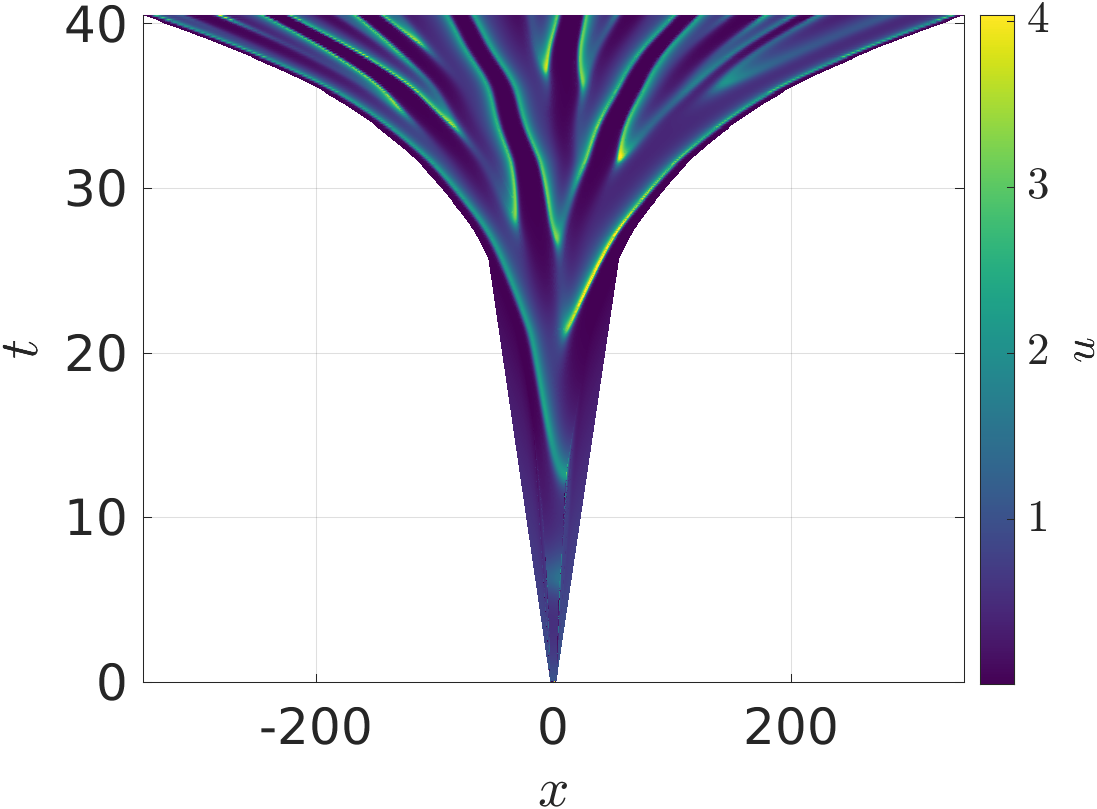}} \\
  \end{tabular}
    \caption{Values of $u$ from 1D simulations of the Schnakenberg kinetics \eqref{Schnack} under different growth scenarios. In all simulations the initial domain length is $L=5$, with $a=0.01$, $b = 1.1$, $D_1=1$, and $D_2=40$. In (d)-(i), we show $u$-dependent growth, whereas in (a)-(c) we show uniform exponential domain growth. Timescales are set so that all domains grew to $\approx 670$. Uniform growth rates were used in (a)-(c) to match the corresponding domain lengths and timescales in (d)-(f), so that the final simulation time and domain sizes are identical.}
    \label{SchnackFig1}
\end{figure}

\begin{figure}
      \begin{tabular}{ccc}
    \subfloat[]{\includegraphics[width=0.3\textwidth]{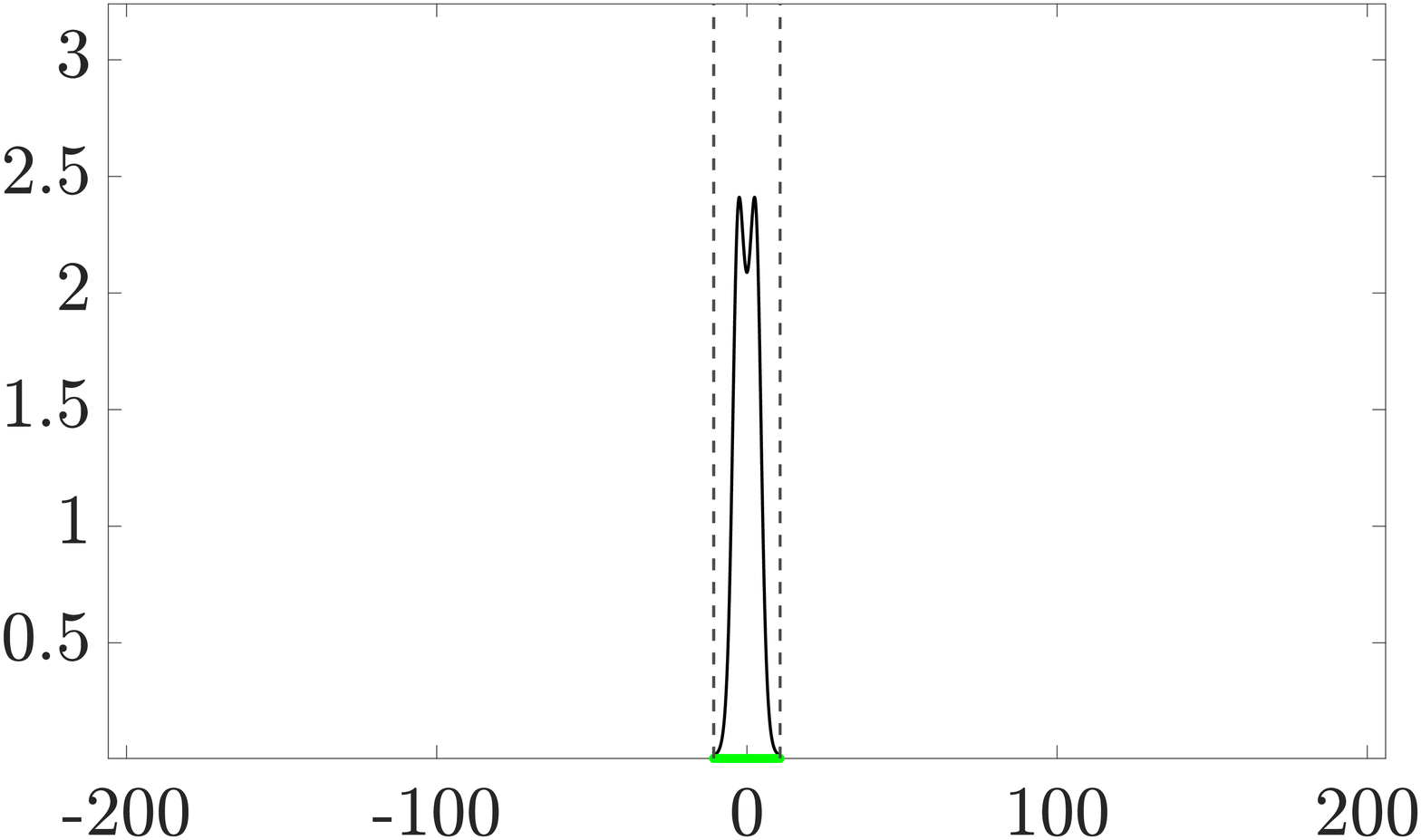}} &
    \subfloat[]{\includegraphics[width=0.3\textwidth]{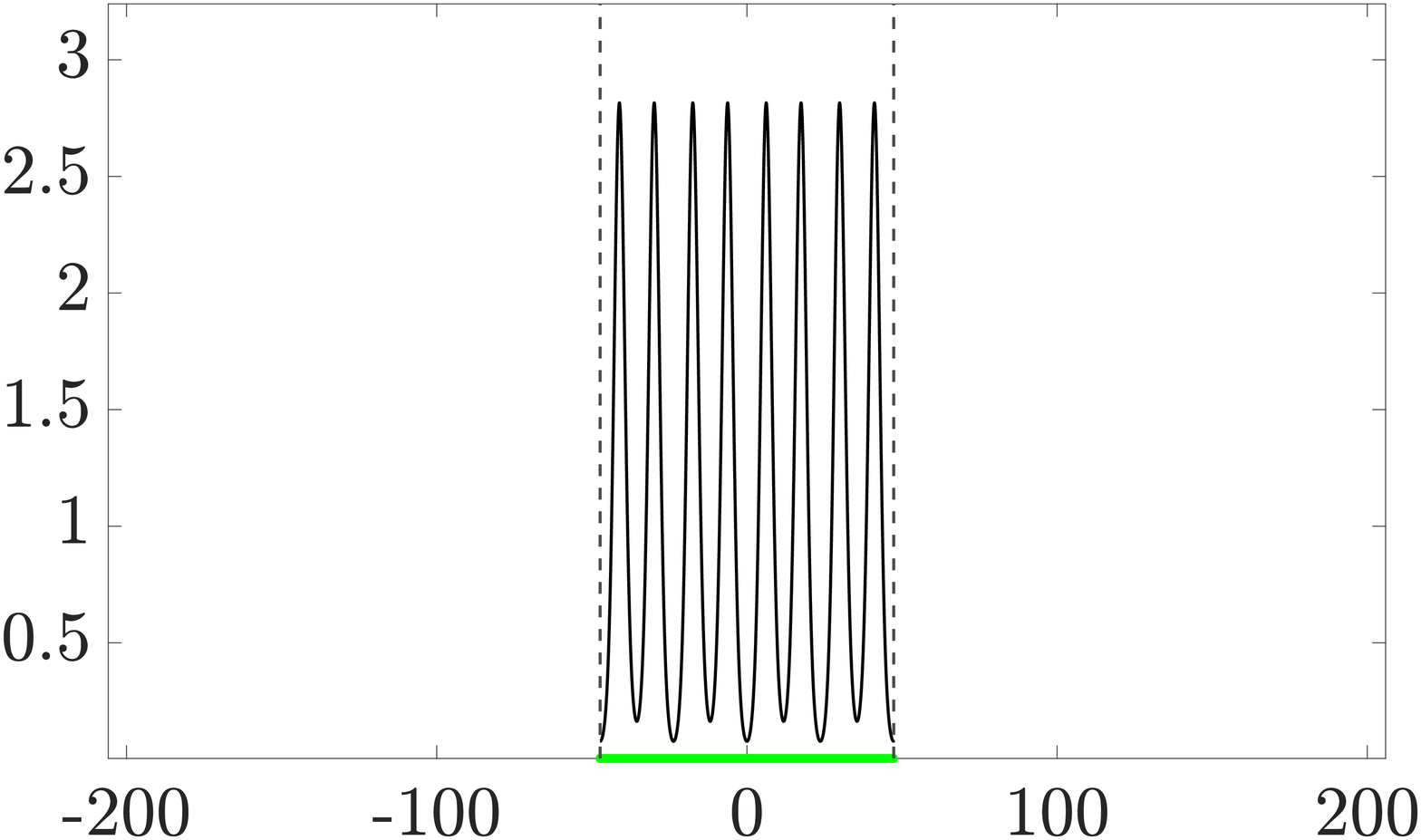}} &
    \subfloat[]{\includegraphics[width=0.3\textwidth]{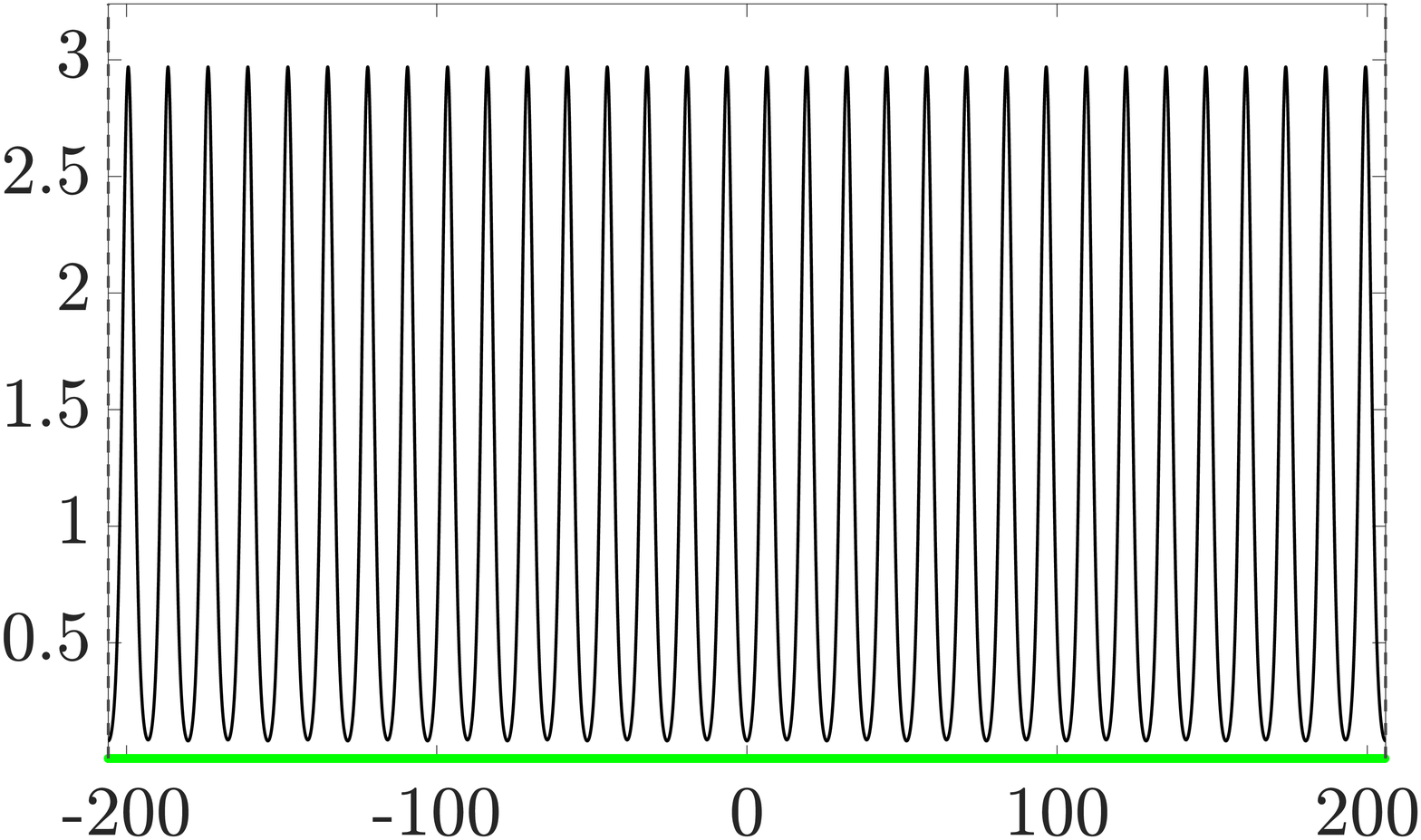}}\\
    \subfloat[]{\includegraphics[width=0.3\textwidth]{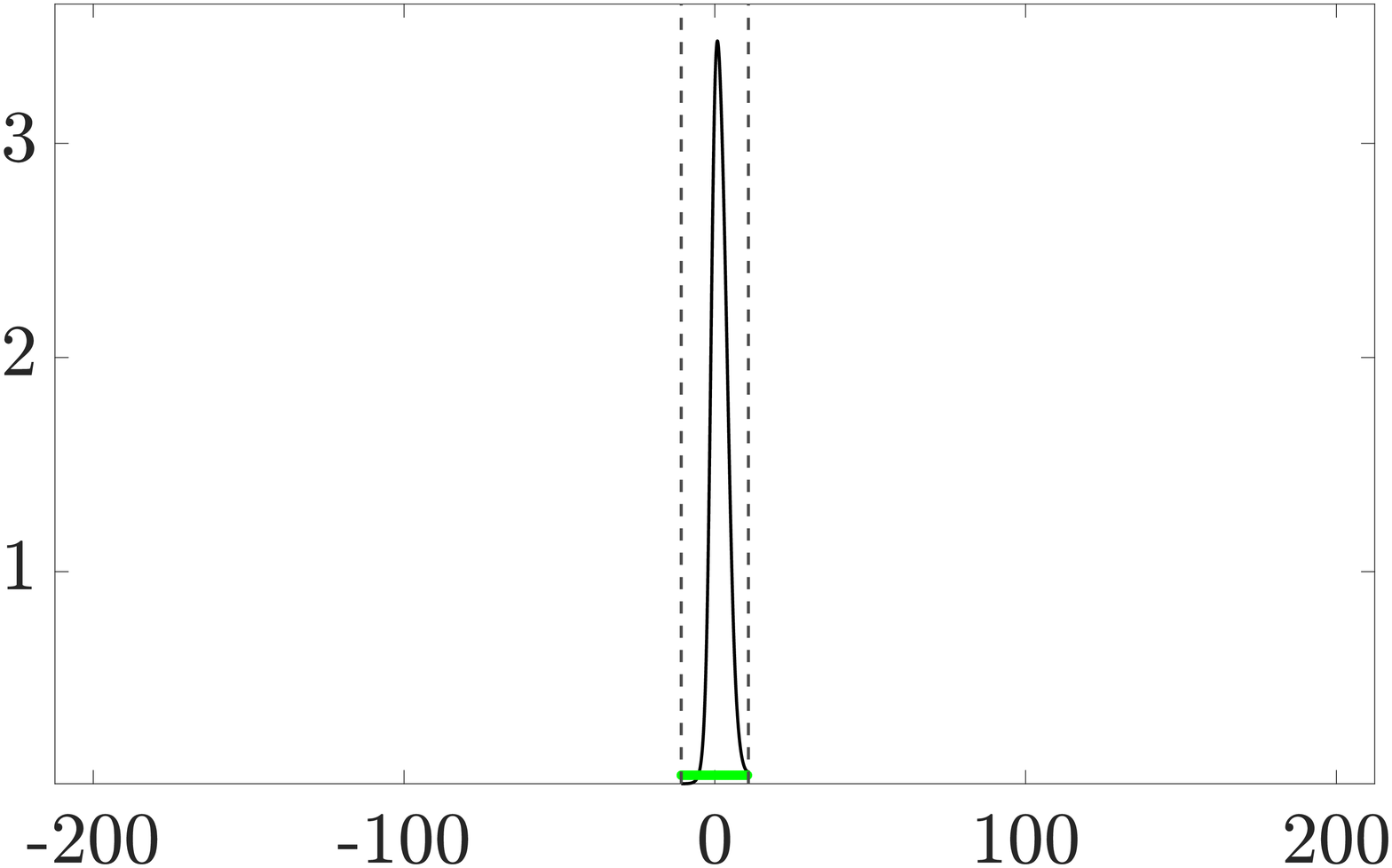}} &
    \subfloat[]{\includegraphics[width=0.3\textwidth]{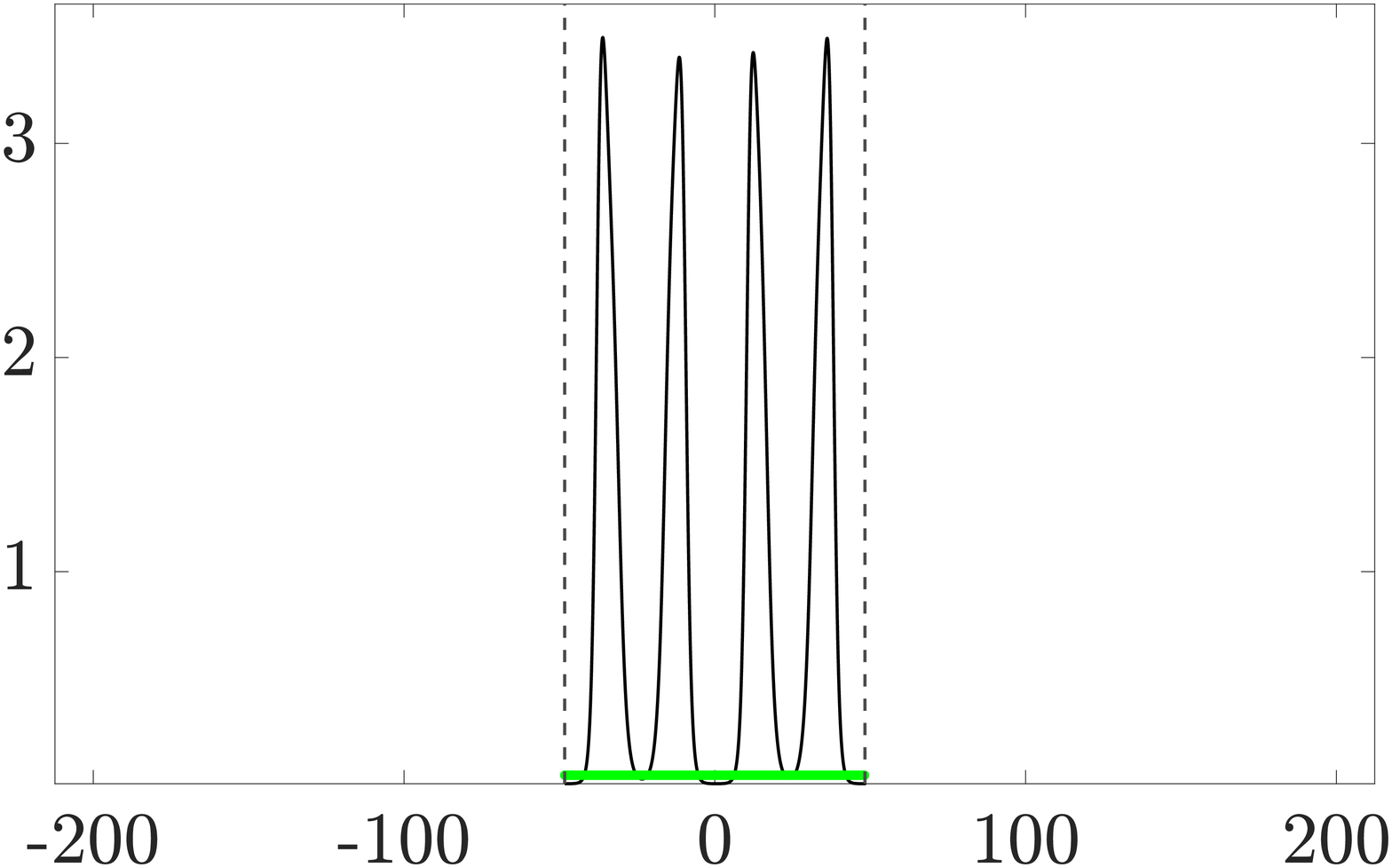}} &
    \subfloat[]{\includegraphics[width=0.3\textwidth]{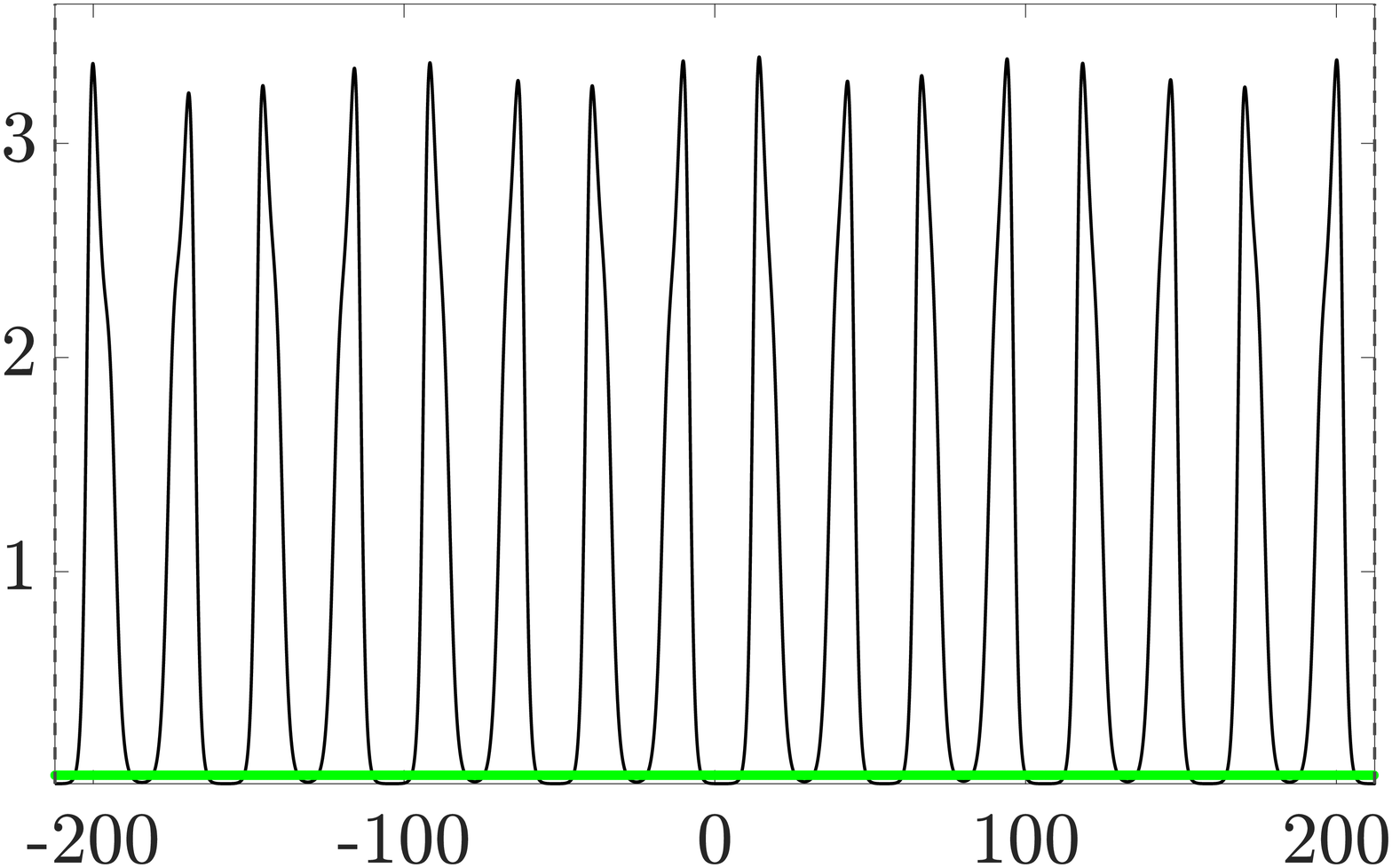}} \\
    \subfloat[]{\includegraphics[width=0.3\textwidth]{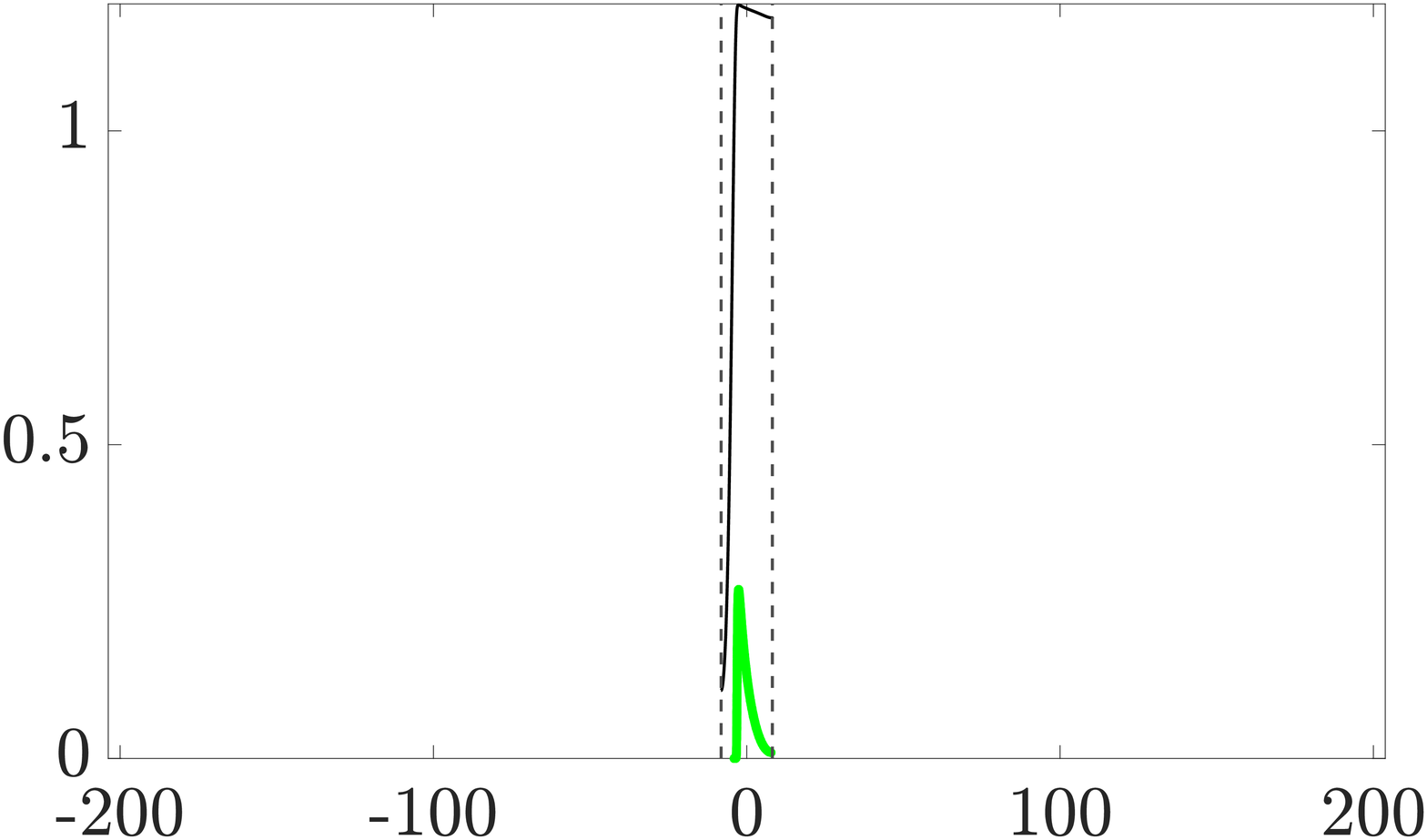}} &
    \subfloat[]{\includegraphics[width=0.3\textwidth]{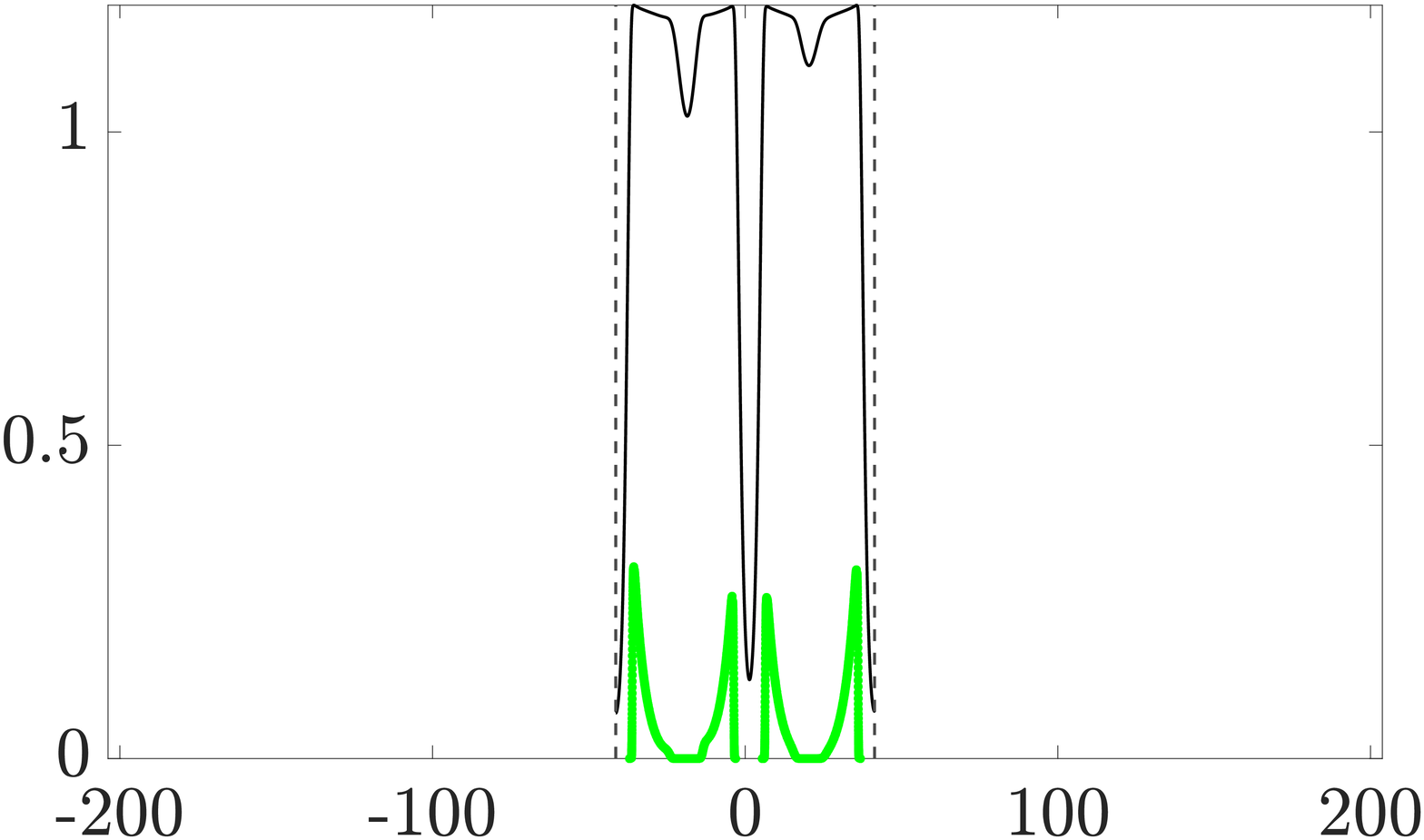}} &
    \subfloat[]{\includegraphics[width=0.3\textwidth]{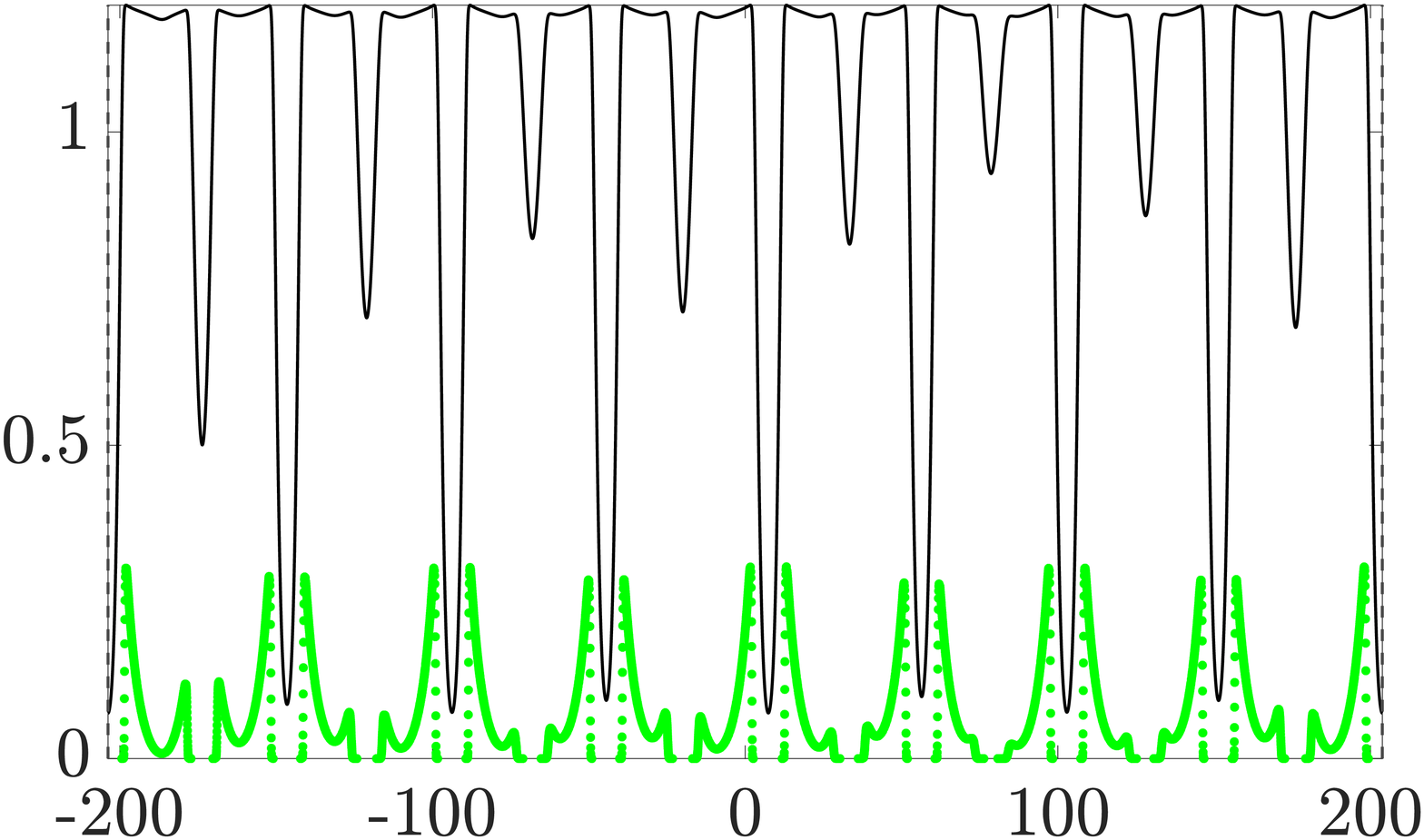}} \\
    \subfloat[]{\includegraphics[width=0.3\textwidth]{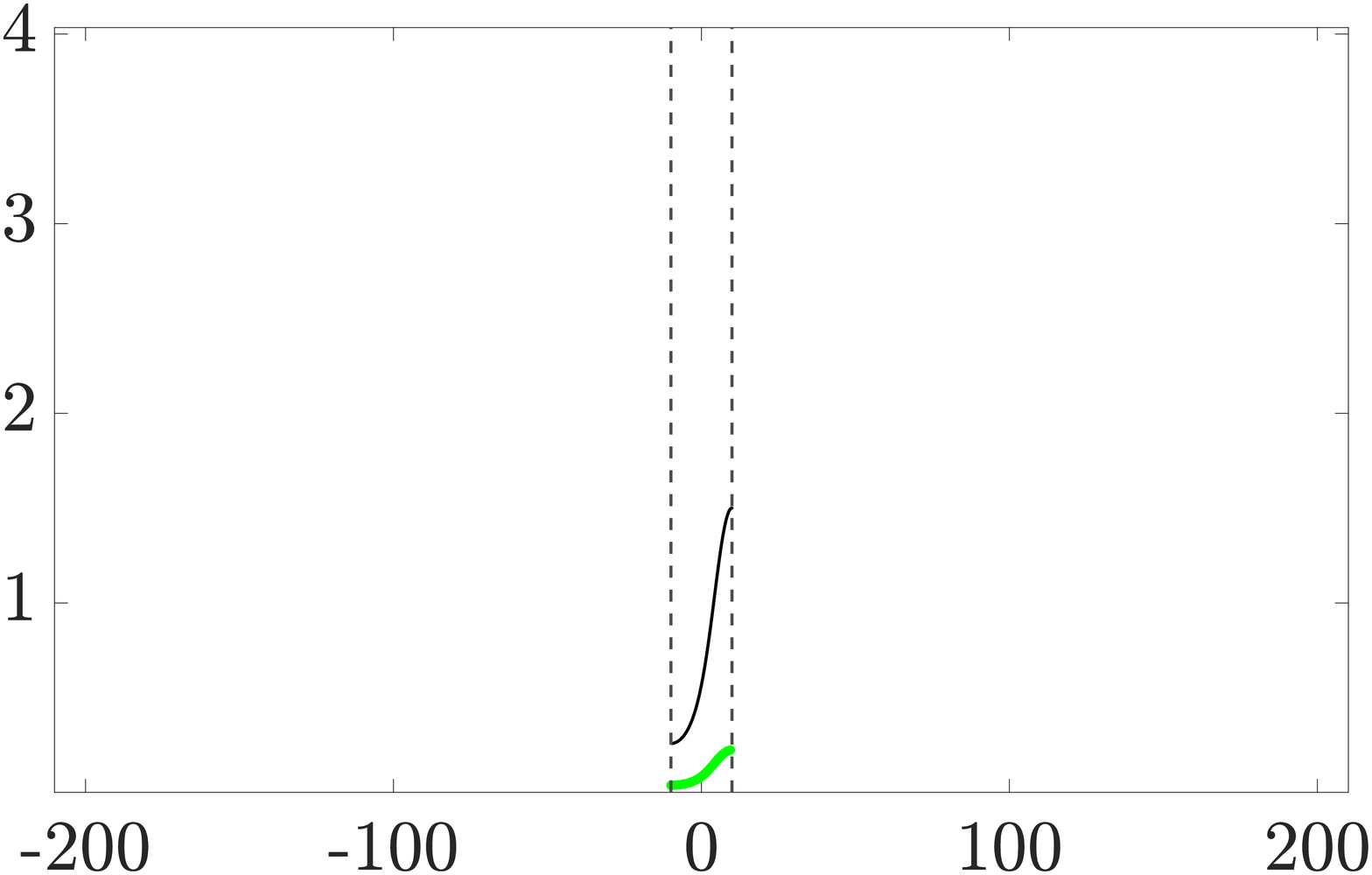}} &
    \subfloat[]{\includegraphics[width=0.3\textwidth]{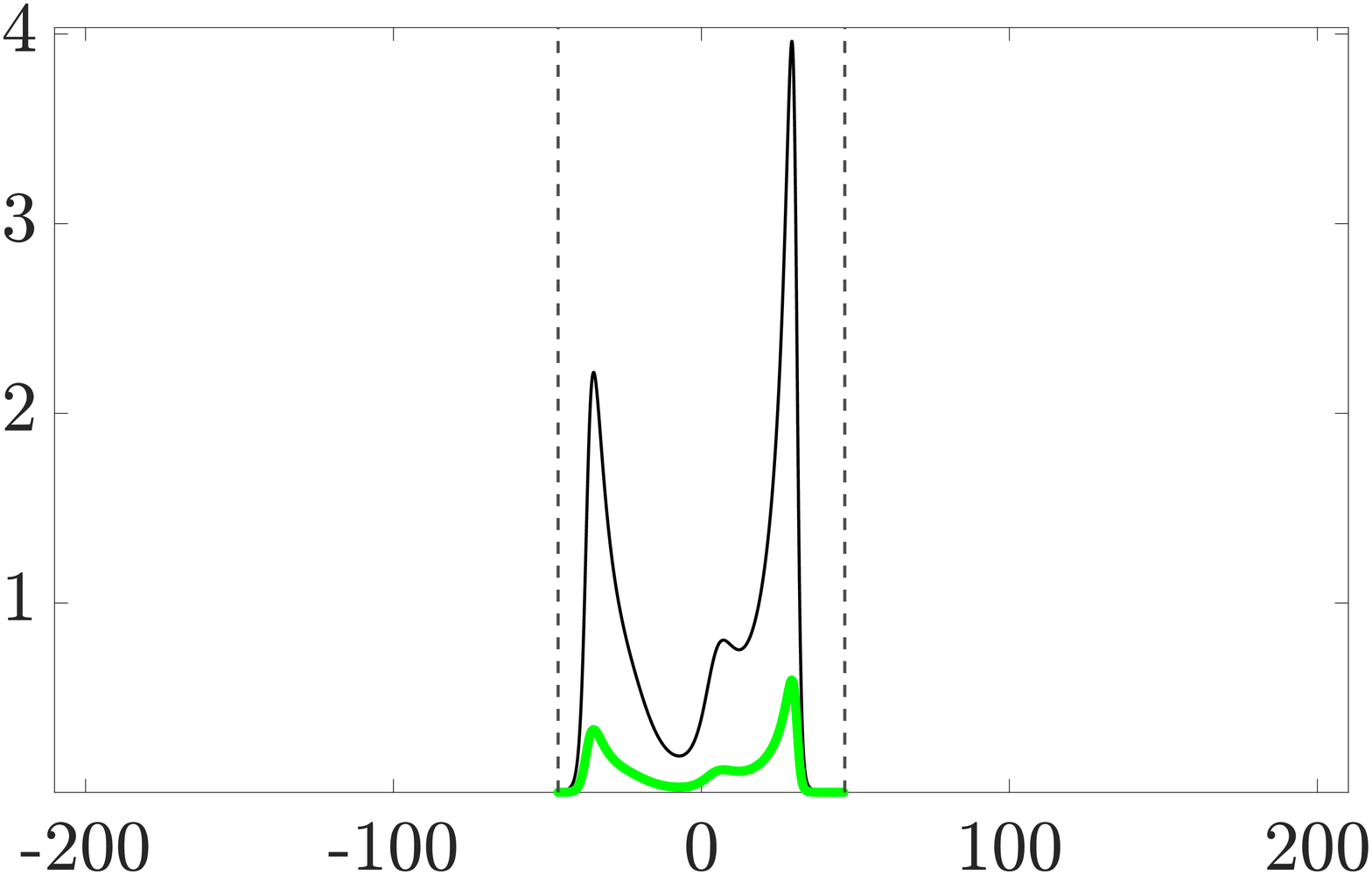}} &
    \subfloat[]{\includegraphics[width=0.3\textwidth]{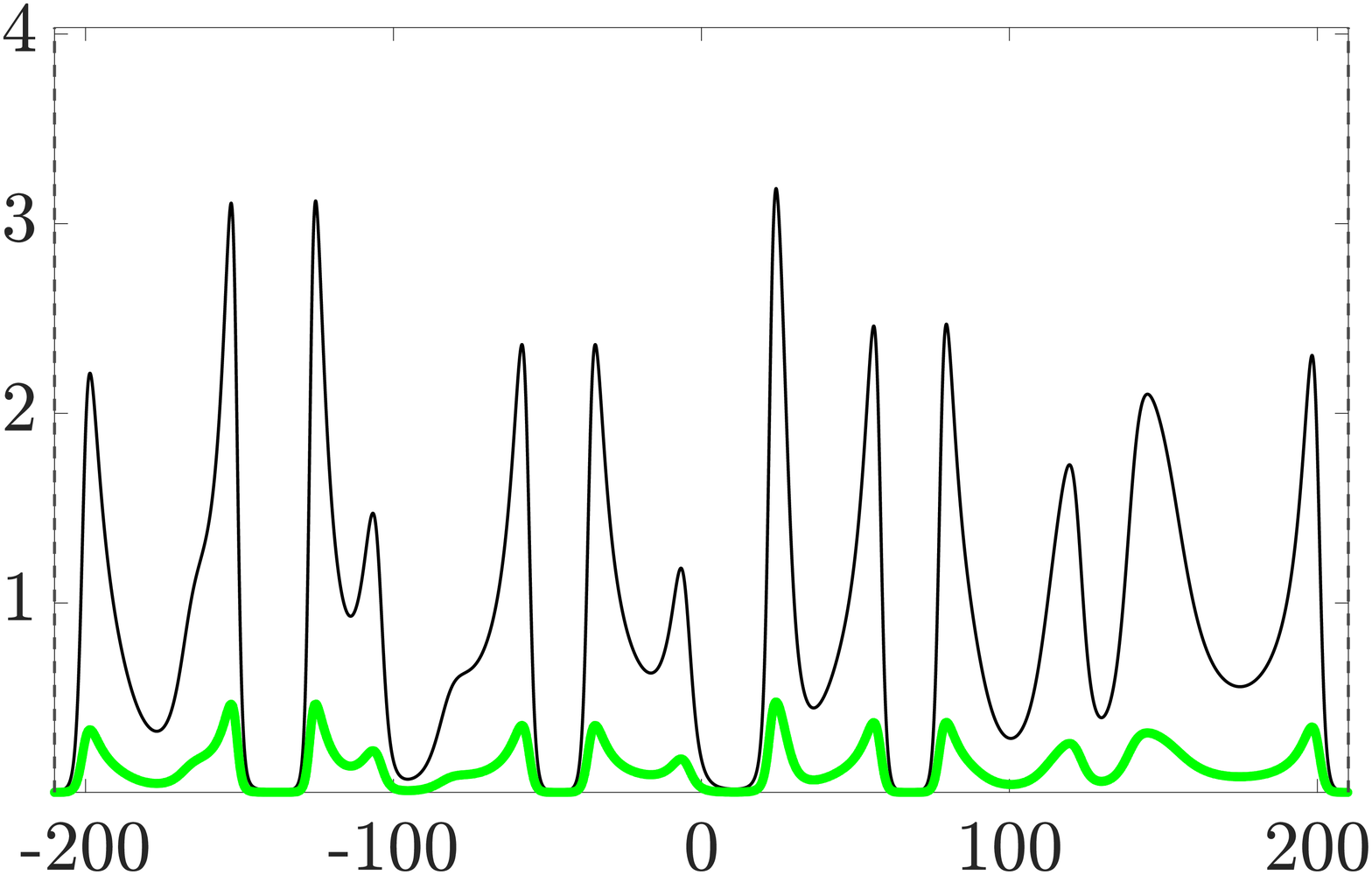}} \\
  \end{tabular}
    \caption{Values of $u$ as black curves and $S$ as green curves at specific times from the same simulations in Fig.~\ref{SchnackFig1}. Row (a)-(c) corresponds to panel (a) in Fig.~\ref{SchnackFig1}, Row (d)-(f) corresponds to panel (c) in Fig.~\ref{SchnackFig1}, row (g)-(i) corresponds to panel (f) in Fig.~\ref{SchnackFig1}, and Row (j)-(l) corresponds to panel (i) in Fig.~\ref{SchnackFig1}. The first column is taken at $1/3$ of the final simulation time, the second at $2/3$ of the final simulation time, and the last column at the final simulation time.}
    \label{SchnackFig1Slices}
\end{figure}

We give examples of growing domain simulations in Fig.~\ref{SchnackFig1} for the Schnakenberg kinetics \eqref{Schnack}. Panels (a)-(c) are uniformly exponentially growing domains, whereas those in (d)-(f) grow in a `thresholded' manner (that is, regions of space grow approximately where $u>1.2$), and those in (g)-(i) grow at a rate proportional to $u$. The panels are arranged so that the columns moving left to right show increasingly fast growth timescales. Hence, the most drastic impacts of the different kinds of growth can be observed in the last column, where we see in (c) that fast uniform growth has the same qualitative character of peak splitting as for lower growth rates. In contrast, the fast thresholded growth in (f) leads to large regions of $u$ that are saturated just beyond the growth threshold, as local dilution decreases the maximum value of $u$ in the fastest growing regions. Finally, in (i), we see that the very high growth rates in spikes can lead to a complete breakdown of the more usual spike-doubling behaviour observed at low growth rates, and the insertion of irregular spikes. This is consistent with such behaviours observed for fast uniform growth, which has been described analytically in \cite{ueda2012mathematical} \ak{(see also \cite{kolokolnikov2006stability} for an alternative view of the underlying mechanism behind such spike-doubling phenomena)}. However, for the concentration-dependent cases, this deviation from spike-doubling is more pronounced and leads to extremely irregular insertion events. One important observation that was confirmed across many other simulations is that, for sufficiently small growth rates, as long as $S \geq 0$,  similar spike-doubling behaviour was typically observed,  as seen here in the first column.

\begin{figure}
      \begin{tabular}{cc}
    \subfloat[$S = 0.01(u-1.3v)$]{\includegraphics[width=0.33\textwidth]{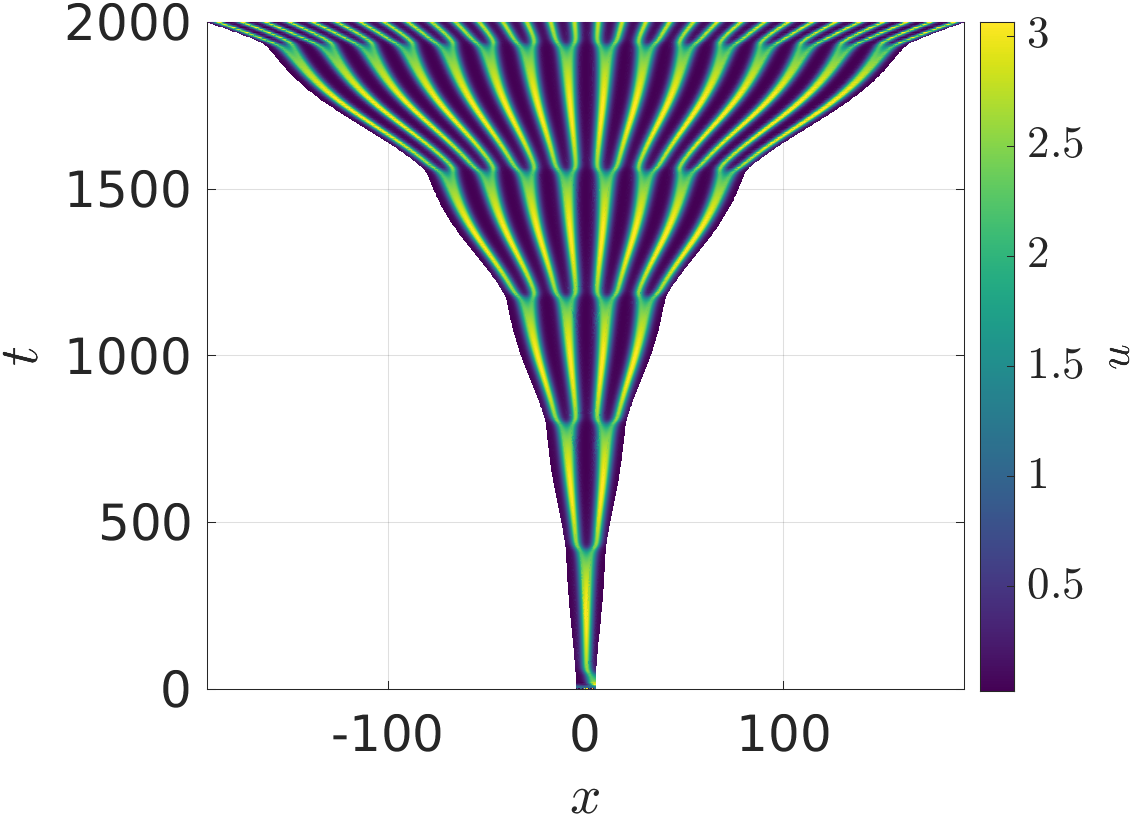}} &
    \subfloat[$S = 0.05(u-1.3v)$]{\includegraphics[width=0.33\textwidth]{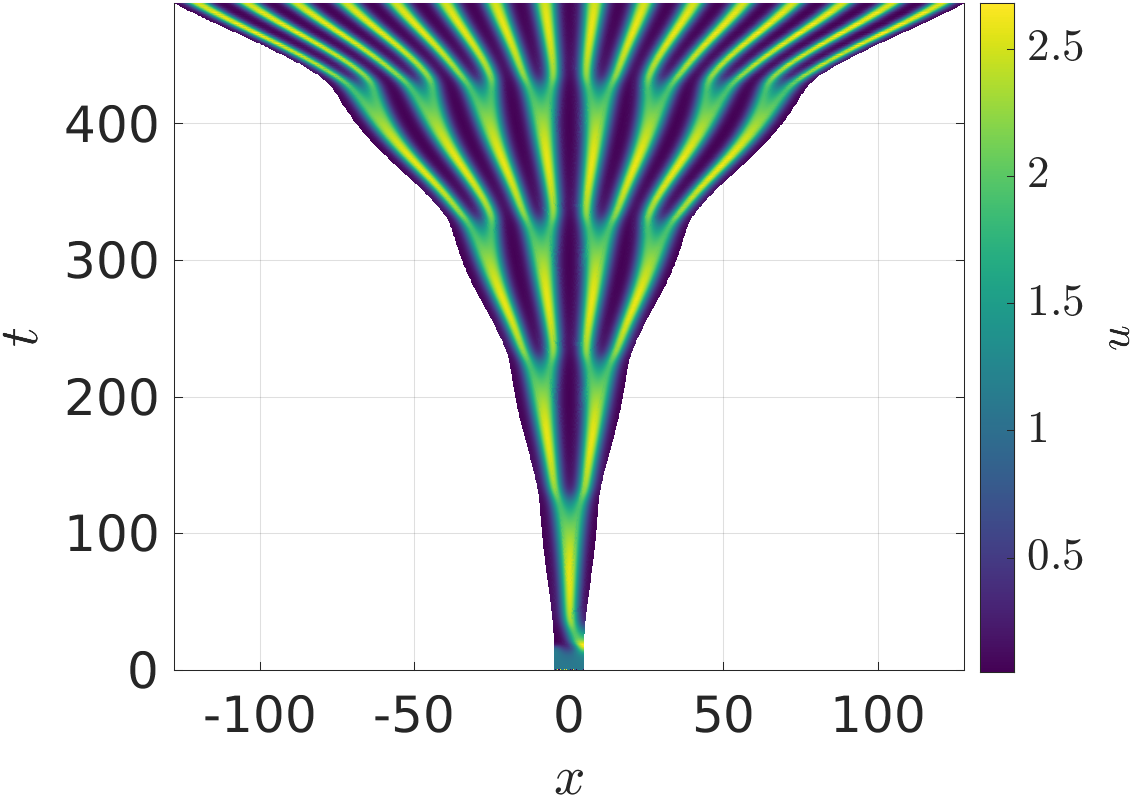}}
    \subfloat[$S =  0.161(u-1.3v)$]{\includegraphics[width=0.33\textwidth]{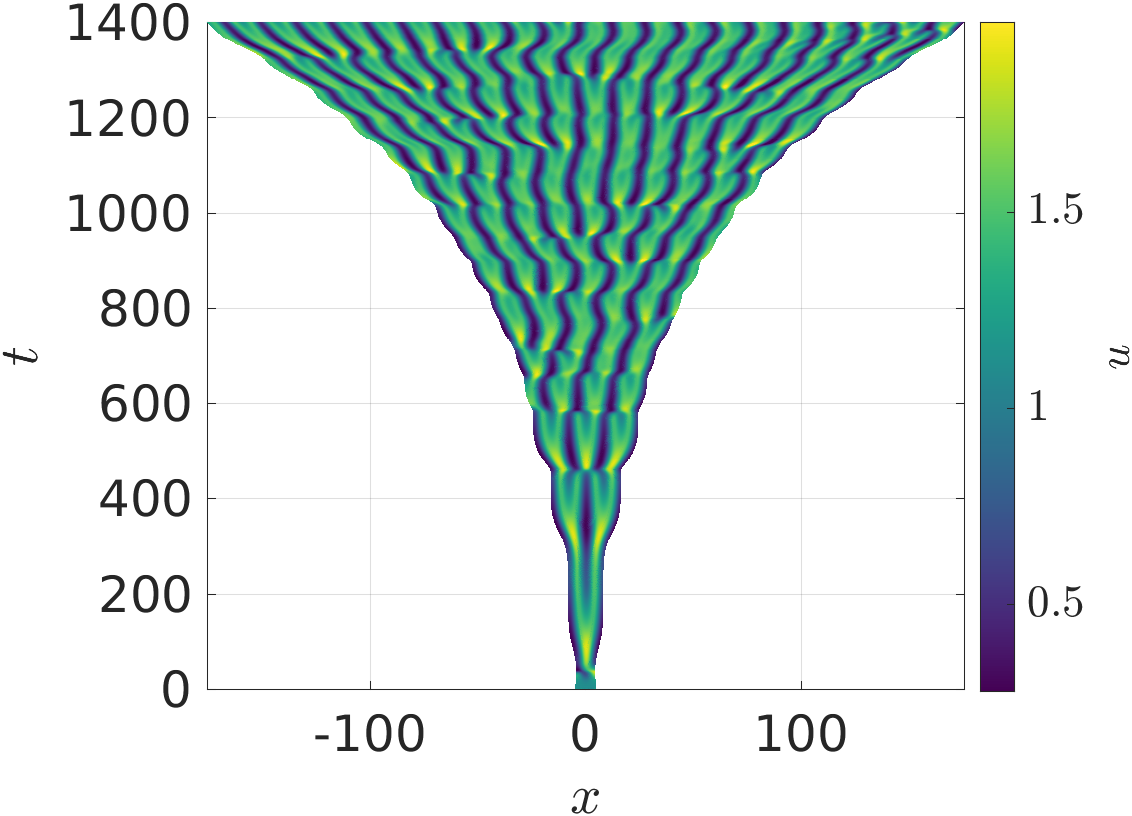}}\\
    \subfloat[$S =  0.162(u-1.3v)$]{\includegraphics[width=0.33\textwidth]{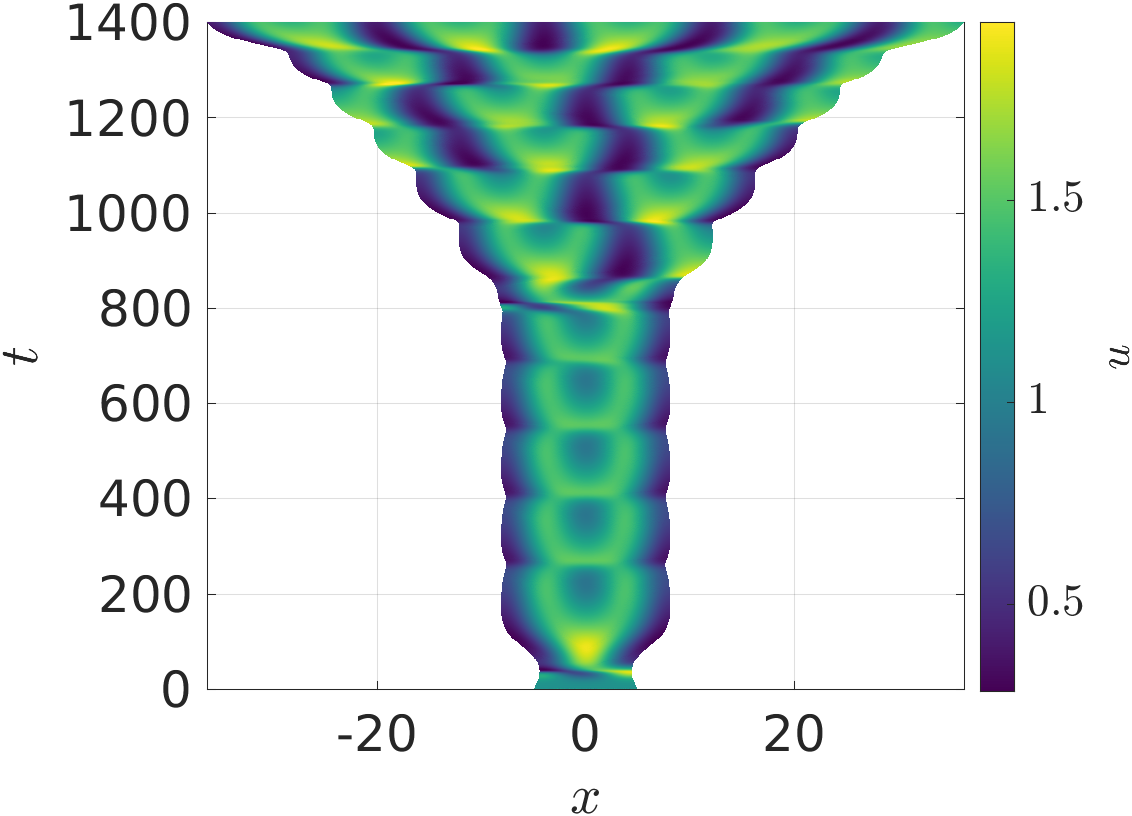}} &
    \subfloat[$S =  0.163(u-1.3v)$]{\includegraphics[width=0.33\textwidth]{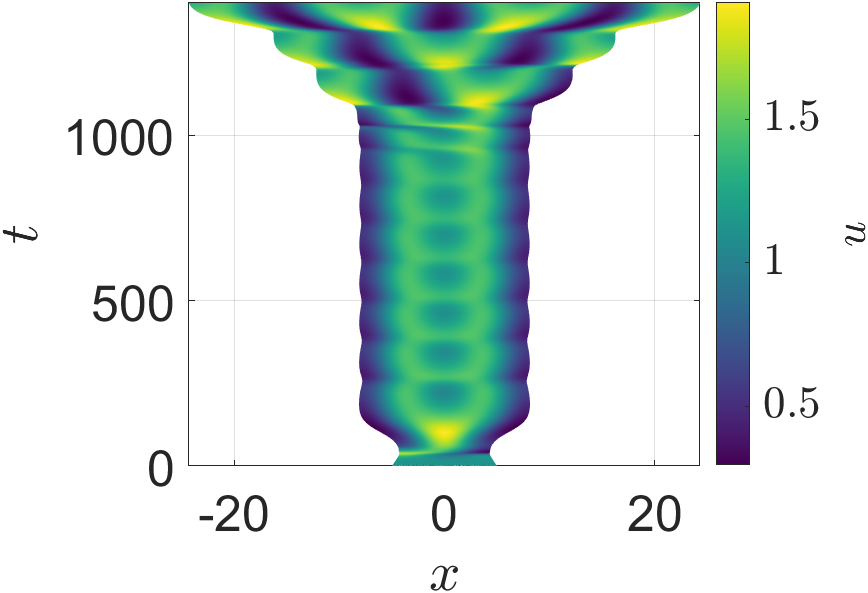}}
    \subfloat[$S =  0.163(u-1.3v)$, $L=5$]{\includegraphics[width=0.33\textwidth]{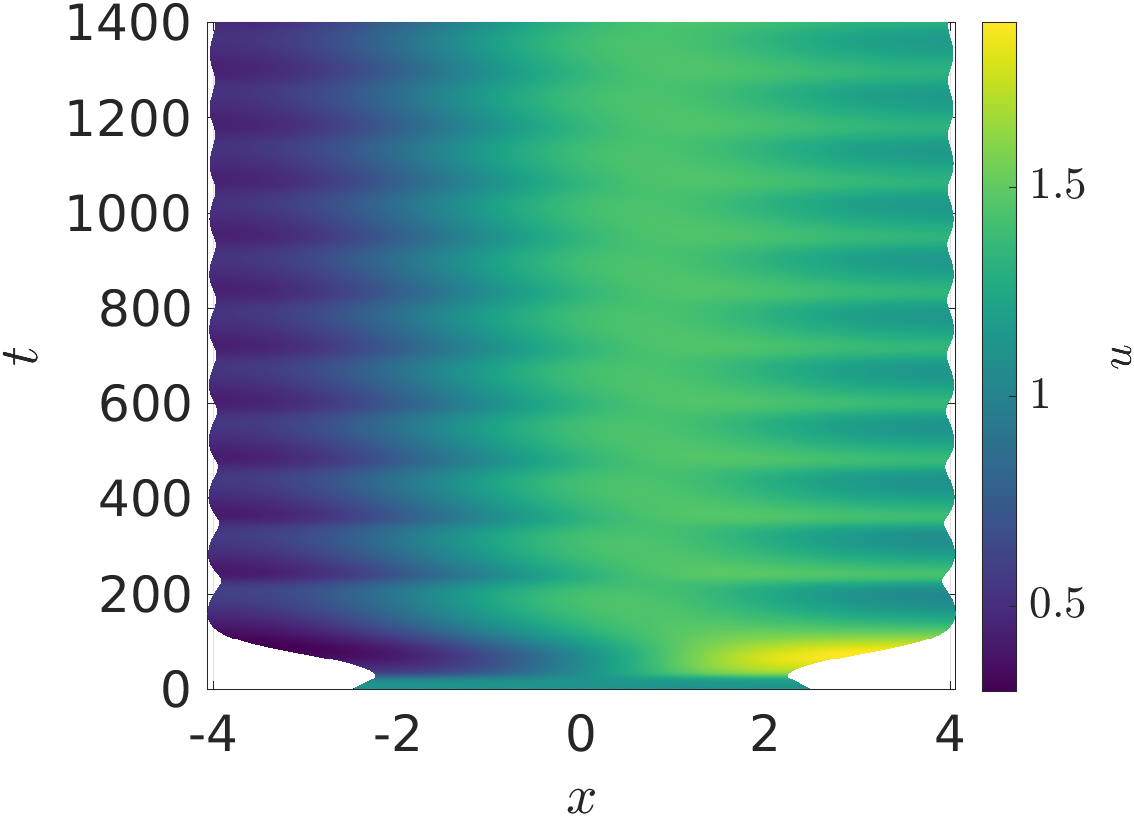}} \\
    \subfloat[$S = 0.165(u-1.3v)$]{\includegraphics[width=0.33\textwidth]{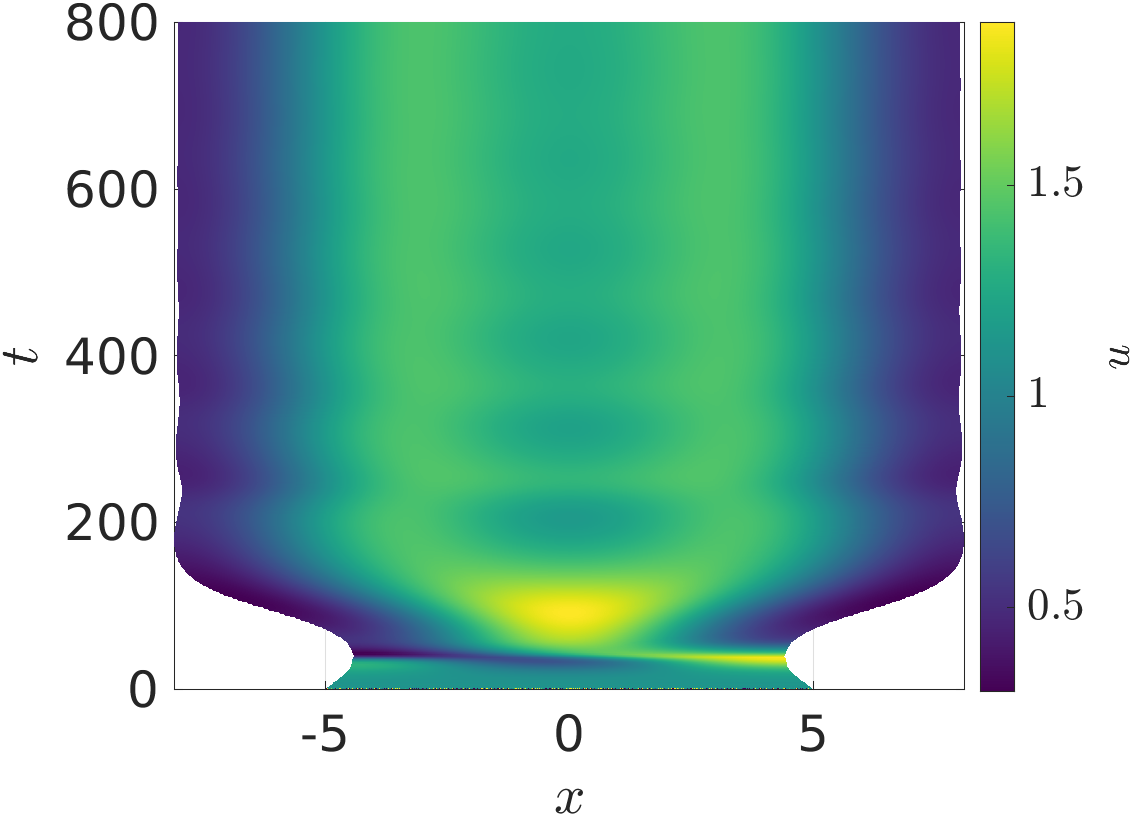}} &
    \subfloat[$S = 0.258(u-1.3v)$]{\includegraphics[width=0.33\textwidth]{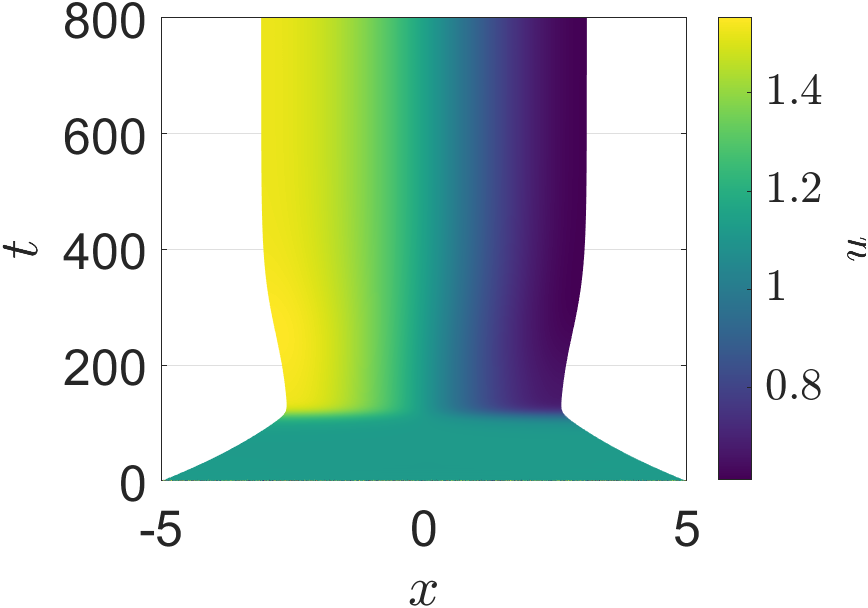}}
    \subfloat[$S =0.259(u-1.3v) $]{\includegraphics[width=0.33\textwidth]{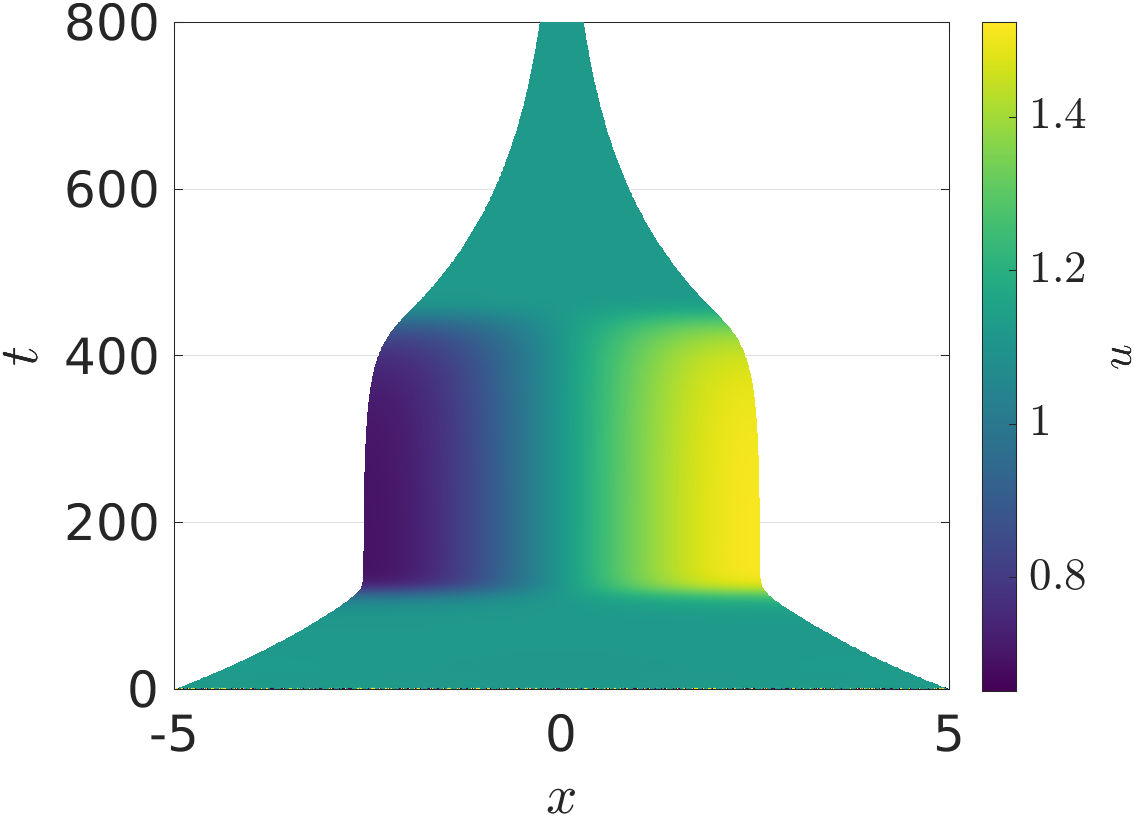}} \\
  \end{tabular}
    \caption{Values of $u$ from 1D simulations of the Schnakenberg kinetics \eqref{Schnack} under different growth scenarios. In all simulations, the initial domain length is $L=10$, except in (f) where $L=5$, with $a=0.01$, $b = 1.1$, $D_1=1$, and $D_2=40$.}
    \label{SchnackFig2}
\end{figure}

In Fig.~\ref{SchnackFig1Slices}, we plot particular profiles of $u$ and $S$ from the preceding simulations at selected time points. As detailed in the caption, the first two rows correspond to the slow and fast uniform growth rates, whereas the third row is the fastest thresholded growth scenario, and the fourth is the fastest growth case with $S \propto u$. The columns are, from left to right, at increasing fractions of the total simulation time. While there are some transient differences between uniform growth at different rates (compare (b) with (e)), broadly the same spike-doubling behaviour is observed, and away from insertion events the spikes maintain an approximately uniform spacing, as they do on fixed domains. In contrast, the thresholded growth shown in (g)-(i) initiates new regions not by splitting sharp peaks, but by growing sharp valleys between plateaus of high $u$ concentration, with growth localized at the ends of these plateaus. The profiles with $S \propto u$ are far more irregular throughout the simulation, with large spikes appearing and moving away from regions where new but smaller spikes initiate. This rapid movement of large spikes is a complex interplay of $u$ and $v$ having different regions of localization (approximately out-of-phase), and the nonlinearity inherent in the dilution term. 

We now explore a scenario where both domain growth and contraction occur simultaneously. In Fig.~\ref{SchnackFig2}, we give examples where $S = r(u-1.3v)$ with $r>0$ a constant. The factor of $1.3$ present in $S$ was chosen as the steady state values satisfy $u^* - 1.3v^*<0$, but integrating this expression for a patterned state generated on a large fixed domain gave a positive net growth rate. Depending on the constant of proportionality and initial domain length $L$, we observe exponential-like (though still complex) growth for $r \leq 0.161$ in (a)-(c), oscillations leading eventually to growth for $0.162 \leq r \leq 0.163$ in (d) and (e), oscillations or transient domain shrinking leading eventually to a fixed domain size for $0.164 \leq r \leq 0.259$ in (g) and (h), and, finally, domain shrinkage for $r \geq 0.259$ as in (i). Additional simulations within each range of $r$ values leads us to believe that, at least for a fixed initial condition and domain size, these qualitatively different regimes can be tuned via the parameter $r$. We did observe different dynamics for $L=5$ where, for $r=0.163$, we see sustained oscillations in panel (f). We confirmed the predictions from panels (f)-(h) appear to be genuine long-time behaviours by simulating these cases over timescales $20$ times longer and checking convergence in space and time steps. While these parameters are intentionally chosen to demonstrate this wide variety of behaviour, similar kinds of dynamics were observed for a range of choices of $S(u)$. One important remark is that while panels (g) and (h) show concentrations and domains approaching a fixed state, the system does not truly reach an equilibrium as the domain will continue expanding and contracting indefinitely. In other words, unlike panel (b) of Fig.~\ref{TWFig}, the value of $\mu$ given by \eqref{growth_law_1D} never reaches an equilibrium.
 
 \subsection{Pattern Formation in Gierer-Meinhardt \& FitzHugh-Nagumo Systems}
 
 \begin{figure}
      \begin{tabular}{cc}
    \subfloat[$S = 0.00001(6v-u^2)$]{\includegraphics[width=0.33\textwidth]{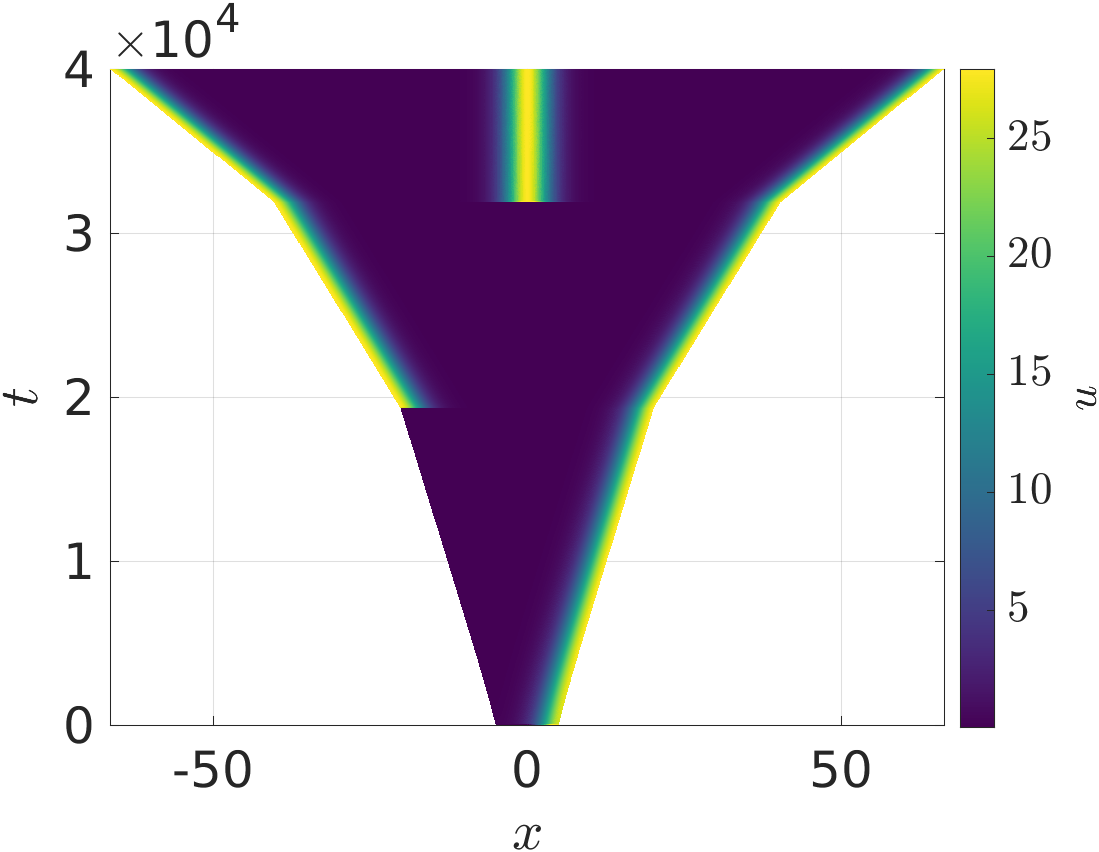}} &
    \subfloat[$S = 0.0002(6v-u^2)$]{\includegraphics[width=0.33\textwidth]{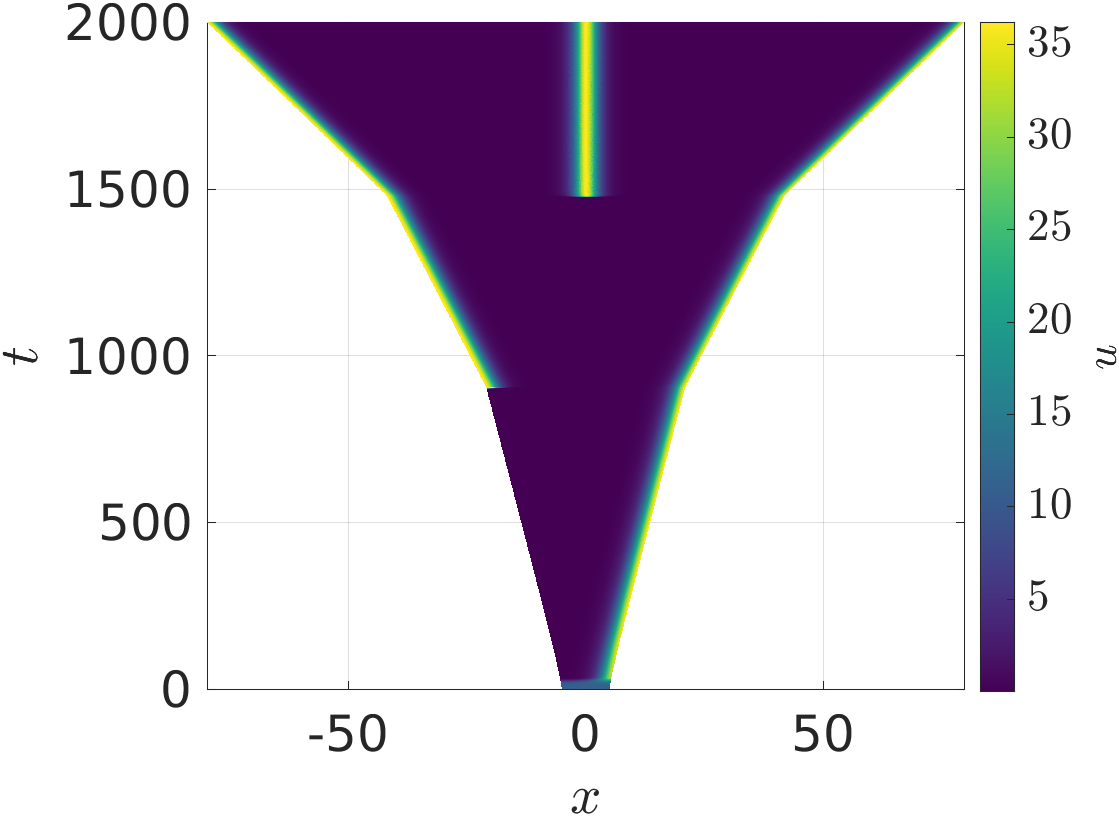}}
    \subfloat[$S =  0.0005(6v-u^2)$]{\includegraphics[width=0.33\textwidth]{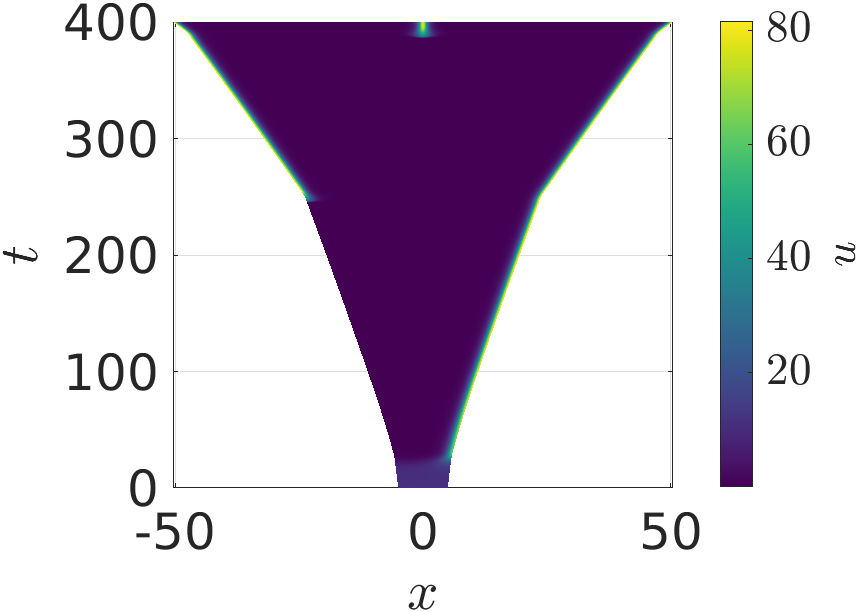}}\\
    \subfloat[$S =  0.000065$]{\includegraphics[width=0.33\textwidth]{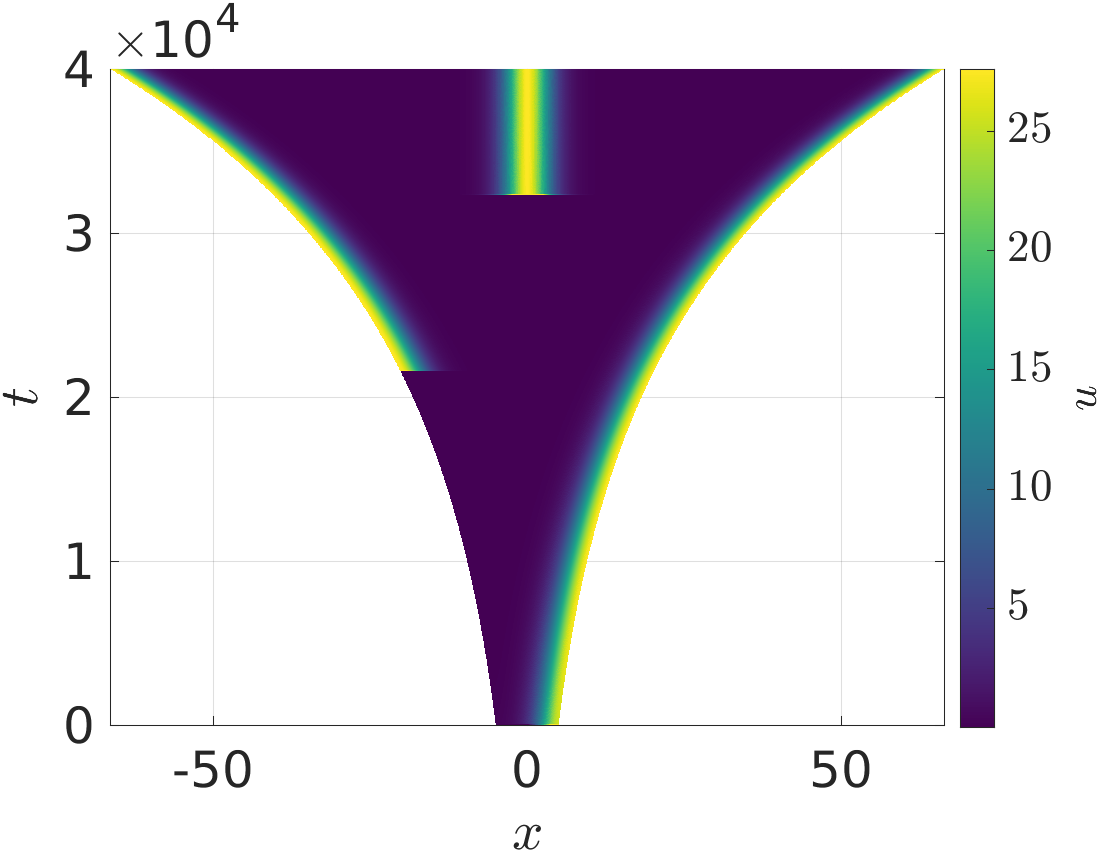}} &
    \subfloat[$S =  0.0014$]{\includegraphics[width=0.33\textwidth]{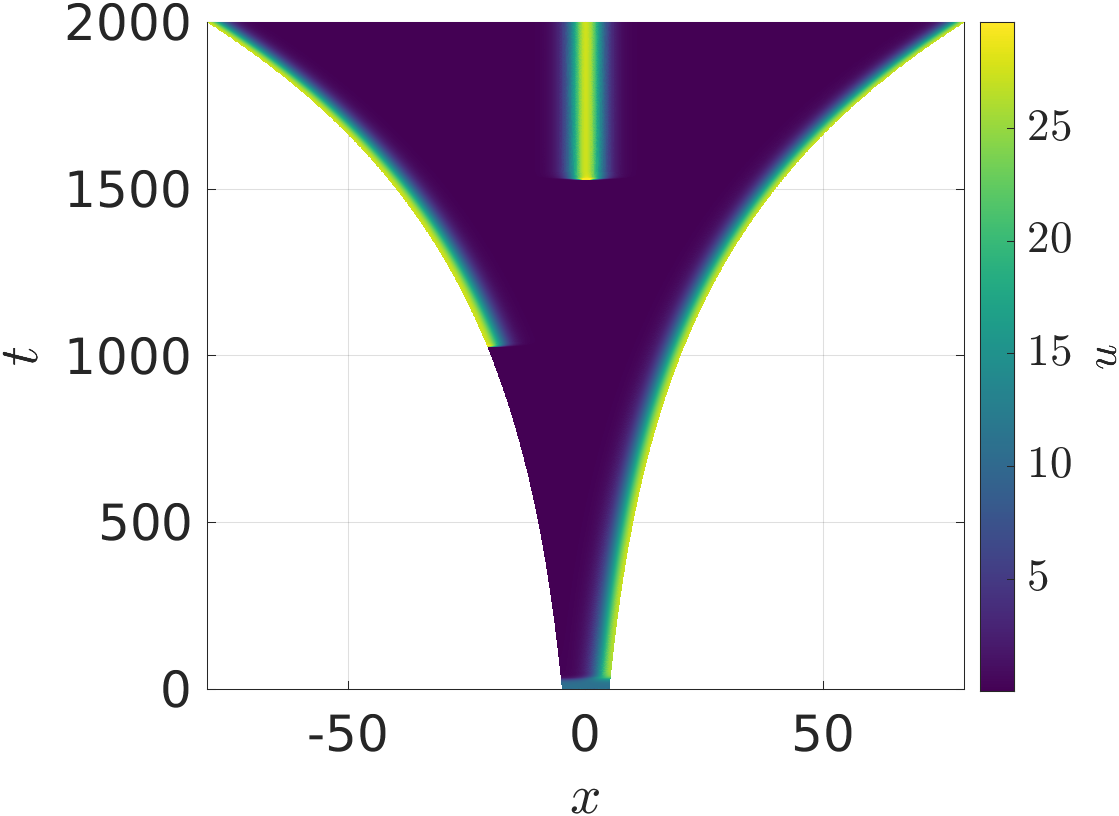}}
    \subfloat[$S =  0.0058$]{\includegraphics[width=0.33\textwidth]{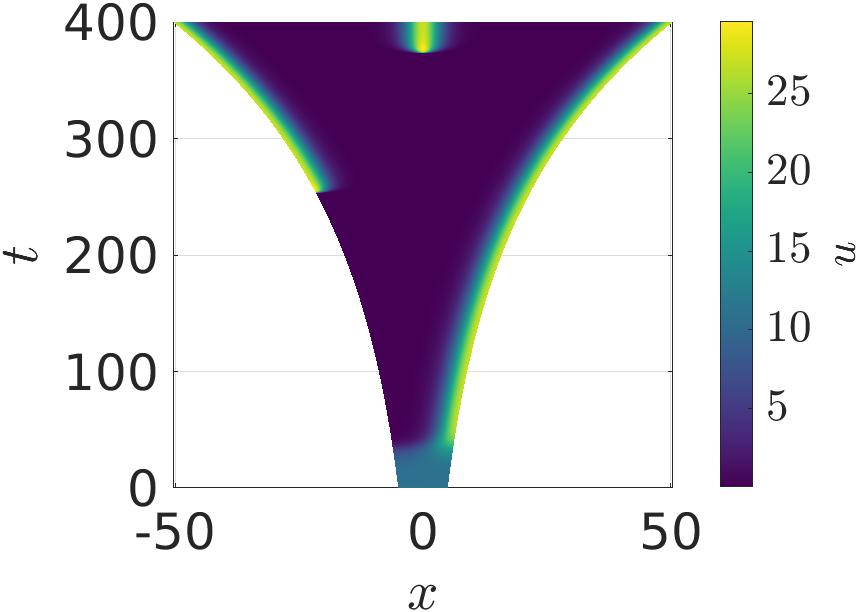}} \\
  \end{tabular}
    \caption{Values of $u$ from 1D simulations of the Gierer-Meinhardt kinetics \eqref{GM} under different growth scenarios. In all simulations the initial domain length is $L=10$, with $a = 0.01$, $b = 0.5$, $c = 5.5$, $D_1=1$ and $D_2=200$. The timescale and growth rates in (a)-(c) are chosen to match those in (e)-(g) so that the final simulation time is on a domain of exactly the same size.}
    \label{GMFig}
\end{figure}

We next consider examples of competing growth and contraction using the Gierer-Meinhardt kinetics \eqref{GM}. As opposed to the preceding subsection, these kinetics have the activator $u$ in phase with the faster-diffusing inhibitor $v$, so that both species become highly localized in the same spike regions, leading to localized regions of contraction and growth within the domain near spikes. We choose the form $S(u,v) = r(6v-u^2)$ for varying $r>0$ so that overall there is a small net positive growth rate despite large local contraction and growth. The particular choice of this functional form is partially inspired by the form of $g(u,v)$ but, as in the choice of the nonlinearity in Fig.~\ref{SchnackFig2}, is chosen primarily to illustrate some of the interesting phenomena that can occur in combining growth and contraction.
    
We plot three examples of this growth in panels (a)-(c) of Fig.~\ref{GMFig}, with panels (d)-(f) giving comparable uniform growth rates. The first thing to notice is that, locally, the domains in (a)-(c) are all growing linearly within the region where only a single spike is stable, independently of the speed of the growth. Such a linear growth is an emergent characteristic of the choice of $S(u,v)$, with $S$ not explicitly depending on time. Overall, the profiles of the cases of uniform exponential growth and concentration-dependent growth are remarkably similar. We also see that the spikes in (b) and (c) are at a much higher amplitude, and are much more localized than in the other simulations. This effect is due to local contraction, which is, in some sense, the opposite of the dilution effect observed in Fig.~\ref{SchnackFig1}(f) and leads to larger amplitudes over smaller regions. 

\begin{figure}
      \begin{tabular}{cc}
    \subfloat[$S = 0.01u$]{\includegraphics[width=0.33\textwidth]{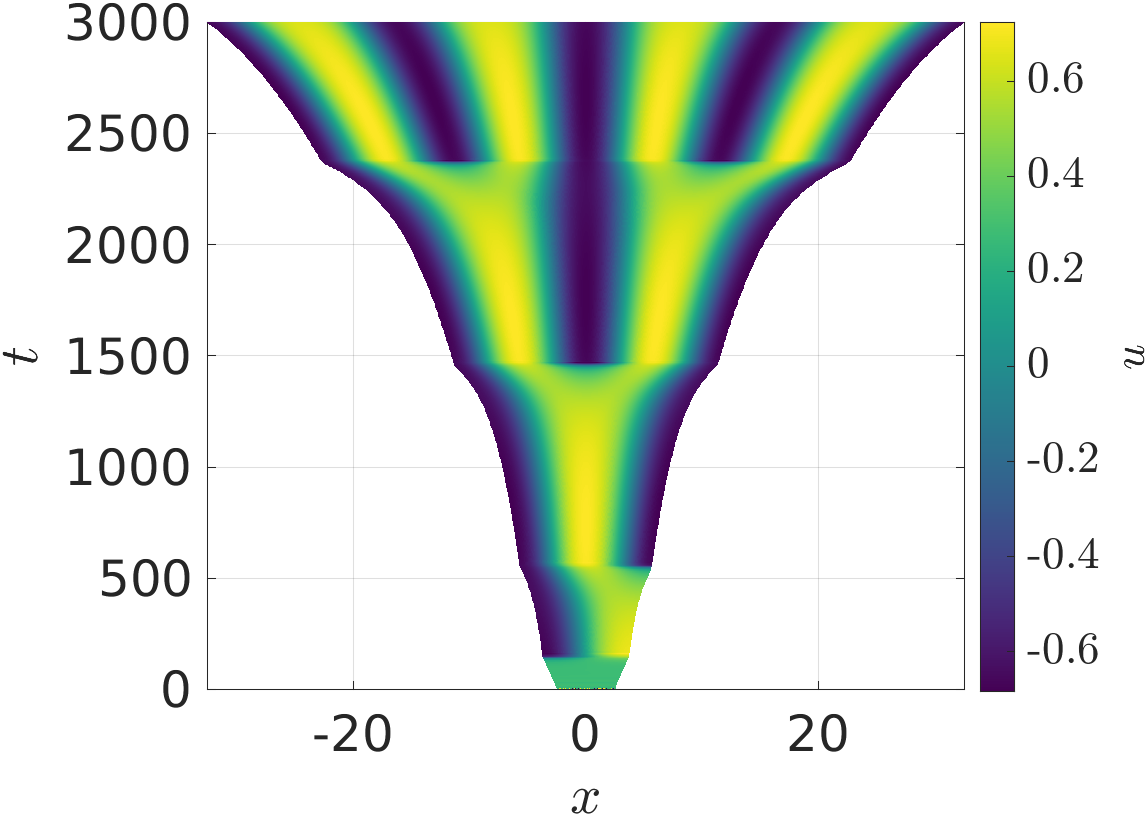}} &
    \subfloat[$S = 0.05u$]{\includegraphics[width=0.33\textwidth]{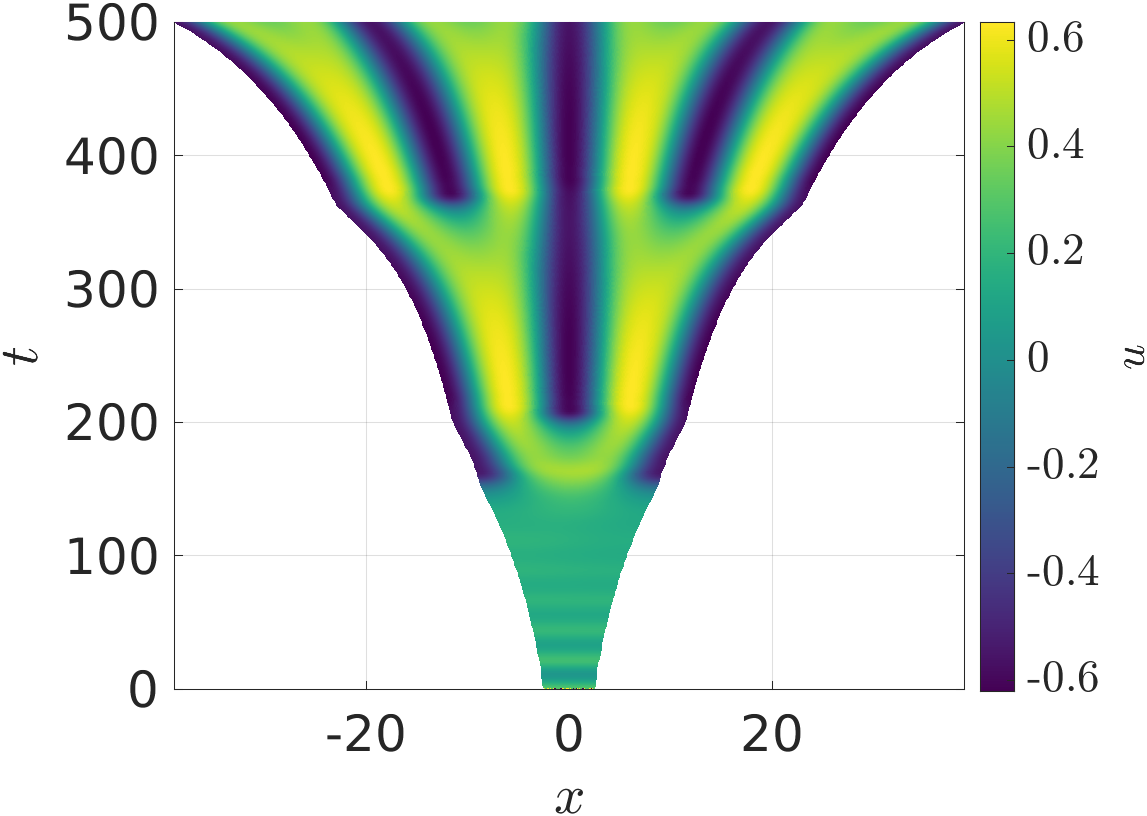}}
    \subfloat[$S =  0.2u$]{\includegraphics[width=0.33\textwidth]{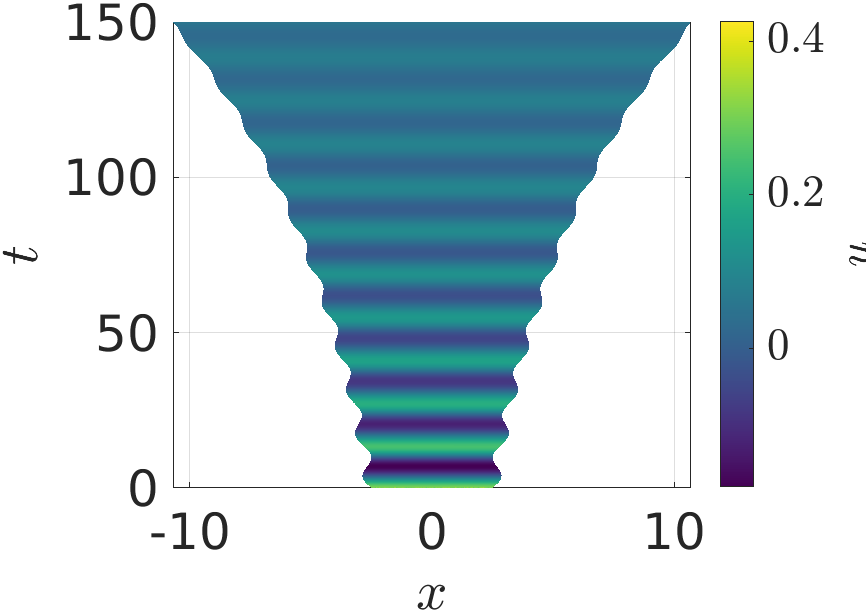}}\\
    \subfloat[$S =  0.00086$]{\includegraphics[width=0.33\textwidth]{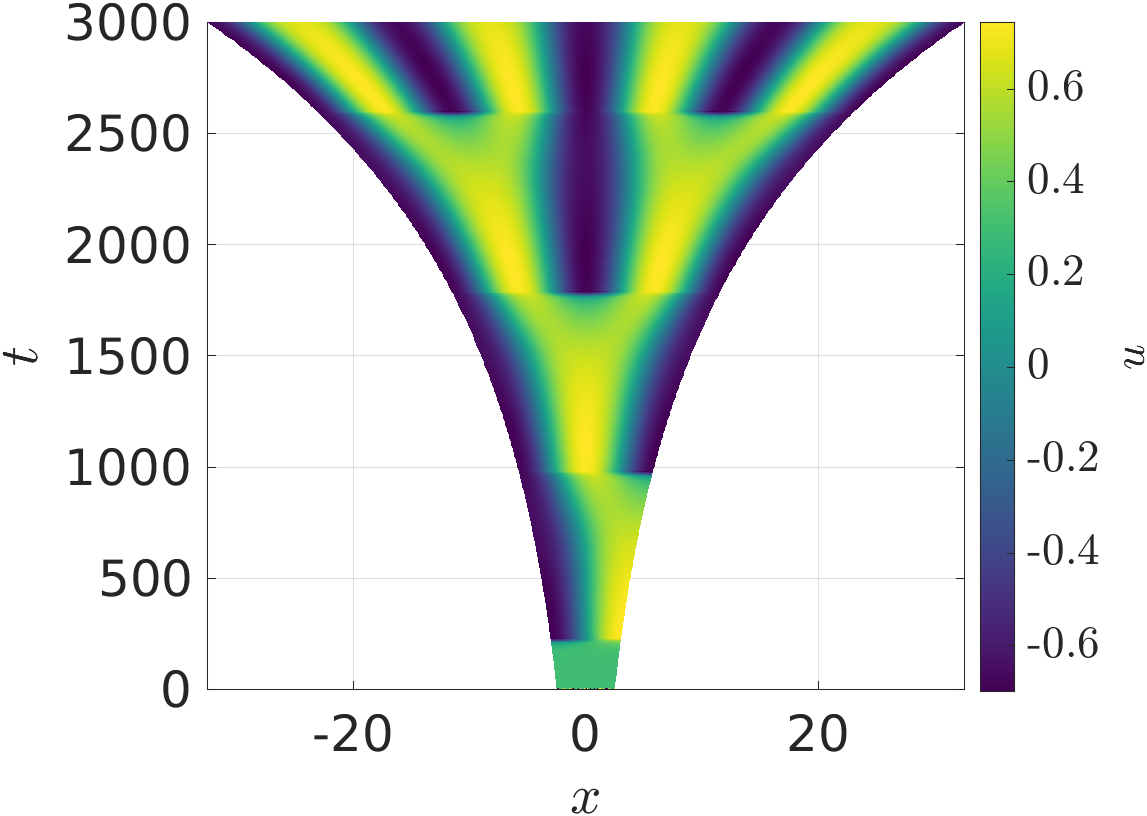}} &
    \subfloat[$S =  0.0055$]{\includegraphics[width=0.33\textwidth]{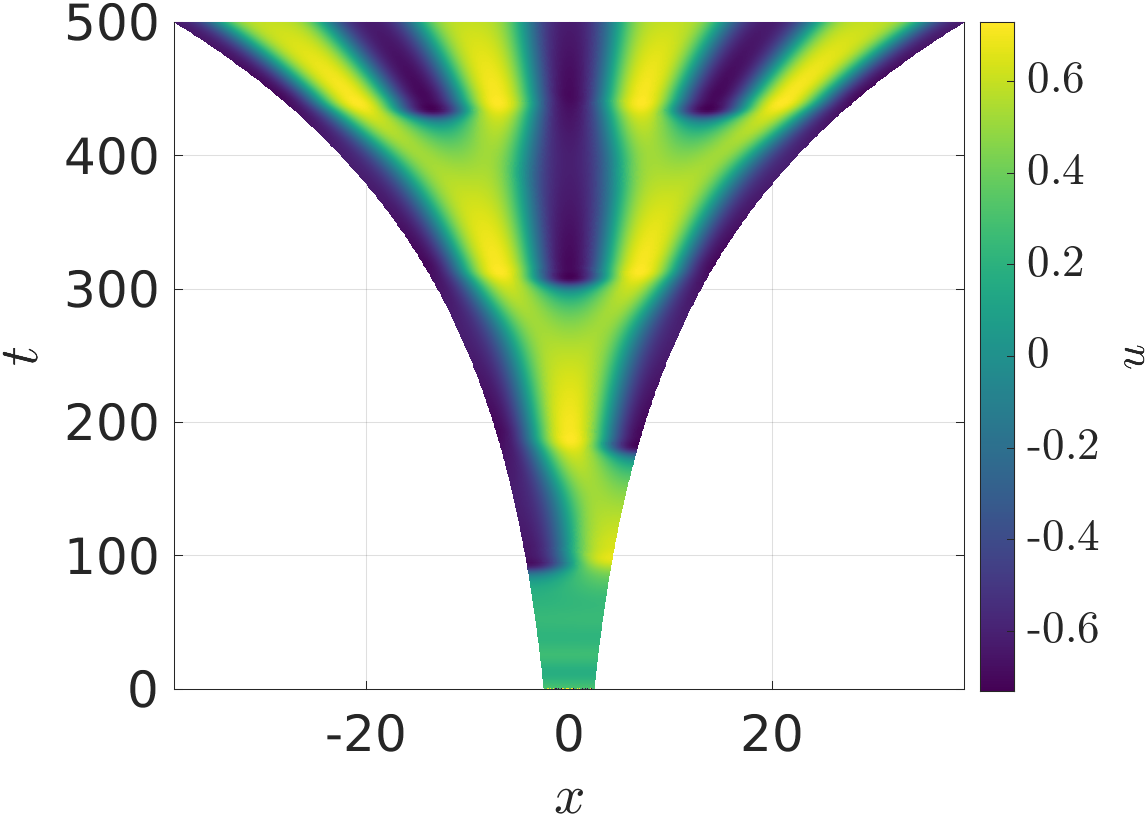}}
    \subfloat[$S =  0.0097$]{\includegraphics[width=0.33\textwidth]{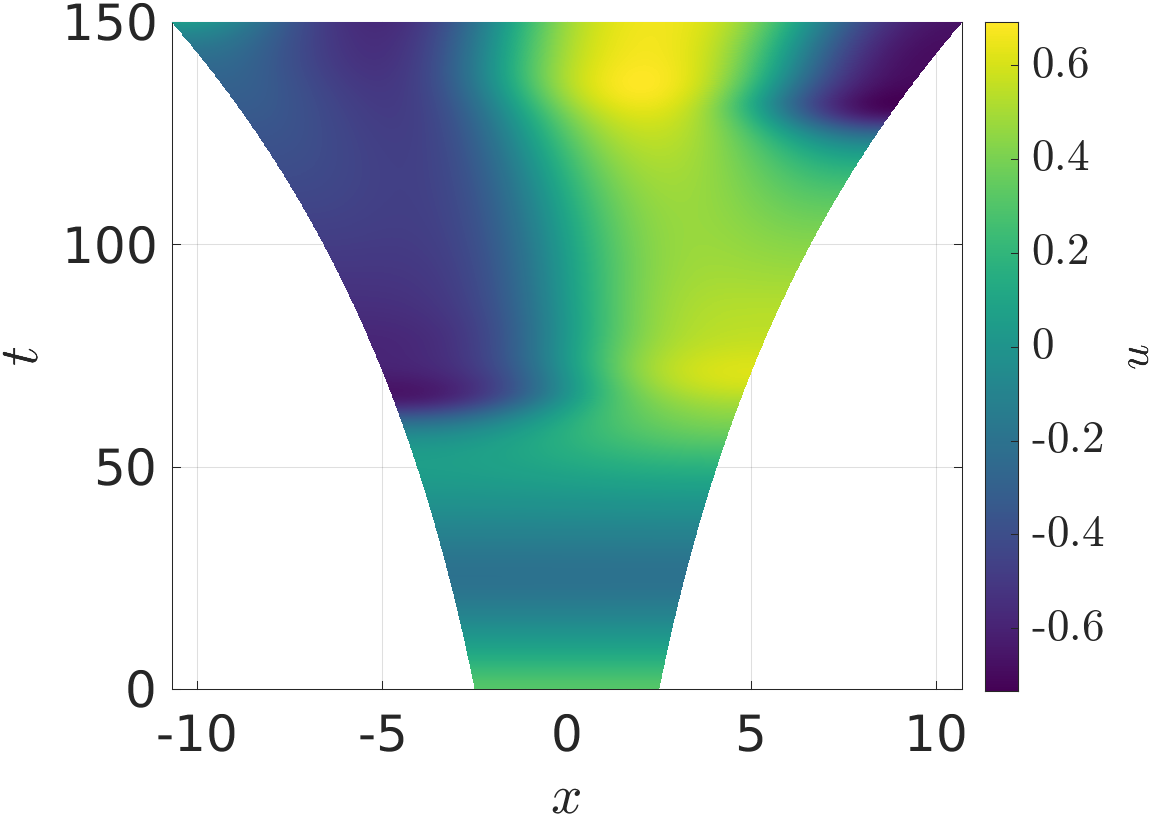}} \\
  \end{tabular}
    \caption{Values of $u$ from 1D simulations of the FitzHugh-Nagumo kinetics \eqref{FHN} under different growth scenarios. In all simulations the initial domain length is $L=5$, with $a = 1.01$, $b = 1$, $c = 1$, $i_0 = 1$, $D_1=1$ and $D_2=2.5$. The timescale and growth rates in (a)-(c) are chosen to match those in (d)-(f), so that the final simulation time is on a domain of exactly the same size.}
    \label{FHNFig}
\end{figure}

Lastly, we consider the FitzHugh-Nagumo kinetics \eqref{FHN} in order to explore the interplay of homogeneous oscillations and pattern formation in the concentration-dependent setting. We give three examples of simple linear functions $S(u)$ in Fig.~\ref{FHNFig}, though note that since $u$ represents a voltage, it can be both positive and negative, and so this results in local domain growth and contraction. For slow growth, panels (a) and (d) show qualitatively similar behaviour, though note that (a) has changes in growth rate as the number or position of spikes change as in Fig.~\ref{GMFig}(a)-(c). For larger growth rates, however, we observe that the concentration-dependent case in (b) undergoes a period of rapid homogeneous oscillations before undergoing pattern formation, whereas the comparable uniform simulation in (e) has a shorter period of longer oscillations before forming patterns.  Finally, in (c) we see that for sufficiently fast growth, rapid homogeneous decaying oscillations lead to periods of growth and contraction (with an overall growing tendency), whereas  in (f) we observe one period of oscillation before the beginning of patterning takes places.
    
    The interactions between homogeneous oscillations, particularly those coming from a Hopf bifurcation of the kinetics \eqref{FHN} are well-studied, e.g.~see \citep{sanchez2019turing} and the references therein. The important contrast between the uniform growth and concentration-dependent cases can be in part explained by considering the  base state given by \eqref{u*}. In particular, for all parameters in Fig.~\ref{FHNFig}, there is a stable spiral  steady state of the kinetics located at  $(u^*,v^*)=(0,v^*)$ for some $S$-dependent $v^*>0$. However for the linear choice of $S$ in panels (a)-(c), this stable spiral has a much larger imaginary eigenvalue, corresponding to higher frequency temporal oscillations, as observed. In fact the concentration-dependent case also undergoes a Hopf bifurcation case for slightly larger growth rates, leading to purely periodic growth and contraction as in the early time shown in (c). Using the inequality in Theorem 1, we find that panel (a) and the uniform growth cases in panels (d)-(f) all exhibit a growing range of unstable wavemodes with $k=0$ stable, whereas panels (b) and (c) exhibit an increasing range of unstable wavemodes including $k=0$ so that the base state is itself unstable, and whether or not a spatial pattern emerges is due to nonlinear competition between modes. While the precise condition determining whether or not an inhomogeneous perturbation leads to sustained amplitude patterns is not precisely determined by the Conditions of Theorem 1 (this is determined by nonlinear mode competition), one can gain insight insight the impact of growth on the dynamics of the spatially homogeneous state given by \eqref{u*}.

\section{Two-Dimensional Simulations}\label{2D_Numerics}

\begin{figure}
    \centering
    \subfloat[]{\includegraphics[width=0.4\textwidth]{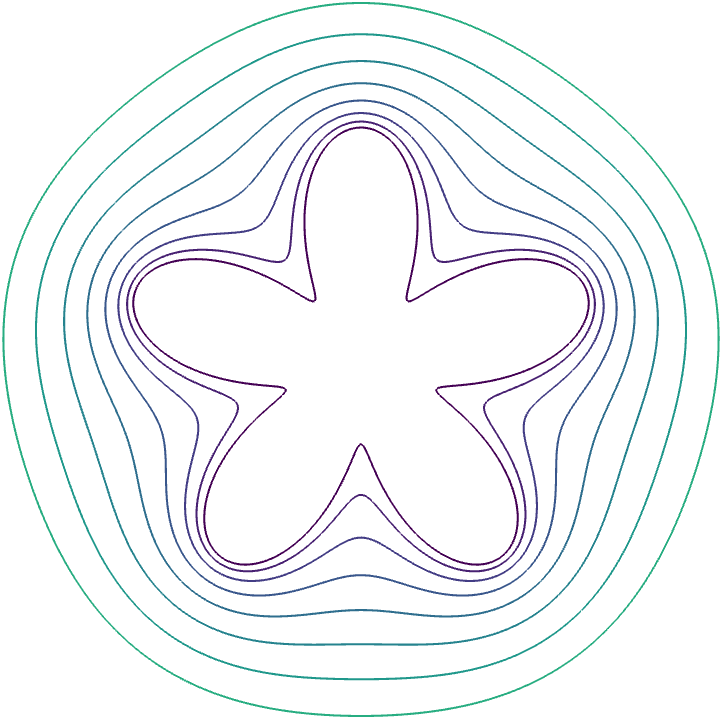}}\hfil
    \subfloat[]{\includegraphics[width=0.4\textwidth]{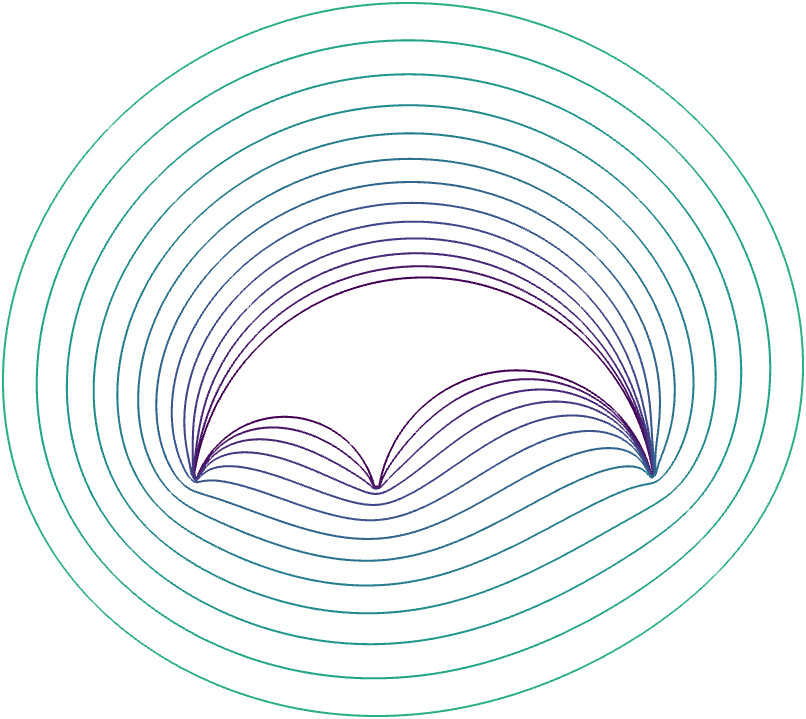}}
    \caption{Two examples of the growth of a boundary with the locally uniform expansion rate $S = 0.001$. In (a) we start with a starfish-like domain given by equation (13) in \cite{krause2020isolating} for $\gamma=0.8$ and $L=1$. In (b) we start with an Arbelos-like domain composed of the boundaries of three semicircles of radius 1, 2/5, and 3/5 respectively, with some truncation done near the lower boundary to prevent issues with extremely small finite elements. Boundary curves shown are arranged so that larger enclosed areas correspond to later times, with times uniformly sampled.} 
    \label{NonUniform2DFig}
\end{figure}

Next we consider 2D simulations of concentration-dependent growth using the formulation given in Section \ref{2D_Model_Sect}. We implemented  Equations \eqref{RD_Eqnu} and \eqref{Poisson}-\eqref{Volume2D} in the finite-element software COMSOL. The domain was discretized using second-order triangular finite elements, with timestepping done via a generalized backwards-difference formula of orders 1 to 5 with a tolerance set at $10^{-4}$. Initial steps within each iteration were taken as $10^{-8}$, with a maximum step constraint taken as $1$ or $0.1$ to confirm accurate resolution of the domain growth. As in the 1D case, each simulation was broken into a number of shorter iterations using automated remeshing via MATLAB LiveLink, with concentrations interpolated onto a new Lagrangian domain at the end of each iteration. In the 2D examples, we fixed the mesh size parameters (maximal and minimal element sizes, growth rates, etc) so that domain growth tended to increase the number of elements used. Checks were carried out in the mesh size parameters, number of iterations used, and maximum timesteps taken to ensure convergence for specific simulations. All code and associated documentation can be found at \citep{Krause_Concentration_Dependent_Growth_2022}.

We begin by showing how the constitutive assumptions on the flow impact uniform growth rates where $S$ is constant. Importantly, unlike in past work on uniformly growing domains \citep{plaza2004effect, krause2019influence, van_gorder_growth_2019}, a locally uniform expansion of the domain does not lead to a uniform dilation given our assumptions on the flow $\bm{a}$ corresponding to normal growth at the boundaries. We give two examples of this in Fig.~\ref{NonUniform2DFig}, where larger enclosed areas correspond to the domain at later times. Here, we see that two highly non-convex domains both grow in a manner where curvature is uniformized at the boundary. This is remarkably similar to curve-shortening flows, which are well-studied in geometric analysis \citep{chou2001curve} and are related to other kinds of manifold evolution, such as the famous Ricci flow \citep{chow2006hamilton}. It is also this quality, that uniform local growth does not lead to isotropic dilation, that prevents the use of the linear stability analysis given in Section \ref{LinStab_Sect} beyond 1D. 

\subsection{Stable Inhomogeneous Solutions in Scalar Bistable Equations}

\begin{figure}
    \centering
    \begin{overpic}[width=0.9\textwidth]{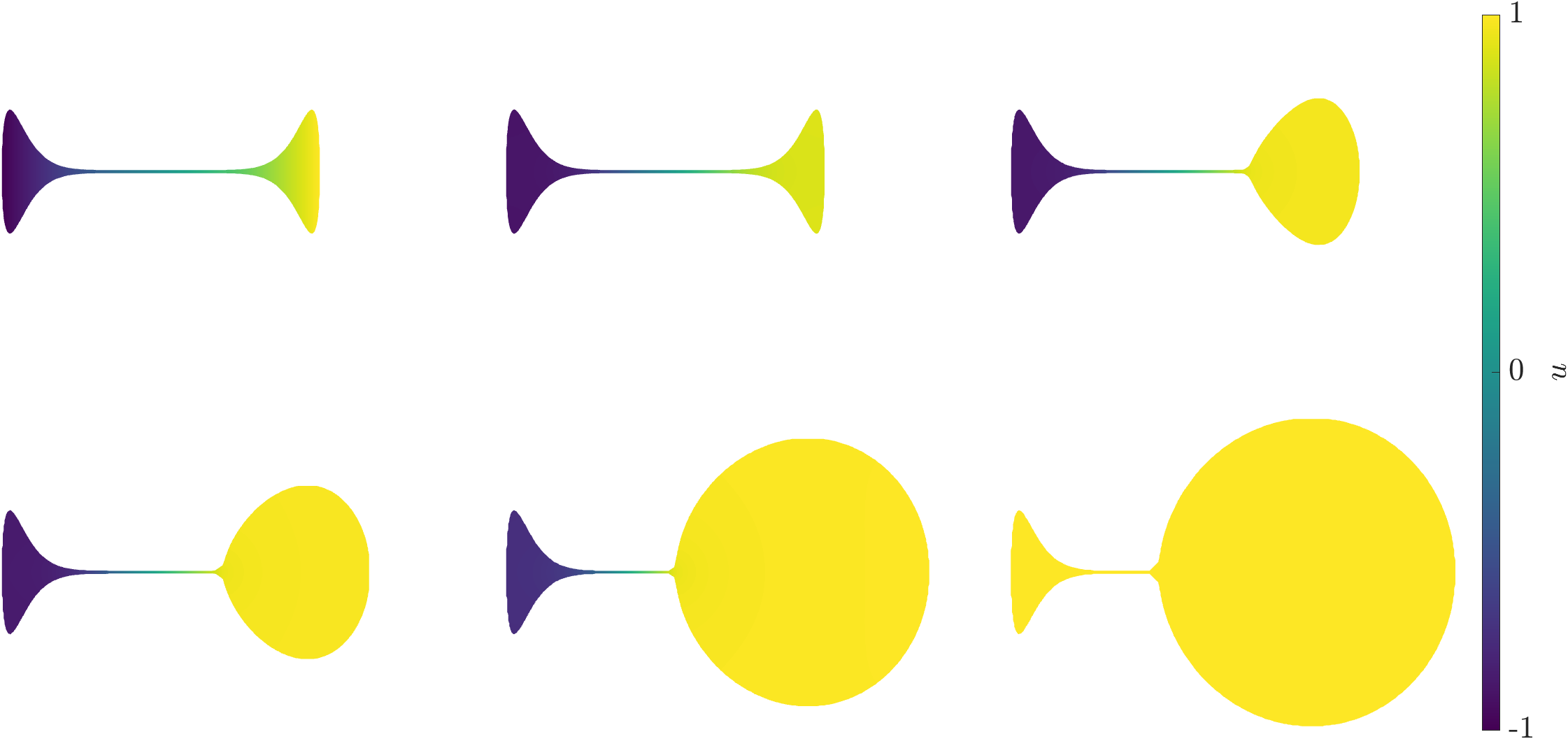}
    \put(9,27){(a)}
    \put(42,27){(b)}
    \put(76,27){(c)}
    \put(9,0){(d)}
    \put(42,0){(e)}
    \put(76,0){(f)}
    \end{overpic}
    \caption{Values of $u$ from 2D simulations of the scalar bistable kinetics \eqref{Bistable} in the dumbbell-shaped domain given by \eqref{DumbbellEqn} with the diffusion parameter $D=1$ and growth rate $S = 0.000125(1+\tanh(50(u-0.9))$. Iterations are shown at times $t=0, 8, 2392, 3192, 4792,$ and $5272$.}
    \label{Dumbbell2DFig}
\end{figure}

\begin{figure}
\centering 
    \begin{overpic}[width=0.75\textwidth]{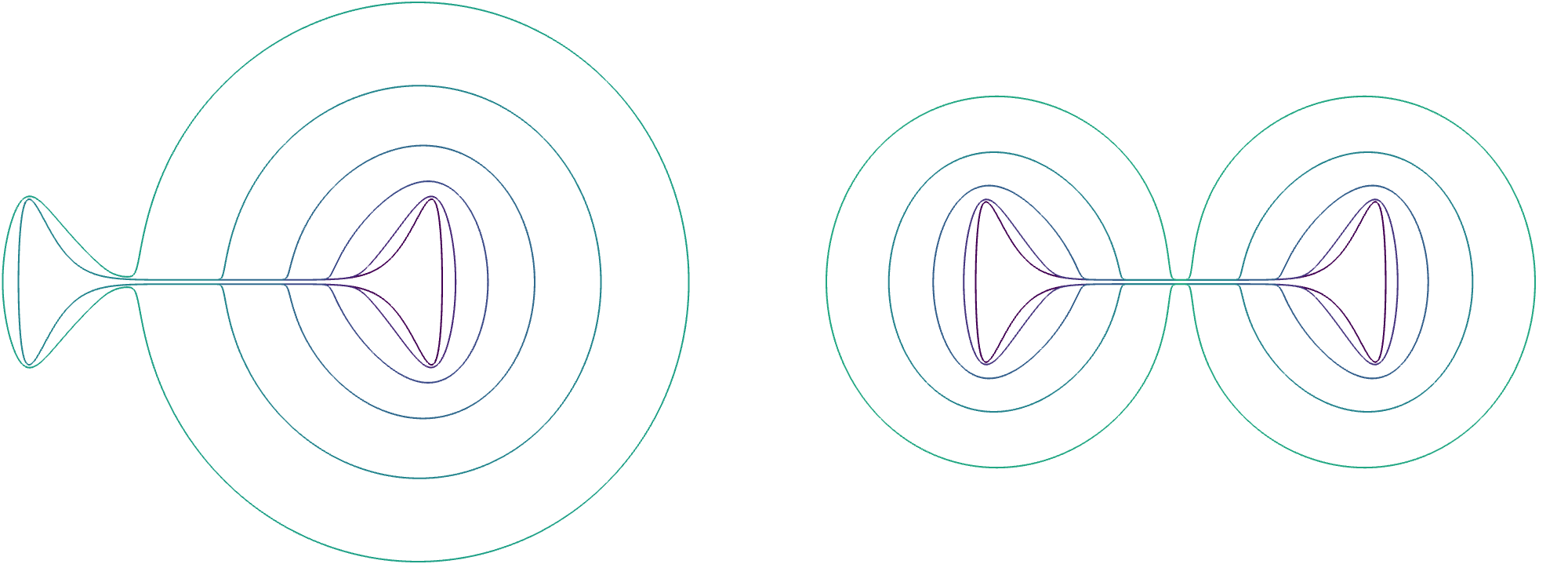}
    \put(25,-4){(a)}
    \put(75,-4){(b)}
    \end{overpic}
    \vspace{1em}
    \caption{Plots of the domain boundary from 2D simulations of the scalar bistable kinetics \eqref{Bistable} in the dumbbell-shaped domain given by \eqref{DumbbellEqn} with the diffusion parameter $D=1$. Panel (a) corresponds to Fig.~\ref{Dumbbell2DFig} with $S = 0.000125(1+\tanh(50(u-0.9))$, and panel (b) to a simulation with $S = 0.000125(1+\tanh(50(|u|-0.9))$. Boundary curves shown are arranged so that larger enclosed areas correspond to later times, with times uniformly sampled.}
    \label{DumbbellBoundary2DFig}
\end{figure}

We next consider a simple example of concentration-dependent growth for a scalar reaction-diffusion equation, namely using the bistable kinetics \eqref{Bistable}. It is known that scalar reaction-diffusion equations with Neumann data cannot admit stable inhomogeneous steady states on 1D or general convex domains \citep{casten1978instability, matano1979asymptotic} and in fact this restriction of long-time behaviour even generalizes to time-dependent semilinear reaction-advection-diffusion problems \citep{hess1989periodic}. In contrast, for non-convex domains bistable (or more generally multistable) kinetics can be used to make `locally' stable regions separated by thin channels which are inhomogeneous, and there is a range of literature exploring these kinds of structures in many settings, for example \citep{matano1979asymptotic, matano1983pattern, ward1999metastable}. We explore a few simple scenarios of concentration-dependent domain evolution to see what happens to such heterogeneous fronts.

We parameterize an initial dumbbell-like domain as,
\begin{equation}\label{DumbbellEqn}
x(s) = (s^8+0.01)\sqrt{1-s^8}, \quad y(s) = -(s^8+0.01)\sqrt{1-s^8}, \quad s \in [-1,1], 
\end{equation}
and take as an initial condition $u(0,x,y) = x$, so that the concentrations approximately equilibrate to the value of $u= -1$ in the left region and $u=1$ in the right, with some diffuse boundary in the thin channel between them and parts of lobular regions near this channel. See Fig.~\ref{Dumbbell2DFig}(a) for this initial condition, which only changes slightly from the quasi-static value $8$ time units later in panel (b). In subsequent panels of Fig.~\ref{Dumbbell2DFig}, we show domain growth that occurs where $u$ is approximately greater than $0.9$, and hence is localized in the rightmost region of the domain. Eventually this region grows so large that the small boundary to diffusion imposed by the thin channel is insufficient to prevent this steady state from overcoming the $u=-1$ state, and the entire domain approaches $u=1$, as would happen on a convex domain. Given the slow growth timescale, we anticipate this destabilization of the inhomogeneous steady state occurs when the geometry destabilizes the heterogeneous front. This boundary could in principle be computed \citep{gokieli2005reaction}, but we do not do so here.

We also consider a growth function $S$ where the domain grows for $|u|>0.9$, and compare this to the simulation in Fig.~\ref{Dumbbell2DFig} by plotting the boundary of these two cases in Fig.~\ref{DumbbellBoundary2DFig}. In panel (a), $S$ is positive during most of the simulation time only in the rightmost region when $u$ is near the value of 1. After the whole domain reaches this value, as shown in Fig.~\ref{Dumbbell2DFig}(f), both regions and the (now quite small) intermediate channel both grow. In contrast, the simulation in panel (b) of Fig.~\ref{DumbbellBoundary2DFig} never reaches a uniform value of $u$, and both regions expand uniformly as shown but maintain the values of $u\approx -1$ on the left and $u \approx 1$ on the right. 

Intriguingly, if the growth is reversed so that the domain locally shrinks when $u>0.9$ (that is, we set $S \to -S$ from the simulation in Fig.~\ref{Dumbbell2DFig}), the rightmost region shrinks slightly, but it eventually reaches a state where $0 < u <0.9$ and the growth appears to halt asymptotically, due to diffusion across the thin channel reducing the value of $u$ in the rightmost region. At $t=120,000$, the maximal value of $u$ is approximately $0.836$. and the overall contraction rate of the domain is substantially smaller -- the ratio of the integral of $S$ at $t=8$ divided by the integral of $S$ at $t=120,000$ is approximately $353$, indicating substantially slower contraction at the later time.

\subsection{Concentration-Dependent Growth in 2D Gierer-Meinhardt Systems}

\begin{figure}
    \centering
    \begin{overpic}[width=0.9\textwidth]{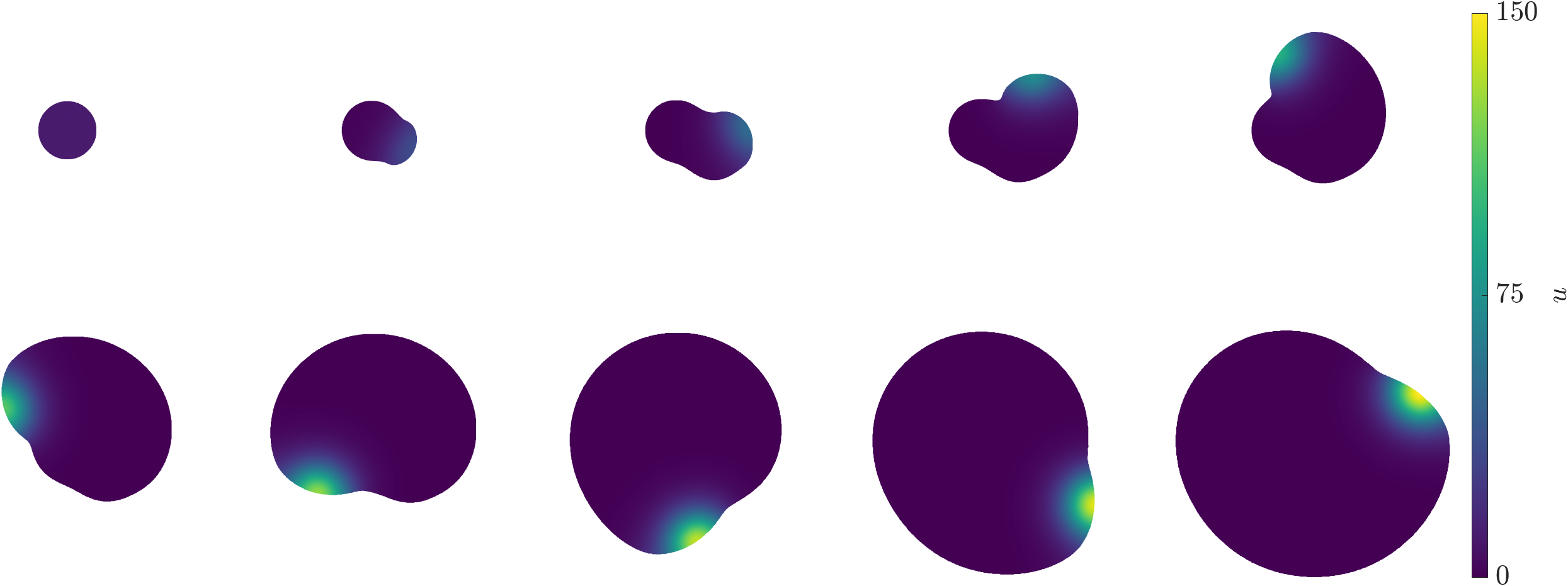}
    \put(3,22){(a)}
    \put(23,22){(b)}
    \put(43,22){(c)}
    \put(63,22){(d)}
    \put(83,22){(e)}
    \put(3,-2){(f)}
    \put(23,-2){(g)}
    \put(43,-2){(h)}
    \put(63,-2){(i)}
    \put(83,-2){(j)}
    \end{overpic}
    \vspace{1em}
    \caption{Values of $u$ from 2D simulations of the Gierer-Meinhardt kinetics \eqref{GM} in an initially circular domain of radius $3$. The parameters used are $D_1=1$, $D_2=1000$, $a=0.01$, $b=0.5$, $c=5.5$, with a growth rate of $S = 0.001(1+tanh(100(u-22))$. Iterations are shown at equally spaced times with all panels using the same spatial scale.}
    \label{Spiral2DFig}
\end{figure}

\begin{figure}
    \centering
    \begin{overpic}[width=0.9\textwidth]{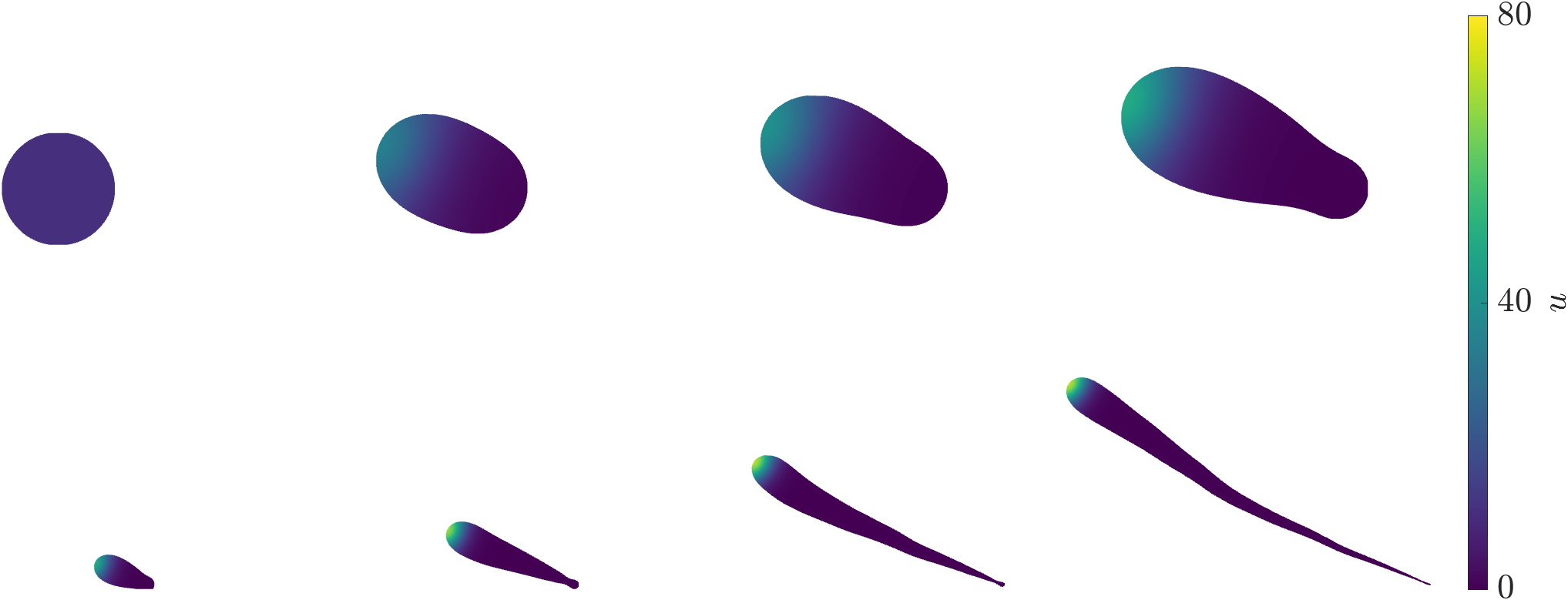}
    \put(2,19){(a)}
    \put(27,19){(b)}
    \put(52,19){(c)}
    \put(77,19){(d)}
    \put(2,-2){(e)}
    \put(27,-2){(f)}
    \put(52,-2){(g)}
    \put(77,-2){(h)}
    \end{overpic}
    \vspace{1em}
    \caption{Values of $u$ from 2D simulations of the Gierer-Meinhardt kinetics \eqref{GM} in an initially circular domain of radius $3$. The parameters used are $D_1=1$, $D_2=1000$, $a=0.01$, $b=0.5$, $c=5.5$, with a growth rate of $S = 0.001((u/u^*)^2-1) = 0.001((u/11.02)^2-1)$. Iterations are shown at equally spaced times, with panels (d) and (e) being the same plot but resized so that panels (a)-(d) are shown on the same scale and panels (e)-(h) are shown on the same scale.}
    \label{GMContract2DFig}
\end{figure}

\begin{figure}
    \centering
    \begin{overpic}[width=0.8\textwidth]{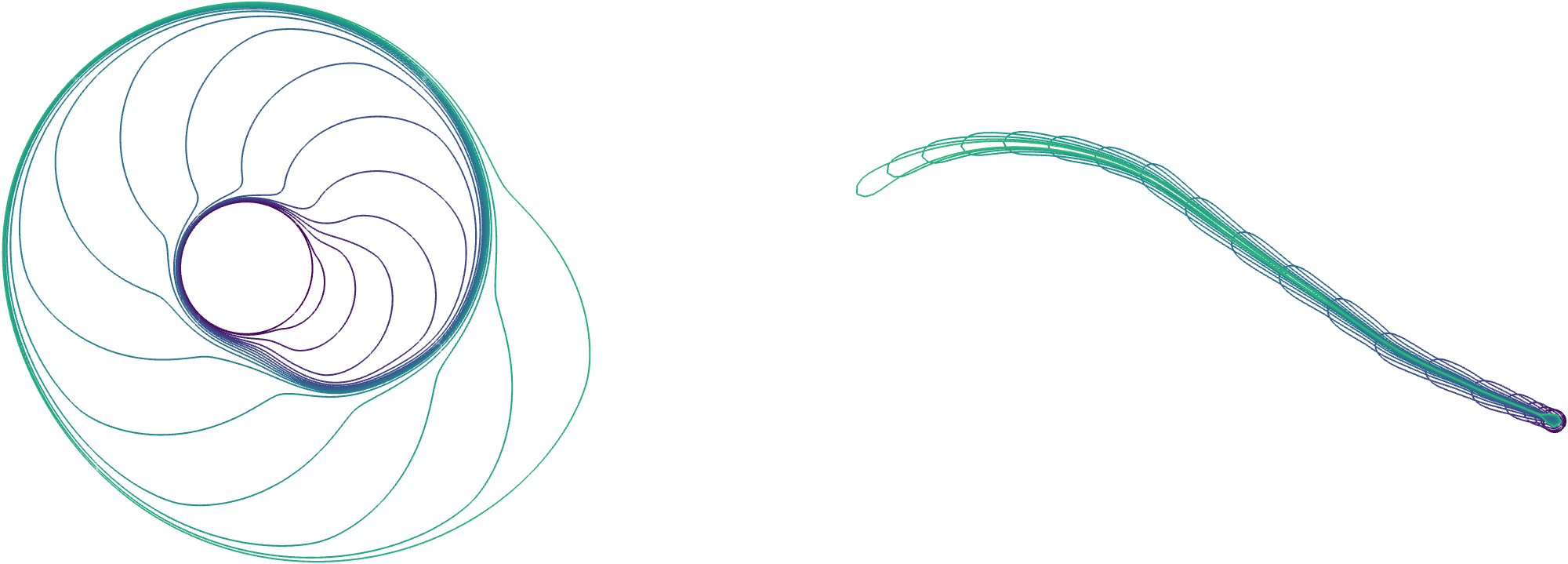}
    \put(15,-3){(a)}
    \put(75,-3){(b)}
    \end{overpic}
    \vspace{1em}
    \caption{Plots of the domain boundary from 2D simulations of the Gierer-Meinhardt kinetics \eqref{GM}. Panel (a) corresponds to Fig.~\ref{Spiral2DFig} and panel (b) to Fig.~\ref{GMContract2DFig}. Boundary curves shown are arranged so that larger enclosed areas correspond to later times, with times uniformly sampled. In panel (b), later boundary curves are to the left but do not enclose larger areas due to contraction towards the right side of the domain.}
    \label{SpiralAndGMBoundary2DFig}
\end{figure}

Next we consider an example of growth in the Gierer-Meinhardt system given by \eqref{RD_Eqnu} and the kinetics \eqref{GM}. We choose parameters so that growth is localized at a peak, and a sufficiently large diffusion ratio so that there is only one low-amplitude peak for a fixed circular domain. We use initial conditions as in the 1D setting of the form $u(0,x,y) = u^*(1+\eta(x,y)), v(0,x,y) = v^*(1+\xi(x,y))$ where for each spatial point $(x,y)$, the random variables $\eta$ and $\xi$ are normally distributed with zero mean and variance $10^{-2}$.

We show the evolution of the domain and concentration of $u$ in Fig.~\ref{Spiral2DFig}. From panels (a) and (b) we can see that there is initially growth at a particular point on the boundary determined by where the localized spot initially forms. From the remaining panels we can see that this spot curves around the domain, presumably following regions of high local curvature as has been shown in Gierer-Meinhardt systems on fixed domains \citep{iron2000dynamics}. The interplay between local growth and curvature-based movement leads to a spiraling motion of the spike around the original circular domain. We show the boundary curves correspinding to this case in Fig.~\ref{SpiralAndGMBoundary2DFig}(a), where we see a strikingly seashell-like pattern of growth. 

Next we consider a case of domain growth and contraction. As before we consider growth near the peak of the activator $u$, but assume the domain contracts away from this peak. We show plots from this case in Fig.~\ref{GMContract2DFig}. The short time dynamics are similar to the previous example, with a spot forming in one part of the domain and leading to a local protrusion between panels (a) and (b). However panels (c) and (d) clearly show that the domain is contracting away from the localized spot, leading to a fairly irregular shape. Over longer timescales the spot continues to grow whereas the `tail' region left behind shrinks leading to the strange shape given in panel (h). Over longer timescales a spot eventually forms in the bottom-right corner of panel (h), which then also begins growing. We show a plot of the boundary curves in this case in Fig.~\ref{SpiralAndGMBoundary2DFig}(b), but note that unlike all previous boundary curves in this paper, these involve intersections rather than monotonically growing regions. These intersections correspond to contraction of the domain with boundaries from earlier times.

\section{Discussion}\label{Discussion_Sect}

Primarily motivated by realistic coupling of domain evolution and morphogen dynamics in reaction-diffusion models of pattern formation, we have presented and explored a class of such models on  domains generated by  concentration-dependent growth. As one might expect, the possible dynamics in these cases can involve rich interactions between the domain and the reaction-diffusion system, leading to intricate bifurcation structures and captivating imagery, such as in the spiral of Fig.~\ref{SpiralAndGMBoundary2DFig}(a). Here we overview some of the key insights of this study, and outline potential fruitful directions for further work.

Considering domain evolution as a local process which expands or contracts space as defined by \eqref{dilution}-\eqref{growth_law}, there is essentially a unique 1D formulation of the model which we gave in Section \ref{1DModel}, and which was previously studied by \citep{seirin2011dynamics}. Here we considered this framework in regimes of large growth, and also in regimes of growth and contraction, highlighting complexity that can emerge due to the interactions of both domain size and local dilution. In particular, we demonstrated that while robust spike-doubling still typically occurs for sufficiently slow growth, fast concentration-dependent evolution, or domain contraction, could lead to a variety of unexpected phenomena. These include: modification of travelling-wave dynamics as in Fig.~\ref{TWFig}; saturating plateaus of activator as in Fig.~\ref{SchnackFig1}(f); bifurcations between growing, oscillating, and shrinking domains as in Fig.~\ref{SchnackFig2}; apparent locally-in-time linear growth as in Fig.~\ref{GMFig}; and additional complexity of the background state, given in Equation \eqref{u*}, as in Fig.~\ref{FHNFig}.  We also performed linear stability analysis of this 1D model, emphasising that it can be useful as a heuristic but provides minimal insight in complex growth regimes. Importantly, unlike in work by \cite{van_gorder_growth_2019} and others, there is no obvious extension of these linear stability results to higher-dimensional growth for these kinds of local-growth models due to the nontrivial choice of constitutive assumptions for $N>1$.

In two and more dimensions, specifying the local dynamics of the domain is insufficient to prescribe the domain's evolution in time. 
Here, we have posed irrotational growth and no tangential movement along the boundary, which may be interpreted as the flow being generated {\it solely} by point sources of density scaling with  $S$ and the impact of the boundary  constraint \citep{Bachelor1967}.  We demonstrated that this leads to domains that reduce boundary curvature over time for constant uniform local growth rates in Fig.~\ref{NonUniform2DFig}. In this 2D setting, we also demonstrated how even simple growth dynamics can give rise to nontrivial domain restructuring in Fig.~\ref{Spiral2DFig}, and that growth and contraction can lead to more exotic phenomena as in Fig.~\ref{GMContract2DFig}.

 We anticipate that many more interesting phenomena can be found, or constructed as in \citep{woolley2021bespoke}, and have made available an open source MATLAB code that can rapidly simulate these systems in 1D \citep{Krause_Concentration_Dependent_Growth_2022}, as well as a COMSOL LiveLInk code in the 2D case. In the 1D setting especially, we anticipate that the autonomous system given by \eqref{RD_Eqnu1D}-\eqref{growth_law_1D} can be investigated directly using various tools from nonlinear dynamics and the analysis of PDEs. If $S$ does not depend explicitly on time, as in all of the simulations reported here, the system is autonomous and in principle not much more complicated than many existing models of nonlinear diffusion and non-diffusible morphogen systems. In principle tools such as shadow-limits, spatial dynamics, and numerical continuation can all be applied to such a system. The travelling-wave case studied briefly in Section \ref{TW_Sect} is one topic of current further work, and highlights several nontrivial aspects of this autonomous yet complicated concentration-dependence. \ak{In addition to past work on spike-splitting events, mesa-splitting patterns like those seen in Figs.~\ref{SchnackFig1Slices}(h)-(i) and \ref{FHNFig}, have also been studied analytically in the slow-growth regime (see Figure 1 and the analysis in \cite{kolokolnikov2007self}).}
 
 While the 1D setting is interesting and plausibly tractable to various kinds of mathematical analyses, we also highlighted stark differences between both the modelling and such analysis in 1D and  models with higher spatial dimensions. These raise important questions regarding appropriate tissue-mechanical constitutive assumptions, which can have nontrivial impacts on the dynamics in 2D and 3D models, and where any mathematical analysis becomes substantially more complicated even if mechanical deformations are essentially neglected, as in our simple constitutive model of the flow $\bm{a}$. Related to this, many of the important insights regarding growth, such as the celebrated spike-doubling robustness shown by \cite{crampin1999reaction}, hold true in 1D models but may not lead to robust patterning in higher-dimensional settings even in slow growth regimes. \ak{For example, 2D stripes can undergo breakup or zigzag instabilities, plausibly ruining simple predictions of pattern-doubling \citep{kolokolnikov2006stability, krause2019influence}.} More work needs to be done in understanding which insights from simple 1D models may be applicable to realistic tissue geometries, especially for scenarios where morphogens influence the geometrical evolution of the domain.

\begin{acknowledgements}
B.J.W. is supported by the Royal Commission for the Exhibition of 1851, and the UK Engineering and Physical Sciences Research
Council (EPSRC), grant EP/N509711/1.
\end{acknowledgements}

\bibliographystyle{plainnat}
\bibliography{refs}

\begin{thebibliography}{92}
\providecommand{\natexlab}[1]{#1}
\providecommand{\url}[1]{\texttt{#1}}
\expandafter\ifx\csname urlstyle\endcsname\relax
  \providecommand{\doi}[1]{doi: #1}\else
  \providecommand{\doi}{doi: \begingroup \urlstyle{rm}\Url}\fi

\bibitem[Adamer et~al.(2020)Adamer, Harrington, Gaffney, and
  Woolley]{adamer2020coloured}
Michael~F Adamer, Heather~A Harrington, Eamonn~A Gaffney, and Thomas~E Woolley.
\newblock Coloured noise from stochastic inflows in reaction--diffusion
  systems.
\newblock \emph{Bulletin of {M}athematical {B}iology}, 82\penalty0
  (4):\penalty0 1--28, 2020.

\bibitem[Baker and Maini(2007)]{baker2007mechanism}
Ruth~E Baker and Philip~K Maini.
\newblock A mechanism for morphogen-controlled domain growth.
\newblock \emph{Journal of mathematical biology}, 54\penalty0 (5):\penalty0
  597--622, 2007.

\bibitem[Bao et~al.(2018)Bao, Du, Lin, and Zhu]{bao2018free}
Wendi Bao, Yihong Du, Zhigui Lin, and Huaiping Zhu.
\newblock Free boundary models for mosquito range movement driven by climate
  warming.
\newblock \emph{Journal of Mathematical Biology}, 76\penalty0 (4):\penalty0
  841--875, 2018.

\bibitem[Barreira et~al.(2011)Barreira, Elliott, and
  Madzvamuse]{barreira2011surface}
Raquel Barreira, Charles~M Elliott, and Anotida Madzvamuse.
\newblock The surface finite element method for pattern formation on evolving
  biological surfaces.
\newblock \emph{Journal of mathematical biology}, 63\penalty0 (6):\penalty0
  1095--1119, 2011.

\bibitem[Batchelor(1967)]{Bachelor1967}
George~K Batchelor.
\newblock \emph{{An Introduction to Fluid Dynamics}}.
\newblock CUP, 1967.

\bibitem[Boehm et~al.(2010)Boehm, Westerberg, Lesnicar-Pucko, Raja, Rautschka,
  Cotterell, Swoger, and Sharpe]{boehm2010role}
Bernd Boehm, Henrik Westerberg, Gaja Lesnicar-Pucko, Sahdia Raja, Michael
  Rautschka, James Cotterell, Jim Swoger, and James Sharpe.
\newblock The role of spatially controlled cell proliferation in limb bud
  morphogenesis.
\newblock \emph{PLoS biology}, 8\penalty0 (7):\penalty0 e1000420, 2010.

\bibitem[Casten and Holland(1978)]{casten1978instability}
Richard~G Casten and Charles~J Holland.
\newblock Instability results for reaction diffusion equations with neumann
  boundary conditions.
\newblock \emph{Journal of Differential Equations}, 27\penalty0 (2):\penalty0
  266--273, 1978.

\bibitem[Chafee and Infante(1974)]{chafee1974bifurcation}
Nathaniel Chafee and Ettore~Ferrari Infante.
\newblock A bifurcation problem for a nonlinear partial differential equation
  of parabolic type.
\newblock \emph{Applicable Analysis}, 4\penalty0 (1):\penalty0 17--37, 1974.

\bibitem[Chen and Lowengrub(2014)]{chen2014tumor}
Ying Chen and John~S Lowengrub.
\newblock Tumor growth in complex, evolving microenvironmental geometries: a
  diffuse domain approach.
\newblock \emph{Journal of theoretical biology}, 361:\penalty0 14--30, 2014.

\bibitem[Chou and Zhu(2001)]{chou2001curve}
Kai-Seng Chou and Xi-Ping Zhu.
\newblock \emph{The curve shortening problem}.
\newblock Chapman and Hall/CRC, 2001.

\bibitem[Chow et~al.(2006)Chow, Lu, and Ni]{chow2006hamilton}
Bennett Chow, Peng Lu, and Lei Ni.
\newblock \emph{Hamilton's Ricci flow}, volume~77.
\newblock American Mathematical Soc., 2006.

\bibitem[Crampin et~al.(1999)Crampin, Gaffney, and Maini]{crampin1999reaction}
Edmund~J Crampin, Eamonn~A Gaffney, and Philip~K Maini.
\newblock Reaction and diffusion on growing domains: scenarios for robust
  pattern formation.
\newblock \emph{Bulletin of {M}athematical {B}iology}, 61\penalty0
  (6):\penalty0 1093--1120, 1999.

\bibitem[Crampin et~al.(2002{\natexlab{a}})Crampin, Gaffney, and
  Maini]{crampin2002mode}
Edmund~J Crampin, Eamonn~A Gaffney, and Philip~K Maini.
\newblock Mode-doubling and tripling in reaction-diffusion patterns on growing
  domains: A piecewise linear model.
\newblock \emph{Journal of mathematical biology}, 44\penalty0 (2):\penalty0
  107--128, 2002{\natexlab{a}}.

\bibitem[Crampin et~al.(2002{\natexlab{b}})Crampin, Hackborn, and
  Maini]{crampin2002pattern}
Edmund~J Crampin, William~W Hackborn, and Philip~K Maini.
\newblock Pattern formation in reaction-diffusion models with nonuniform domain
  growth.
\newblock \emph{Bulletin of {M}athematical {B}iology}, 64\penalty0
  (4):\penalty0 747--769, 2002{\natexlab{b}}.

\bibitem[Diego et~al.(2018)Diego, Marcon, M{\"u}ller, and Sharpe]{diego2018key}
Xavier Diego, Luciano Marcon, Patrick M{\"u}ller, and James Sharpe.
\newblock Key features of {T}uring systems are determined purely by network
  topology.
\newblock \emph{Physical {R}eview {X}}, 8\penalty0 (2):\penalty0 021071, 2018.

\bibitem[Dillon and Othmer(1999)]{dillon1999mathematical}
Robert Dillon and Hans~G Othmer.
\newblock A mathematical model for outgrowth and spatial patterning of the
  vertebrate limb bud.
\newblock \emph{Journal of theoretical biology}, 197\penalty0 (3):\penalty0
  295--330, 1999.

\bibitem[Du and Lin(2010)]{du2010spreading}
Yihong Du and Zhigui Lin.
\newblock Spreading-vanishing dichotomy in the diffusive logistic model with a
  free boundary.
\newblock \emph{SIAM Journal on Mathematical Analysis}, 42\penalty0
  (1):\penalty0 377--405, 2010.

\bibitem[Dziuk and Elliott(2013)]{dziuk2013finite}
Gerhard Dziuk and Charles~M Elliott.
\newblock Finite element methods for surface pdes.
\newblock \emph{Acta Numerica}, 22:\penalty0 289--396, 2013.

\bibitem[El-Hachem et~al.(2019)El-Hachem, McCue, Jin, Du, and
  Simpson]{el2019revisiting}
Maud El-Hachem, Scott~W McCue, Wang Jin, Yihong Du, and Matthew~J Simpson.
\newblock Revisiting the fisher--kolmogorov--petrovsky--piskunov equation to
  interpret the spreading--extinction dichotomy.
\newblock \emph{Proceedings of the Royal Society A}, 475\penalty0
  (2229):\penalty0 20190378, 2019.

\bibitem[Erban and Chapman(2019)]{erban2019stochastic}
Radek Erban and S~Jonathan Chapman.
\newblock \emph{Stochastic modelling of reaction--diffusion processes},
  volume~60.
\newblock Cambridge {U}niversity {P}ress, 2019.

\bibitem[Fisher(1937)]{fisher1937wave}
Ronald~Aylmer Fisher.
\newblock The wave of advance of advantageous genes.
\newblock \emph{Annals of eugenics}, 7\penalty0 (4):\penalty0 355--369, 1937.

\bibitem[FitzHugh(1955)]{fitzhugh1955mathematical}
Richard FitzHugh.
\newblock Mathematical models of threshold phenomena in the nerve membrane.
\newblock \emph{The {B}ulletin of {M}athematical {B}iophysics}, 17\penalty0
  (4):\penalty0 257--278, 1955.

\bibitem[Fletcher and Osborne(2022)]{fletcher2022seven}
Alexander~G Fletcher and James~M Osborne.
\newblock Seven challenges in the multiscale modeling of multicellular tissues.
\newblock \emph{WIREs mechanisms of disease}, 14\penalty0 (1):\penalty0 e1527,
  2022.

\bibitem[Gaffney and Monk(2006)]{gaffney2006gene}
Eamonn~A Gaffney and NAM Monk.
\newblock Gene expression time delays and {T}uring pattern formation systems.
\newblock \emph{Bulletin of {M}athematical {B}iology}, 68\penalty0
  (1):\penalty0 99--130, 2006.

\bibitem[Gierer and Meinhardt(1972)]{gierer1972theory}
A.~Gierer and H.~Meinhardt.
\newblock A theory of biological pattern formation.
\newblock \emph{Kybernetik}, 12\penalty0 (1):\penalty0 30--39, 1972.

\bibitem[Gokieli and Varchon(2005)]{gokieli2005reaction}
Maria Gokieli and Nicolas Varchon.
\newblock The reaction-diffusion problem on dumbbell domains.
\newblock \emph{Gakuto International Series, Mathematical Sciences and
  Applications}, 2:\penalty0 55--72, 2005.

\bibitem[Green and Sharpe(2015)]{green2015positional}
Jeremy B~A Green and James Sharpe.
\newblock Positional information and reaction-diffusion: two big ideas in
  developmental biology combine.
\newblock \emph{Development}, 142\penalty0 (7):\penalty0 1203--1211, 2015.

\bibitem[Groves et~al.(2020)Groves, Placzek, and Fletcher]{groves2020mitogens}
Ian Groves, Marysia Placzek, and Alexander~G Fletcher.
\newblock Of mitogens and morphogens: modelling sonic hedgehog mechanisms in
  vertebrate development.
\newblock \emph{Philosophical Transactions of the Royal Society B},
  375\penalty0 (1809):\penalty0 20190660, 2020.

\bibitem[Hadeler(2016)]{hadeler2016stefan}
Karl~P Hadeler.
\newblock Stefan problem, traveling fronts, and epidemic spread.
\newblock \emph{Discrete \& Continuous Dynamical Systems-B}, 21\penalty0
  (2):\penalty0 417, 2016.

\bibitem[Hess(1989)]{hess1989periodic}
Peter Hess.
\newblock On periodic-parabolic boundary value problems.
\newblock In \emph{Differential Equations: Proceedings of the 1987 Equadiff
  Conference}, volume 118, page 311. CRC Press, 1989.

\bibitem[Ide et~al.(2016)Ide, Izuhara, and Machida]{ide2016turing}
Yusuke Ide, Hirofumi Izuhara, and Takuya Machida.
\newblock Turing instability in reaction--diffusion models on complex networks.
\newblock \emph{Physica {A}:{S}Statistical {M}echanics and its {A}pplications},
  457:\penalty0 331--347, 2016.

\bibitem[Iron and Ward(2000)]{iron2000dynamics}
David Iron and Michael~J Ward.
\newblock The dynamics of boundary spikes for a nonlocal reaction-diffusion
  model.
\newblock \emph{European Journal of Applied Mathematics}, 11\penalty0
  (5):\penalty0 491--514, 2000.

\bibitem[Jepson et~al.(2022)Jepson, Fadai, and O’Dea]{jepson2022travelling}
Jacob~M Jepson, Nabil~T Fadai, and Reuben~D O’Dea.
\newblock Travelling-wave and asymptotic analysis of a multiphase moving
  boundary model for engineered tissue growth.
\newblock \emph{Bulletin of Mathematical Biology}, 84\penalty0 (8):\penalty0
  1--19, 2022.

\bibitem[Keener(2021)]{keener2021biology}
James~P Keener.
\newblock \emph{Biology in Time and Space: A Partial Differential Equation
  Modeling Approach}, volume~50.
\newblock American Mathematical Soc., 2021.

\bibitem[Kolokolnikov et~al.(2007)Kolokolnikov, Ward, and
  Wei]{kolokolnikov2007self}
T~Kolokolnikov, MJ~Ward, and J~Wei.
\newblock Self-replication of mesa patterns in reaction--diffusion systems.
\newblock \emph{Physica D: Nonlinear Phenomena}, 236\penalty0 (2):\penalty0
  104--122, 2007.

\bibitem[Kolokolnikov et~al.(2006)Kolokolnikov, Sun, Ward, and
  Wei]{kolokolnikov2006stability}
Theodore Kolokolnikov, Wentao Sun, Michael Ward, and Juncheng Wei.
\newblock The stability of a stripe for the gierer--meinhardt model and the
  effect of saturation.
\newblock \emph{SIAM Journal on Applied Dynamical Systems}, 5\penalty0
  (2):\penalty0 313--363, 2006.

\bibitem[Konow et~al.(2021)Konow, Dolnik, and Epstein]{konow2021insights}
Christopher Konow, M~Dolnik, and IR~Epstein.
\newblock Insights from chemical systems into turing-type morphogenesis.
\newblock \emph{Philosophical Transactions of the Royal Society A},
  379\penalty0 (2213):\penalty0 20200269, 2021.

\bibitem[Krause and Van~Gorder(2020)]{krause2020non}
Andrew~L Krause and Robert~A Van~Gorder.
\newblock A non-local cross-diffusion model of population dynamics ii: Exact,
  approximate, and numerical traveling waves in single-and multi-species
  populations.
\newblock \emph{Bulletin of Mathematical Biology}, 82\penalty0 (8):\penalty0
  1--30, 2020.

\bibitem[Krause et~al.(2018)Krause, Klika, Woolley, and
  Gaffney]{krause2018heterogeneity}
Andrew~L Krause, V.~Klika, Thomas~E Woolley, and Eamonn~A Gaffney.
\newblock Heterogeneity induces spatiotemporal oscillations in
  reaction-diffusion systems.
\newblock \emph{Physical {R}eview {E}}, 97\penalty0 (5):\penalty0 052206, 2018.

\bibitem[Krause et~al.(2019)Krause, Ellis, and Van~Gorder]{krause2019influence}
Andrew~L Krause, M.~A. Ellis, and Robert~A Van~Gorder.
\newblock Influence of curvature, growth, and anisotropy on the evolution of
  {T}uring patterns on growing manifolds.
\newblock \emph{Bulletin of {M}athematical {B}iology}, 81\penalty0
  (3):\penalty0 759--799, 2019.

\bibitem[Krause et~al.(2020)Krause, Klika, Woolley, and Gaffney]{krause_WKB}
Andrew~L Krause, V.~Klika, Thomas~E Woolley, and Eamonn~A Gaffney.
\newblock From one pattern into another: Analysis of {T}uring patterns in
  heterogeneous domains via {W}{K}{B}{J}.
\newblock \emph{Journal of the {R}oyal {S}ociety {I}nterface}, 17:\penalty0
  20190621, 2020.

\bibitem[Krause et~al.(2021{\natexlab{a}})Krause, Gaffney, Maini, and
  Klika]{krause_near_2021}
Andrew~L Krause, Eamonn~A Gaffney, Philip~K Maini, and V\'{a}clav Klika.
\newblock Modern perspectives on near-equilibrium analysis of {T}uring systems.
\newblock \emph{Philosophical Transactions of the Royal Society A:
  Mathematical, Physical and Engineering Sciences}, 379\penalty0 (2213),
  2021{\natexlab{a}}.

\bibitem[Krause et~al.(2021{\natexlab{b}})Krause, Klika, Maini, Headon, and
  Gaffney]{krause2020isolating}
Andrew~L Krause, V.~Klika, Philip~K Maini, D.~Headon, and Eamonn~A Gaffney.
\newblock Isolating patterns in open reaction-diffusion systems.
\newblock \emph{Bulletin of {M}athematical {B}iology}, 83\penalty0
  (7):\penalty0 1--35, 2021{\natexlab{b}}.

\bibitem[Krause et~al.(2022)Krause, Gaffney, and
  Walker]{Krause_Concentration_Dependent_Growth_2022}
Andrew~L Krause, Eamonn~A Gaffney, and Benjamin Walker.
\newblock {Concentration Dependent Growth Simulations}, 2022.
\newblock URL
  \url{https://github.com/AndrewLKrause/Concentration-Dependent-Growth-Simulations}.

\bibitem[Landman et~al.(2003)Landman, Pettet, and
  Newgreen]{landman2003mathematical}
Kerry~A Landman, Graeme~J Pettet, and Donald~F Newgreen.
\newblock Mathematical models of cell colonization of uniformly growing
  domains.
\newblock \emph{Bulletin of Mathematical Biology}, 65\penalty0 (2):\penalty0
  235--262, 2003.

\bibitem[Liu et~al.(2022)Liu, Maini, and Baker]{liu2022control}
Yue Liu, Philip~K Maini, and Ruth~E Baker.
\newblock Control of diffusion-driven pattern formation behind a wave of
  competency.
\newblock \emph{Physica D: Nonlinear Phenomena}, page 133297, 2022.

\bibitem[MacDonald et~al.(2016)MacDonald, Mackenzie, Nolan, and
  Insall]{macdonald2016computational}
G~MacDonald, John~A Mackenzie, M~Nolan, and RH~Insall.
\newblock A computational method for the coupled solution of
  reaction--diffusion equations on evolving domains and manifolds: Application
  to a model of cell migration and chemotaxis.
\newblock \emph{Journal of computational physics}, 309:\penalty0 207--226,
  2016.

\bibitem[MacKenzie et~al.(2021)MacKenzie, Rowlatt, and
  Insall]{mackenzie2021conservative}
John MacKenzie, Christopher Rowlatt, and Robert Insall.
\newblock A conservative finite element ale scheme for mass-conservative
  reaction-diffusion equations on evolving two-dimensional domains.
\newblock \emph{SIAM Journal on Scientific Computing}, 43\penalty0
  (1):\penalty0 B132--B166, 2021.

\bibitem[Madzvamuse et~al.(2010)Madzvamuse, Gaffney, and
  Maini]{madzvamuse2010stability}
Anotida Madzvamuse, Eamonn~A Gaffney, and Philip~K Maini.
\newblock Stability analysis of non-autonomous reaction-diffusion systems: the
  effects of growing domains.
\newblock \emph{Journal of {M}athematical {B}iology}, 61\penalty0 (1):\penalty0
  133--164, 2010.

\bibitem[Maini(1995)]{maini1995hierarchical}
Philip~K Maini.
\newblock Hierarchical models for spatial pattern formation in biology.
\newblock \emph{Journal of Biological Systems}, 3\penalty0 (04):\penalty0
  987--997, 1995.

\bibitem[Maini et~al.(2012)Maini, Woolley, Baker, Gaffney, and
  Lee]{maini2012turing}
Philip~K Maini, Thomas~E Woolley, Ruth~E Baker, Eamonn~A Gaffney, and Seirin~S
  Lee.
\newblock Turing's model for biological pattern formation and the robustness
  problem.
\newblock \emph{Interface {F}ocus}, 2\penalty0 (4):\penalty0 487--496, 2012.

\bibitem[Matano(1979)]{matano1979asymptotic}
Hiroshi Matano.
\newblock Asymptotic behavior and stability of solutions of semilinear
  diffusion equations.
\newblock \emph{Publications of the Research Institute for Mathematical
  Sciences}, 15\penalty0 (2):\penalty0 401--454, 1979.

\bibitem[Matano and Mimura(1983)]{matano1983pattern}
Hiroshi Matano and Masayasu Mimura.
\newblock Pattern formation in competition-diffusion systems in nonconvex
  domains.
\newblock \emph{Publications of the Research Institute for Mathematical
  Sciences}, 19\penalty0 (3):\penalty0 1049--1079, 1983.

\bibitem[McCullen and Wagenknecht(2016)]{mccullen2016pattern}
Nick McCullen and Thomas Wagenknecht.
\newblock Pattern formation on networks: From localised activity to {T}uring
  patterns.
\newblock \emph{Scientific {R}eports}, 6\penalty0 (1):\penalty0 1--8, 2016.

\bibitem[Metzcar et~al.(2019)Metzcar, Wang, Heiland, and
  Macklin]{metzcar2019review}
John Metzcar, Yafei Wang, Randy Heiland, and Paul Macklin.
\newblock A review of cell-based computational modeling in cancer biology.
\newblock \emph{JCO clinical cancer informatics}, 2:\penalty0 1--13, 2019.

\bibitem[Murphy et~al.(2021)Murphy, Buenzli, Baker, and
  Simpson]{murphy2021travelling}
Ryan~J Murphy, Pascal~R Buenzli, Ruth~E Baker, and Matthew~J Simpson.
\newblock Travelling waves in a free boundary mechanobiological model of an
  epithelial tissue.
\newblock \emph{Applied Mathematics Letters}, 111:\penalty0 106636, 2021.

\bibitem[Murray(2004)]{murray2004mathematical}
James~D Murray.
\newblock \emph{Mathematical {B}iology. {II}. Spatial models and biomedical
  applications}.
\newblock Interdisciplinary applied mathematics. Springer, New York, 2004.

\bibitem[Murray(2007)]{murray2007mathematical}
James~D Murray.
\newblock \emph{Mathematical biology: I. An introduction}.
\newblock Springer Science \& Business Media, 2007.

\bibitem[Murray and Oster(1984{\natexlab{a}})]{murray1984cell}
James~D Murray and George~F Oster.
\newblock Cell traction models for generating pattern and form in
  morphogenesis.
\newblock \emph{Journal of mathematical biology}, 19\penalty0 (3):\penalty0
  265--279, 1984{\natexlab{a}}.

\bibitem[Murray and Oster(1984{\natexlab{b}})]{murray1984generation}
James~D Murray and George~F Oster.
\newblock Generation of biological pattern and form.
\newblock \emph{Mathematical Medicine and Biology: A Journal of the IMA},
  1\penalty0 (1):\penalty0 51--75, 1984{\natexlab{b}}.

\bibitem[Myerscough and Murray(1992)]{myerscough1992analysis}
Mary~R Myerscough and James~D Murray.
\newblock Analysis of propagating pattern in a chemotaxis system.
\newblock \emph{Bulletin of mathematical biology}, 54\penalty0 (1):\penalty0
  77--94, 1992.

\bibitem[Nagumo et~al.(1962)Nagumo, Arimoto, and Yoshizawa]{nagumo1962active}
Jinichi Nagumo, Suguru Arimoto, and Shuji Yoshizawa.
\newblock An active pulse transmission line simulating nerve axon.
\newblock \emph{Proceedings of the {I}{R}{E}}, 50\penalty0 (10):\penalty0
  2061--2070, 1962.

\bibitem[Neville et~al.(2006)Neville, Matthews, and
  Byrne]{neville2006interactions}
Alex~A Neville, Paul~C Matthews, and Helen~M Byrne.
\newblock Interactions between pattern formation and domain growth.
\newblock \emph{Bulletin of mathematical biology}, 68\penalty0 (8):\penalty0
  1975--2003, 2006.

\bibitem[Osborne et~al.(2017)Osborne, Fletcher, Pitt-Francis, Maini, and
  Gavaghan]{osborne2017comparing}
James~M Osborne, Alexander~G Fletcher, Joe~M Pitt-Francis, Philip~K Maini, and
  David~J Gavaghan.
\newblock Comparing individual-based approaches to modelling the
  self-organization of multicellular tissues.
\newblock \emph{PLoS computational biology}, 13\penalty0 (2):\penalty0
  e1005387, 2017.

\bibitem[Oster et~al.(1985)Oster, Murray, and Maini]{oster1985model}
George~F Oster, James~D Murray, and Philip~K Maini.
\newblock A model for chondrogenic condensations in the developing limb: the
  role of extracellular matrix and cell tractions.
\newblock \emph{Journal of Embryology and Experimental Morphology}, 1985.

\bibitem[Page et~al.(2003)Page, Maini, and Monk]{page2003pattern}
Karen~M Page, Philip~K Maini, and Nicholas A~M Monk.
\newblock Pattern formation in spatially heterogeneous {T}uring
  reaction--diffusion models.
\newblock \emph{Physica {D}: {N}onlinear {P}henomena}, 181\penalty0
  (1-2):\penalty0 80--101, 2003.

\bibitem[Page et~al.(2005)Page, Maini, and Monk]{page2005complex}
Karen~M Page, Philip~K Maini, and Nicholas A~M Monk.
\newblock Complex pattern formation in reaction--diffusion systems with
  spatially varying parameters.
\newblock \emph{Physica {D}: {N}onlinear {P}henomena}, 202\penalty0
  (1-2):\penalty0 95--115, 2005.

\bibitem[Plaza et~al.(2004)Plaza, Sanchez-Garduno, Padilla, Barrio, and
  Maini]{plaza2004effect}
Ram{\'o}n~G Plaza, Faustino Sanchez-Garduno, Pablo Padilla, Rafael~A Barrio,
  and Philip~K Maini.
\newblock The effect of growth and curvature on pattern formation.
\newblock \emph{Journal of {D}ynamics and {D}ifferential {E}quations},
  16\penalty0 (4):\penalty0 1093--1121, 2004.

\bibitem[Ritchie et~al.(2022)Ritchie, Krause, and
  Van~Gorder]{ritchie_hyperbolic_2020}
Joshua~S Ritchie, Andrew~L Krause, and Robert~A Van~Gorder.
\newblock Turing and wave instabilities in hyperbolic reaction-diffusion
  systems: The role of second-order time derivatives and cross-diffusion terms
  on pattern formation.
\newblock \emph{Annals of Physics}, 444:\penalty0 169033, 2022.
\newblock ISSN 0003-4916.

\bibitem[S{\'a}nchez-Garduno et~al.(2019)S{\'a}nchez-Garduno, Krause, Castillo,
  and Padilla]{sanchez2019turing}
Faustino S{\'a}nchez-Garduno, Andrew~L Krause, Jorge~A Castillo, and Pablo
  Padilla.
\newblock Turing--{H}opf patterns on growing domains: the torus and the sphere.
\newblock \emph{Journal of {T}heoretical {B}iology}, 481:\penalty0 136--150,
  2019.

\bibitem[Sargood et~al.(2022)Sargood, Gaffney, and Krause]{sargood2022fixed}
Alec Sargood, Eamonn~A Gaffney, and Andrew~L Krause.
\newblock Fixed and distributed gene expression time delays in
  reaction-diffusion systems.
\newblock \emph{Bulletin of Mathematical Biology}, 83, 2022.

\bibitem[Schnakenberg(1979)]{schnakenberg1979simple}
J~Schnakenberg.
\newblock Simple chemical reaction systems with limit cycle behaviour.
\newblock \emph{Journal of {T}heoretical {B}iology}, 81\penalty0 (3):\penalty0
  389--400, 1979.

\bibitem[Scholes et~al.(2019)Scholes, Schnoerr, Isalan, and
  Stumpf]{scholes2019comprehensive}
Natalie~S Scholes, David Schnoerr, Mark Isalan, and Michael~PH Stumpf.
\newblock A comprehensive network atlas reveals that {T}uring patterns are
  common but not robust.
\newblock \emph{Cell systems}, 9\penalty0 (3):\penalty0 243--257, 2019.

\bibitem[Seirin~Lee et~al.(2010)Seirin~Lee, Gaffney, and
  Monk]{seirin2010influence}
S~Seirin~Lee, Eamonn~A Gaffney, and Nicholas A~M Monk.
\newblock The influence of gene expression time delays on gierer--meinhardt
  pattern formation systems.
\newblock \emph{Bulletin of mathematical biology}, 72\penalty0 (8):\penalty0
  2139--2160, 2010.

\bibitem[Seirin~Lee et~al.(2011{\natexlab{a}})Seirin~Lee, Gaffney, and
  Baker]{lee2011dynamics}
S~Seirin~Lee, Eamonn~A Gaffney, and Ruth~E Baker.
\newblock The dynamics of {T}uring patterns for morphogen-regulated growing
  domains with cellular response delays.
\newblock \emph{Bulletin of {M}athematical {B}iology}, 73\penalty0
  (11):\penalty0 2527--2551, 2011{\natexlab{a}}.

\bibitem[Seirin~Lee et~al.(2011{\natexlab{b}})Seirin~Lee, Gaffney, and
  Baker]{seirin2011dynamics}
S~Seirin~Lee, Eamonn~A Gaffney, and Ruth~E Baker.
\newblock The dynamics of turing patterns for morphogen-regulated growing
  domains with cellular response delays.
\newblock \emph{Bulletin of mathematical biology}, 73\penalty0 (11):\penalty0
  2527--2551, 2011{\natexlab{b}}.

\bibitem[Shampine and Reichelt(1997)]{Shampine1997}
Lawrence~F. Shampine and Mark~W. Reichelt.
\newblock {The MATLAB ODE Suite}.
\newblock \emph{SIAM Journal on Scientific Computing}, 18\penalty0
  (1):\penalty0 1--22, jan 1997.
\newblock ISSN 1064-8275.
\newblock \doi{10.1137/S1064827594276424}.
\newblock URL \url{http://epubs.siam.org/doi/10.1137/S1064827594276424}.

\bibitem[Sharma and Morgan(2016)]{sharma2016global}
Vandana Sharma and Jeff Morgan.
\newblock Global existence of solutions to reaction-diffusion systems with mass
  transport type boundary conditions.
\newblock \emph{SIAM Journal on Mathematical Analysis}, 48\penalty0
  (6):\penalty0 4202--4240, 2016.

\bibitem[Sharma and Prajapat(2021)]{sharma2021global}
Vandana Sharma and Jyotshana~V Prajapat.
\newblock Global existence of solutions to reaction diffusion systems with mass
  transport type boundary conditions on an evolving domain.
\newblock \emph{arXiv preprint arXiv:2102.00165}, 2021.

\bibitem[Sharpe(2017)]{sharpe2017computer}
James Sharpe.
\newblock Computer modeling in developmental biology: growing today, essential
  tomorrow.
\newblock \emph{Development}, 144\penalty0 (23):\penalty0 4214--4225, 2017.

\bibitem[Spiess et~al.(2022)Spiess, Fulton, Hwang, Toh, Saunders, Paige,
  Steventon, and Verd]{spiess2022approximated}
Kay Spiess, Timothy Fulton, Seogwon Hwang, Kane Toh, Dillan Saunders, Brooks
  Paige, Benjamin Steventon, and Berta Verd.
\newblock Approximated gene expression trajectories (agets) for gene regulatory
  network inference on cell tracks.
\newblock \emph{bioRxiv}, 2022.

\bibitem[Tam and Simpson(2022)]{tam2022pattern}
Alexander~KY Tam and Matthew~J Simpson.
\newblock Pattern formation and front stability for a moving-boundary model of
  biological invasion and recession.
\newblock \emph{arXiv preprint arXiv:2207.03053}, 2022.

\bibitem[Tauriello and Koumoutsakos(2013)]{tauriello2013coupling}
Gerardo Tauriello and Petros Koumoutsakos.
\newblock Coupling remeshed particle and phase field methods for the simulation
  of reaction-diffusion on the surface and the interior of deforming
  geometries.
\newblock \emph{SIAM Journal on Scientific Computing}, 35\penalty0
  (6):\penalty0 B1285--B1303, 2013.

\bibitem[Turing(1952)]{turing1952chemical}
Alan~M Turing.
\newblock The chemical basis of morphogenesis.
\newblock \emph{Philosophical Transactions of the Royal Society of London.
  Series B, Biological Sciences}, 237\penalty0 (641):\penalty0 37--72, 1952.

\bibitem[Ueda and Nishiura(2012)]{ueda2012mathematical}
Kei-Ichi Ueda and Yasumasa Nishiura.
\newblock A mathematical mechanism for instabilities in stripe formation on
  growing domains.
\newblock \emph{Physica {D}: {N}onlinear {P}henomena}, 241\penalty0
  (1):\penalty0 37--59, 2012.

\bibitem[Van~Gorder(2020)]{van2020turing}
Robert~A Van~Gorder.
\newblock Turing and {B}enjamin--{F}eir instability mechanisms in
  non-autonomous systems.
\newblock \emph{Proceedings of the {R}oyal {S}ociety {A}: {M}athematical,
  {P}hysical and {E}ngineering {S}ciences}, 476\penalty0 (2238):\penalty0
  20200003, 2020.

\bibitem[Van~Gorder et~al.(2021)Van~Gorder, Klika, and
  Krause]{van_gorder_growth_2019}
Robert~A Van~Gorder, V.~Klika, and Andrew~L Krause.
\newblock Turing conditions for pattern forming systems on evolving manifolds.
\newblock \emph{Journal of {M}athematical {B}iology}, 82\penalty0 (4), 2021.

\bibitem[Vaughan~Jr et~al.(2013)Vaughan~Jr, Baker, Kay, and
  Maini]{vaughan2013modified}
Benjamin~L Vaughan~Jr, Ruth~E Baker, David Kay, and Philip~K Maini.
\newblock A modified oster--murray--harris mechanical model of morphogenesis.
\newblock \emph{Siam Journal on Applied Mathematics}, 73\penalty0 (6):\penalty0
  2124--2142, 2013.

\bibitem[Ward and King(1997)]{ward1997mathematical}
John~P Ward and John~R King.
\newblock Mathematical modelling of avascular-tumour growth.
\newblock \emph{Mathematical Medicine and Biology: A Journal of the IMA},
  14\penalty0 (1):\penalty0 39--69, 1997.

\bibitem[Ward and Stafford(1999)]{ward1999metastable}
Michael~J Ward and Douglas Stafford.
\newblock Metastable dynamics and spatially inhomogeneous equilibria in
  dumbbell-shaped domains.
\newblock \emph{Studies in Applied Mathematics}, 103\penalty0 (1):\penalty0
  51--73, 1999.

\bibitem[Woolley et~al.(2011)Woolley, Baker, Gaffney, and
  Maini]{woolley2011stochastic}
Thomas~E Woolley, Ruth~E Baker, Eamonn~A Gaffney, and Philip~K Maini.
\newblock Stochastic reaction and diffusion on growing domains: understanding
  the breakdown of robust pattern formation.
\newblock \emph{Physical {R}eview {E}}, 84\penalty0 (4):\penalty0 046216, 2011.

\bibitem[Woolley et~al.(2021)Woolley, Krause, and Gaffney]{woolley2021bespoke}
Thomas~E Woolley, Andrew~L Krause, and Eamonn~A Gaffney.
\newblock Bespoke {T}uring systems.
\newblock \emph{Bulletin of {M}athematical {B}iology}, 83\penalty0
  (5):\penalty0 1--32, 2021.

\end{thebibliography}

\end{document}